\documentclass{amsart}

\usepackage{amsmath}
\usepackage{amssymb}
\usepackage{amsthm}
\usepackage{dsfont}

\usepackage{graphicx}
\usepackage{subfigure}
\graphicspath{{figures//}}

\usepackage{siunitx}

\usepackage{enumerate}

\usepackage{enumitem}

\usepackage{bbold}

\usepackage{bm}

\usepackage[colorlinks = true, linkcolor = blue, citecolor = blue, urlcolor = blue]{hyperref}

\usepackage{mathtools}

\usepackage{algorithm}
\usepackage{algorithmic}

\usepackage{ragged2e}

\usepackage{tikz}
\usetikzlibrary{fit}
\usetikzlibrary{matrix}

\usepackage{multirow}

\usepackage{xcolor}

\theoremstyle{remark}

\theoremstyle{definition}

\newcommand{\Z}{\mathbb{Z}}

\newcommand{\bO}{\mathcal{O}}

\newcommand{\bq}{\begin{equation}}
\newcommand{\eq}{\end{equation}}

\newcommand{\abs}[1]{\left\vert#1\right\vert}

\newcommand{\widthtwofigures}{0.48\textwidth}
\newcommand{\widththreefigures}{0.28\textwidth}

\usepackage{caption}

\newcommand{\dx}{\:dx}

\newcommand{\tr}{\text{tr}}

\newcommand{\boxalgorithm}[2]{
\begin{algorithm}#1
\begin{justifying}\noindent
#2
\end{justifying}
\end{algorithm}}

\makeatletter
\renewcommand*\env@matrix[1][*\c@MaxMatrixCols c]{%
  \hskip -\arraycolsep
  \let\@ifnextchar\new@ifnextchar
  \array{#1}}
\makeatother

\setlist[enumerate,1]{label={(\roman*)}}

\graphicspath{{figures/}{figures/Histograms2Dcomparison/}{figures/Histograms3Dcomparison/}{figures/Microstructures/}}

\begin{document}

\title{The Role of Surface Tension and Mobility Model\\ in Simulations of Grain Growth}

\author{Tiago Salvador}
\address{Department of Mathematics, Michigan University, 530 Church St. Ann Arbor, MI 48105 ({\tt saldanha@umich.edu}).}

\author{Selim Esedo\={g}lu}
\address{Department of Mathematics, Michigan University, 530 Church St. Ann Arbor, MI 48105 ({\tt esedoglu@umich.edu}).}

\date{\today}

\begin{abstract}
We explore the effects of surface tension and mobility models in simulations of grain growth using threshold dynamics algorithms that allow performing large scale simulations, while naturally capturing the Herring angle condition at junctions and automatically handling topological transitions.
The results indicate that in two dimensions, the different surface tension / mobility models considered do not play a significant role in the stationary grain size distribution.
However, in three dimensions, there is a substantial difference between the distributions obtained from the same three models, depending on whether the reduced mobilities are isotropic or anisotropic.
Additional results show that in three dimensions, the misorientation distribution function of a grain network with random orientation texture returns to the close vicinity of the Mackenzie distribution even if started very far from it.

\end{abstract}



\maketitle

\setcounter{tocdepth}{1}
\tableofcontents

\section{Introduction}

The shapes and sizes of grains constituting polycrystalline materials, such as most metals and ceramics, influence important physical attributes of these materials, for example their conductivity and yield strength.
Therefore, there has been much interest in modeling and simulation of grain boundary motion during common manufacturing processes, such as annealing.
In particular, the evolution of grain size distribution (GSD), the grain boundary character distribution (GBCD), as well as a simpler version of the latter, the misorientation distribution function (MDF), have been a central focus of research in materials science.

In this paper, we report simulation results for grain boundary motion at the mesoscale, using a recent, very concise and flexible numerical algorithm developed for the continuum model of grain boundaries that was introduced by Mullins \cite{MullinsGrainBoundaryMotion} and earlier authors.
According to this well known model, interfaces in the grain network evolve via steepest descent for an energy that penalizes surface areas of the interfaces, often weighted by a factor, known as the surface tension, that depends on the misorientation across each interface.
The resulting dynamics is described by a system of partial differential equations (PDEs), coupled to each other along junctions by what are known as Herring angle conditions \cite{HerringAngleCondition}, and allows specifying in addition a mobility factor for each interface.
The results we present are the first large scale simulations using the new algorithm proposed in \cite{SalvadorEsedogluSimplified} for this PDE system.

We investigate several questions that have repeatedly come up in previous literature concerning certain statistics of the grain network.
In particular, we study, in both two and three dimensions:
\begin{itemize}
\item The effect of surface tension and mobility model used on statistics such as the GSD. 
\item The asymptotic behavior of the MDF and its dependence on initial configuration.
\end{itemize}
Specifically, we simulate in both 2D and 3D Mullins' mesoscale model with 
\begin{enumerate}
\item Read-Shockley surface tensions \cite{ReadShockley} (and its extension \cite{HolmMisorientationAngles} to 3D anisotropy) along with isotropic mobilities, 
\item Read-Shockley surface tensions with ``reciprocal'' mobilities resulting in isotropic {\it reduced} mobilities but asymmetric junction angles, and
\item Isotropic surface tensions and mobilities, resulting in isotropic reduced mobilities as well as symmetric (all $120^\circ$) junction angles.
\end{enumerate}
We compare to experimental data whenever it is available in existing literature.
Some of our results and conclusions differ substantially from earlier studies.
Highlights of our findings are:
\begin{itemize}
\item In 2D simulations, the stationary GSD that is reached appears to not depend on the surface tension and mobility models used: it is essentially the same in all three surface tension / mobility combinations (i), (ii), (iii) used in our simulations.
This is in agreement with a variety of previous studies, e.g. \cite{upmanyu,Hallberg2014}.
\item However, in 3D, the stationary GSD reached is substantially different when Read-Shockley surface tensions along with isotropic mobilities (i) are used, compared to (ii) and (iii) that result in isotropic reduced mobilities.
This finding appears to be distinct from earlier studies, e.g. \cite{GruberMCPart1}.
\item Moreover, the different GSD reached in 3D by models (i) and (ii) that differ only in their choice of mobilities constitutes a contrary point to previous studies that suggest mobilities do not play a significant role in the long time behavior of the GSD.
\item The GSD reached by (ii) and (iii) are almost identical in both 2D and 3D, even though junction angles in these simulations are very different, suggesting that perhaps the isotropy vs. anisotropy of reduced mobilities is what is significant.
\item The GSD reached in our 3D simulations with surface tension / mobility combination (i) is not less consistent with experimental measurements \cite{rowenhurst,zhang,groeber}, which have substantial variation themselves.
\item Statistics of grain {\em shapes}, such as that of isoperimetric ratios (eccentricities), also display considerable dependence on the choice of surface tension / mobility model.
\item Models (i) \& (ii) allow us to study the asymptotic behavior of the MDF.
Confirming earlier studies, the two drastically different choices of mobility in (i) \& (ii) lead to almost identical behavior, both in 2D and 3D.
Distinct from  earlier studies, we exhibit 3D initial data with MDFs perturbed substantially away from the Mackenzie distribution and observe the MDF return to a very close vicinity of the Mackenzie distribution, using either (i) or (ii) as the surface tension / mobility model.
\end{itemize}

The code for the simulations is publicly available and can be found at \url{https://github.com/tiagosalvador/statisticsGBM}.

\section{Previous Work}

The exising literature on the statistics of grain shapes and sizes during annealing is absolutely vast, both from the experimental and the modeling and simulation side.
Our discussion will therefore be necessarily limited to a few representative studies closest to the questions addressed in the present paper.

There are a plethora of numerical algorithms developed for the simulation of Mullins' mesoscale model of grain boundary motion.
They include the Monte Carlo Potts, front tracking, phase field, level set, and threshold dynamics methods.
In principle, all these methods should converge to the same continuum model described by the same system of PDEs, provided that various parameters appearing in them are scaled appropriately, at least until topological changes start to take place (even then, one hopes there would be agreement in the vast majority of situations).
Nevertheless, there isn't always good agreement even among simulations.

Very large, well resolved simulations of normal grain growth were carried out by Mason et. al. \cite{MasonStatisticsGBM} using the front tracking method in both 2D and 3D.
The authors catalogue a large number of grain size and shape statistics,  for the two dimensional model, for the three dimensional model, and for two dimensional slices of the three dimensional model.
However, this work is restricted to the equal surface tension, equal mobility, i.e. the fully isotropic, version of Mullins' model.
That in particular means the exclusion of such statistics as the GBCD or the MDF which are predicated on variation of surface tension between different interfaces in the network.
Furthermore, left unadressed is the question of how sensitively these shape and size statistics depend on the surface tension and mobility model used in conjunction with Mullins' continuum model.

Upmanyu et. al. \cite{upmanyu} study the dependence of grain boundary statistics on the surface tension and mobility models used via Monte Carlo Potts and phase field simulations in 2D.
The GSDs they report, including when the surface tension, the mobility, and both are anisotropic, show little deviation from that of the isotropic model; they all follow a more or less log-normal distribution.
Their investigation of the MDF in the same vein reveals strong dependence on the surface tension model used, but not so much on mobilities:
The network is driven towards a prevalence of interfaces with low misorientations and therefore low surface tensions.
Other statistics, such as the number of neighbors (or faces), are also observed to have a much more pronounced dependence on surface tensions than mobilities.
Similar MC Potts simulations in \cite{HolmMisorientationAngles} also show little difference between the GSDs of isotropic vs. Read-Shockley surface tensions.

Gruber et. al. \cite{GruberMCPart1} carry out MC Potts simulations in 3D, using isotropic as well as Read-Shockley (anisotropic) surface tension models.
Among the grain properties they document are the GSD and the MDF.
They observe no difference between the stationary GSDs reached using isotropic vs. anisotropic surface tension models.
As regards the MDF, this study suggests surface tension anisotropy has a measurable effect on the steady state MDF reached in simulations that start from an initial data with random orientation texture, regardless of the dimension.
When the initial orientation texture is not random, no steady state for the MDF is observed in the case of anisotropic surface tensions: instead, a proliferation of low angle grain boundaries results in a gradual but persistent concentration of the MDF at the origin.

Kim et. al. \cite{Kim2006} carry out phase field simulations in 2D and 3D using isotropic surface tensions and mobilities. In 3D, the stationary GSD they reach (given in terms of the reduced equivalent radii) follows very closely the Hillert distribution.
On the other hand, the Monte-Carlo Potts simulations of Zoellner et. al. \cite{zoellner} in 3D, using isotropic surface tensions and mobilities, and starting from an initial data with random texture, results in a stationary GSD that is quite far from the Hillert distribution.
The more recent very large scale simulations of Miyoshi et. al. \cite{Miyoshi2017} in 3D using the phase field method also yielded a GSD similar to that of \cite{zoellner} that deviates substantially from the Hillert distribution.

The recent 2D simulations reported in \cite{MiyoshiAnisotropic} using a phase field method treat surface tension / mobility models (i) and (iii), as well as a third model that combines Read-Shockley surface tensions with mobilities that have a sigmoidal dependence on the misorientation angle as proposed by Humphreys \cite{Humphreys}.
For this third surface tension / mobility model, the stationary GSD they obtain is significantly different from those with models (i) \& (iii).
The stationary GSDs of models (i) \& (iii) are reported to be very similar.
\section{Our model and algorithm}\label{sec:model}

Our starting point is the curvature driven grain boundary motion model that goes back at least to Mullins \cite{MullinsGrainBoundaryMotion}.
According to this model, grain boundaries evolve via steepest descent for the energy
\bq\label{eq:energy}
E = \sum_{i<j} \sigma_{ij}\text{Area}(\Gamma_{ij}) 
\eq
where $\Gamma_{i,j}$ denotes the boundary between two adjacent grains $\Sigma_i$ and $\Sigma_j$. 
In particular we will neglect the normal dependence of the energy density, but allow each grain boundary $\Gamma_{i,j}$ to have a distinct surface tension $\sigma_{i,j}$.
The very special case $\sigma_{i,j} = 1$ for all $i\not= j$ is often referred to as the {\em isotropic} surface tension model.
We will also be interested in the surface tension model due to Read \& Shockley \cite{ReadShockley} and its extension \cite{HolmMisorientationAngles} to 3D crystallography:
\bq\label{eq:RSmodel}
\sigma_{ij} = \begin{cases}
\frac{\theta_{ij}}{\theta_*}\left(1-\log\left(\frac{\theta_{ij}}{\theta_*}\right)\right)	& \text{if } \theta_{ij} \in [0,\theta_*],\\
1	& \text{if } \theta_{ij} \geq \theta_*.
\end{cases}
\eq
where $\theta_{i,j}$ is the misorientation angle between $\Sigma_i$ and $\Sigma_j$ and $\theta_*$ is the  Brandon angle \cite{BrandonHighAngle} , often taken to be between $15^\circ$ and $30^\circ$.
Following the convention in materials science literature, we will refer to (\ref{eq:RSmodel}) as an {\em anisotropic} surface tension model, even though there is still no dependence on the normal to the interface.

In two dimensions, each grain $\Sigma_i$ is modeled as a square lattice and is assigned an orientation angle $\theta_i$ of a clockwise rotation about the origin that maps it back to the standard two-dimensional lattice $\Z^2$. Thus, the misorientation between two grains $\Sigma_i$ and $\Sigma_j$ is given by
\bq\label{eq:misorentation_angle2D}
\theta_{i,j} = \min_{k\in\Z} \abs{\theta_i-\theta_j+k\frac{\pi}{2}}.
\eq
In three dimensions, the orientation of a grain with cubic lattice can be described (nonuniquely) by a matrix $g \in SO(3)$ that corresponds to the rotation required to obtain the lattice of the grain from the standard integer lattice $\Z^3$. The misorientation angle between two grains $\Sigma_i$ and $\Sigma_j$ is defined as
\bq\label{eq:misorentation_angle3D}
\theta_{i,j} = \min_{r\in\bO} \arccos\left(\frac{1}{2}(\tr( rg_ig_j^T)-1)\right),
\eq
where $\bO$ denotes the octahedral group (of symmetries of the cube in the three dimensions).

According to Mullins, the dynamics associated with energy \eqref{eq:energy} is given by $L^2$ gradient descent for the interfaces, leading to the normal speed
\bq\label{eq:normalspeed}
v_{i,j} = \mu_{i,j} \sigma_{i,j} \kappa_{i,j}
\eq
for interface  $\Gamma_{i,j}$, where $\kappa_{i,j}$ denotes the mean curvature of $\Gamma_{i,j}$ and $\mu_{i,j}>0$ is a mobility factor. In addition, a condition known as the \emph{Herring angle condition} \cite{HerringAngleCondition} holds along triple junctions. At a junction formed by the meeting of the three distinct grains $\Sigma_i$, $\Sigma_j$ and $\Sigma_k$, this condition reads
\bq\label{eq:YoungsLaw}
\sigma_{i,j} n_{i,j} + \sigma_{j,k} n_{j,k} + \sigma_{k,i} n_{k,i} = 0.
\eq
As a consequence the angles formed by normals $n_{i,j}$, $n_{j,k}$ and $n_{k,i}$ to the three interfaces $\Gamma_{i,j}$, $\Gamma_{j,k}$ and $\Gamma_{k,i}$ along the triple junction are determined by their associated surface tensions; this relation is also known as Young's law.

The algorithms used in this study are obtained from a non-local approximation to energy \eqref{eq:energy}
\bq\label{eq:approximate_energy}
\frac{1}{\sqrt{\delta t}} \sum_{i<j} \sigma_{i,j} \int_{\Sigma_i} G_{\sqrt{\delta t}} \ast \mathbb{1}_{\Sigma_j} \dx
\eq
where $G_t$ denotes the Gaussian kernel
\[
G_{\sqrt{t}}(x) = \frac{1}{(4\pi t)^{d/2}}\exp\left(-\frac{\abs{x}^2}{4t}\right)
\]
and $\mathbb{1}_\Sigma(x)$ denotes the characteristic function for a set $\Sigma$:
\[
\mathbb{1}_\Sigma(x) = \begin{cases}
1 & \text{if } x\in\Sigma,\\
0 & \text{otherwise}.
\end{cases}
\]
The width $\delta t$ of the Gaussian kernel appearing in \eqref{eq:approximate_energy} ends up playing the role of the time step size for our scheme, described below, that approximates gradient decent of \eqref{eq:energy} in $L^2$ sense, as described by \cite{MullinsGrainBoundaryMotion}. Energy \eqref{eq:approximate_energy} has been proposed in \cite{SelimFelix} and has been shown to converge in a very precise sense (namely, that of Gamma-convergence) to energy \eqref{eq:energy} in the limit $\delta t\to 0^+$. Intuitively, we have
\[
\frac{1}{\sqrt{\delta t}} \int_{\Sigma_i} G_{\sqrt{\delta t}} \ast \mathbb{1}_{\Sigma_j} \dx \approx \frac{1}{\sqrt{\pi}}\text{Area}(\Gamma_{i,j})
\]
since the function $\frac{1}{\sqrt{\delta t}} \mathbb{1}_{\Sigma_i} G_{\sqrt{\delta t}} \ast \mathbb{1}_{\Sigma_j}$ approximates a delta function concentrating near $\Gamma_{i,j}$ as $\delta t \to 0^+$. The reason for our interest in this specific - perhaps unusual - approximation of Mullins' energy is that it generates exceptionally simple and efficient algorithms for simulating gradient descent dynamics associated with \eqref{eq:energy}. Indeed, it has been shown in \cite{SelimFelix} to lead, in a systematic way, to the correct multiphase, arbitrary surface tension analogue of a very fast algorithm known as \emph{threshold dynamics} that was originally proposed in \cite{MBO92,MBO94} for networks with all equal surface tensions (i.e., $\theta_{i,j} = 1$ for all $i$ and $j$) and mobilities.

We start by recalling Algorithm \ref{alg:EO}, the simplest version of the generalization of threshold dynamics to arbitrary surface tensions given in \cite{SelimFelix}. Its benefits include its unconditional stability (time step size $\delta t$ can be chosen arbitrarily large, constrained only by accuracy considerations), seamless handling of topological changes in any dimension and very low per time step cost $\bO(N\log N)$ on a uniform grid of $N$ points, using the FFT to compute the convolutions. Yet, despite its extreme simplicity, it handles automatically and correctly all the essential features of the dynamics, including Herring angle conditions \eqref{eq:YoungsLaw} along junctions, and countless types of topological changes that may occur during the evolution. Note that Algorithm \ref{alg:EO} can be seen as a type of level set method, where the function
\[
-\psi_i^n(x) + \min_{j\neq i} \psi_j^n(x)
\]
plays the role of the level set function depicting the shape of the $i$-th grain at the $(n+1)$-th time step. It turns out that this very simple algorithm results in a misorientation dependent mobility $\mu_{i,j}$, associated with the boundary $\Gamma_{i,j}$ between grains $i$ and $j$ that is given by
\[
\mu_{i,j} = \frac{1}{\sigma_{i,j}}
\]
leading to the normal speed
\[
v_{i,j} = \mu_{i,j}\sigma_{i,j}\kappa_{i,j}(x) = \kappa_{i,j}(x)
\]
at any point $x\in\Gamma_{i,j}$.
In other words, using the Gaussian as the convolution kernel, Algorithm \ref{alg:EO} naturally prefers  equal (isotropic) {\em reduced mobilities}: Each interface in the network moves with the same (equal) multiple of its (mean) curvature, and the only difference from the isotropic model are the junction angles.

\boxalgorithm{\caption{(in \cite{SelimFelix})}\label{alg:EO}}{Given the initial grain shapes $\Sigma_1^0, \ldots, \Sigma_N^0$ and a time step size $\delta t$, obtain the grain shapes $\Sigma_1^{n+1}, \ldots, \Sigma_N^{n+1}$ at the $(n+1)$-th time from the grain shapes $\Sigma_1^n, \ldots, \Sigma_N^n$ at the end of the $n$-th times as follows:
\begin{enumerate}
	\item Compute the convolutions:
\[
\phi_i^n = G_{\sqrt{\delta t}} \ast \bm{1}_{\Sigma_i^k}.
\]
	\item Form the comparison functions:
\[
\psi^n_i = \sum_{j \not= i} \sigma_{i,j} \phi_j^n.
\]
	\item Update the grain shapes:
\[
\Sigma_i^{n+1} = \left\{x: \psi_i^n(x) < \min_{j\neq i} \psi_j^n(x)\right\}.
\]
\end{enumerate}
}

A more general algorithm was in fact given in \cite{SelimFelix} that allows arbitrary choice of surface tensions {\em and} mobilities, but is considerably more complicated and deviates from the simplicity of Algorithm \ref{alg:EO}.
In this paper, we use for the first time on large scale simulations a more recent version, namely Algorithm \ref{alg:ES}, that was proposed in \cite{SalvadorEsedogluSimplified}. 
It is simpler and closer in spirit to Algorithm \ref{alg:EO} while allowing for both arbitrary surface tensions and arbitrary mobilities.
This is achieved by merely replacing the convolution kernel used in Algorithm \ref{alg:EO} by the positive sum of two distinct Gaussians, thus preserving the efficiency and simplicity of the original version.
As shown in \cite{SalvadorEsedogluSimplified}, in the special but important and very common case of {\em equal mobilities} (i.e., $\mu_{i,j} = 1$ for all $i\not= j$), a careful choice of the widths of the two Gaussians, denoted by the parameters $\alpha$ and $\beta$ in what follows, guarantees the retention of  the major benefits of the original Algorithm \ref{alg:EO}, namely unconditional stability and convergence to the correct Mullins' energy \eqref{eq:energy}, while allowing its extension to e.g. Read-Shockley surface tensions \eqref{eq:RSmodel}.

\boxalgorithm{\caption{}\label{alg:ES}}{Given the initial grain shapes $\Sigma_1^0, \ldots, \Sigma_N^0$ and a time step size $\delta t$, obtain the grain shapes $\Sigma_1^{n+1}, \ldots, \Sigma_N^{n+1}$ at the $(n+1)$-th time from the grain shapes $\Sigma_1^n, \ldots, \Sigma_N^n$ at the end of the $n$-th times as follows:
\begin{enumerate}
	\item Compute the convolutions:
\[
\phi^n_{1,i} = G_{\sqrt{\alpha\delta t}} * \mathbf{1}_{\Sigma_i^n} \mbox{ and } 
\phi^n_{2,i} = G_{\sqrt{\beta\delta t}} * \mathbf{1}_{\Sigma_i^n}.
\]
	\item Form the comparison functions:
\[
\psi^n_i = \sum_{j \not= i} a_{i,j} \phi_{1,j}^n + b_{i,j} \phi_{2,j}^n
\]
where $a_{i,j}$ and $b_{i,j}$ are given by
\[
a_{i,j} = \frac{\sqrt{\pi}\sqrt{\alpha}}{\alpha-\beta}\left(\sigma_{i,j}-\beta\mu_{i,j}^{-1}\right) \quad \text{and} \quad b_{i,j} = \frac{\sqrt{\pi}\sqrt{\beta}}{\alpha-\beta}\left(-\sigma_{i,j}+\alpha\mu_{i,j}^{-1}\right).
\]
	\item Update the grain shapes:
\[
\Sigma_i^{n+1} = \left\{x: \psi_i^n(x) < \min_{j\neq i} \psi_j^n(x)\right\}.
\]
\end{enumerate}
}

\section{Large scale simulations}

In this section, we use Algorithms \ref{alg:EO} and \ref{alg:ES} to study the evolution of large networks of grains, in both 2D and 3D.
The 2D simulations are initialized with approximately 100,000 well-resolved grains and are repeated 3 times, while the 3D simulations are initialized with 10,000 grains and repeated 10 times.
Recall from the Introduction the following surface tension and mobility models that will be explored in both cases:
\begin{enumerate}
\item Read-Shockley surface tensions with constant (isotropic) mobilities: $\sigma_{i,j}$ given by (\ref{eq:RSmodel}) and $\mu_{i,j} = 1$.
\item Read-Shockley surface tensions and reciprocal mobilities: $\sigma_{i,j}$ given by (\ref{eq:RSmodel}) and $\mu_{i,j} = \sigma_{i,j}^{-1}$.
This results in equal (isotropic) {\em reduced mobilities}.
\item Equal (isotropic) surface tensions and mobilities: $\sigma_{i,j} = \mu_{i,j} = 1$.
This also results in equal (isotropic) reduced mobilities.
\end{enumerate}
Whenever Read-Shockley surface tensions are used, we take the Brandon angle $\theta_*$ in (\ref{eq:RSmodel}) to be $\theta_* = 30^\circ$ to agree with simulations performed in Refs. \cite{GruberMCPart1,HolmMisorientationAngles} and to lie within the experimentally observed range \cite{SuttonBalluffiBook}.

In all simulations, the initial grain configuration is taken to be the Voronoi diagram of a set of points chosen uniformly at random from a periodic box.
\subsection{Two spatial dimensions}
In these 2D simulations, the initial grain configuration of around 100,000 well-resolved grains is assigned a random fiber texture: All grain orientations are obtained from a reference configuration by a rotation about the axis normal to the 2D domain, and the angle of the rotation is taken from $[0,2\pi)$ uniformly at random.
We present results at two distinct times for each model: model (i) at $t_i = 1.907 \times 10^{-5}$ and $t_f = 8.345 \times 10^{-5}$, model (ii) at $t_i =  1.550 \times 10^{-5}$ and $t_f = 6.259 \times 10^{-5}$ and model (iii) at $t_i =  1.788 \times 10^{-5}$ and $t_f = 7.033 \times 10^{-5}$. The different stopping times ensure that at the intermediate time $t_i$, around 30\% of all grains remain in each one of our runs, while at the final time $t_f$ around 10\% of all grains remain.
The simulations are performed on a $4096 \times 4096$ grid discretizing $[0,1]^2$.

Algorithm \ref{alg:ES} requires choosing the parameters $\alpha$ and $\beta$, the widths of the two Gaussian kernels. 
They need to satisfy
\[
\alpha > \max_{i\neq j} \sigma_{ij} \quad \text{and} \quad \beta < \min_{i\neq j} \sigma_{ij}
\]
to guarantee the properties described in \autoref{sec:model}.
However, from a practical point of view,  we also need the quotient $\alpha/\beta$ to remain moderate for accuracy considerations so that the Gaussians have comparable widths and thus may be sampled well without requiring unduly high spatial resolution.
Since both $\alpha$ and $\beta$ depend on the surface tensions which in turn depend on the misorientation angle, this can be accomplished by imposing a minimum misorientation angle between any two grains to be $1^\circ$. 
This reduces the effective size of the surface tension matrix, while keeping the same number of distinct grains, and ultimately results in a small quotient $\alpha/\beta$. 

We achieve this in the following way. Once the grain orientations are assigned, we partition the domain $[0,2\pi)$ into equally spaced bins of $1^\circ$ length. Then all grains with orientations in the same bin are reassigned a new grain orientation, the midpoint of the bin. This is equivalent to assigning grain orientations from the uniform distribution on $\{0.5^\circ, 1.5^\circ,\ldots,359.5^\circ\}$. 

\begin{figure}[h]
\centering
\begin{tabular}{ccc}
\includegraphics[width=\widththreefigures] {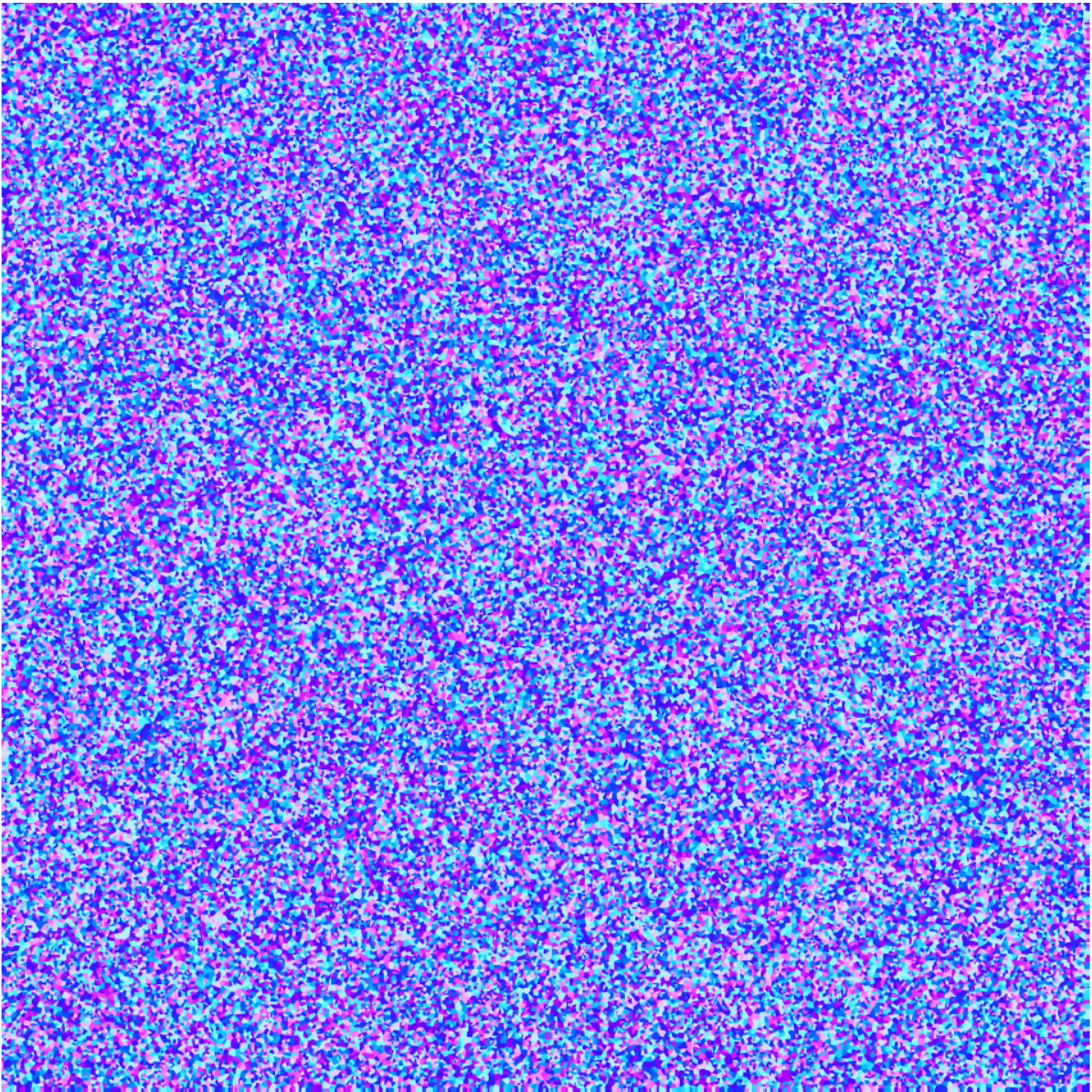} & \includegraphics[width=\widththreefigures]{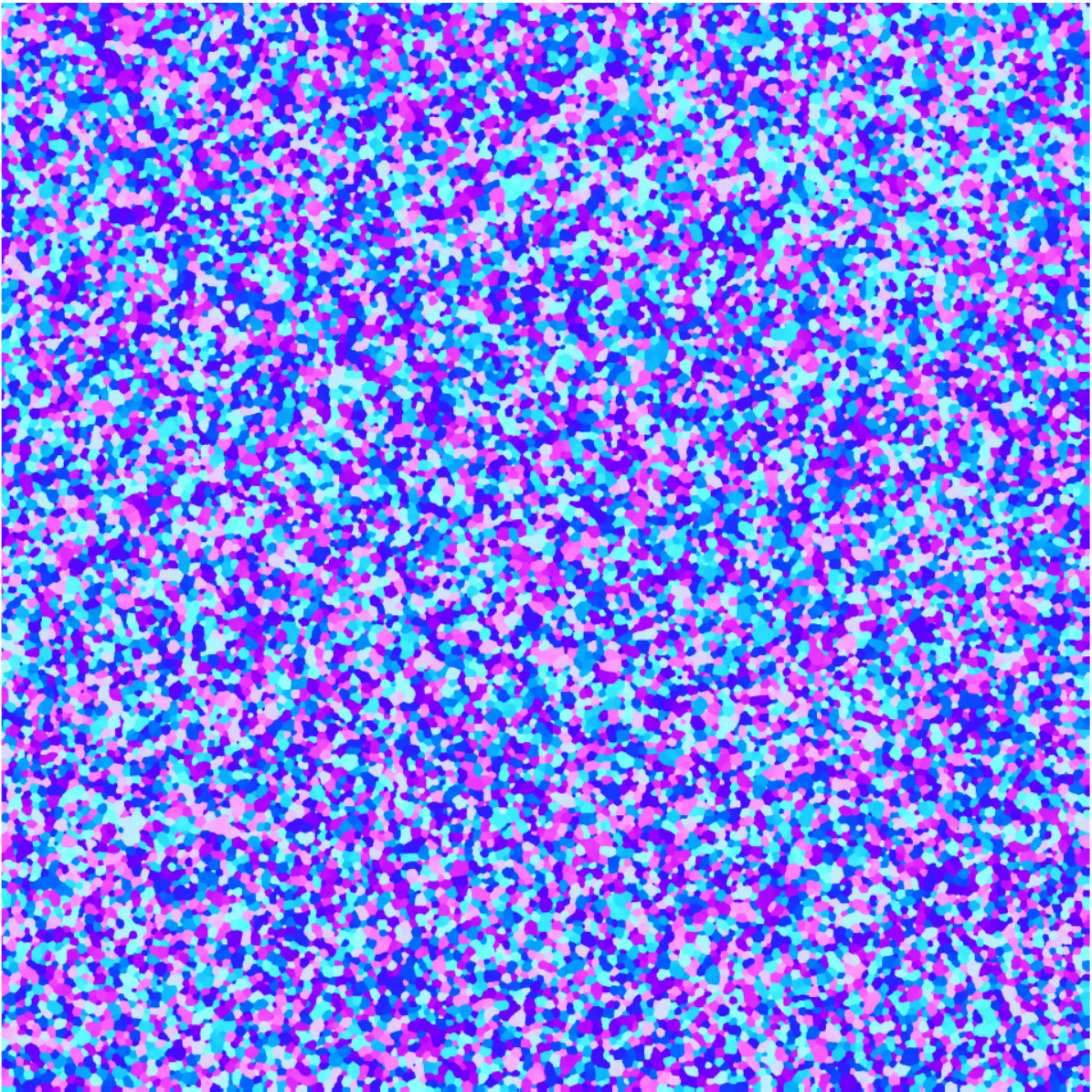} & \includegraphics[width=\widththreefigures]{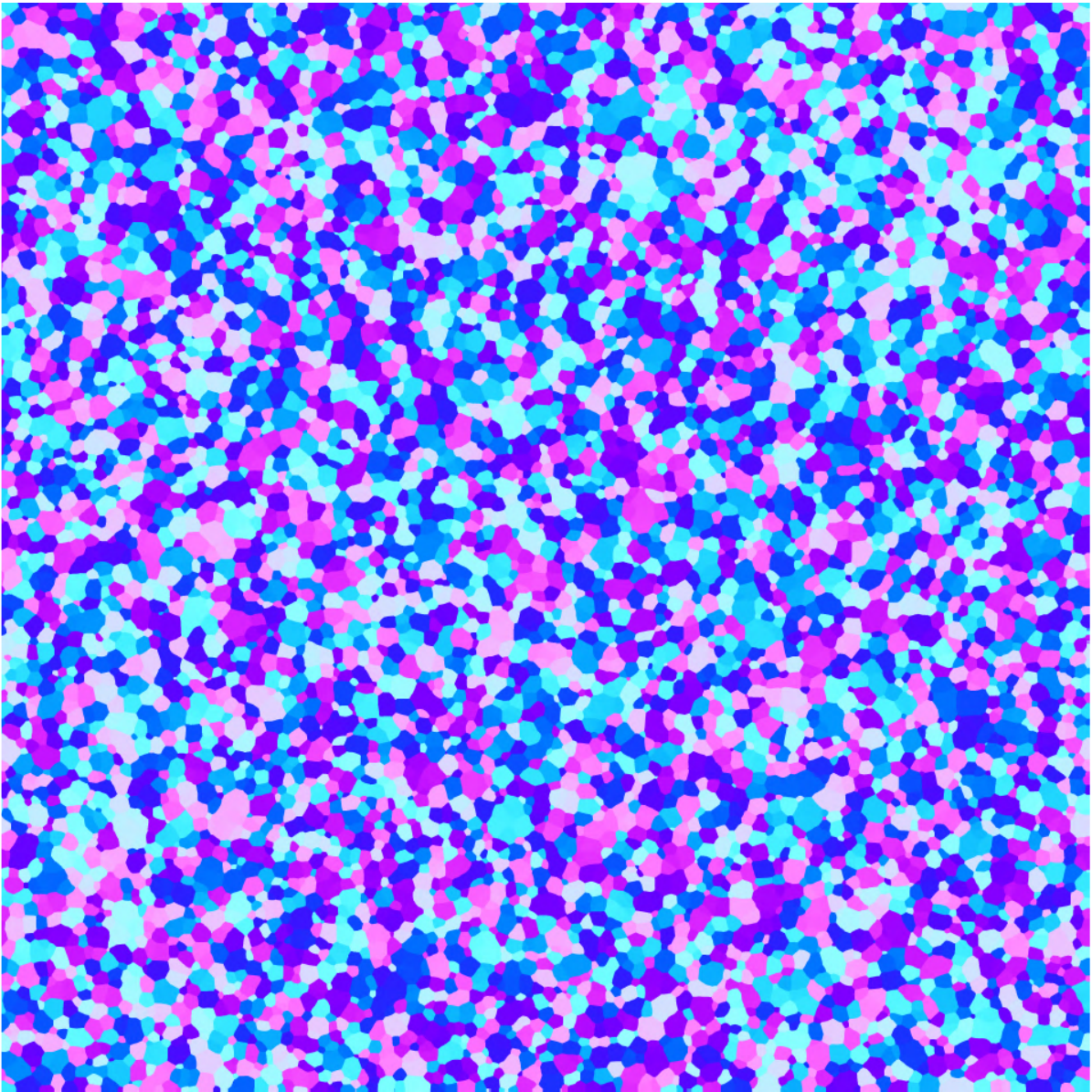}
\end{tabular}
\caption{The initial microstructure contains $99,130$ grains (left). At the intermediate time $t_i = 1.907 \times 10^{-5}$, $29,841$ grains remain in full microstructure while at the final time $t_f = 8.345 \times 10^{-5}$, $10,003$ grains remain (right). These microstructures were obtained with model (i).}
\label{fig:microstructures2dzoomin}
\end{figure}

\begin{figure}[h]
\centering
\begin{tabular}{ccc}
\includegraphics[width=\widthtwofigures]{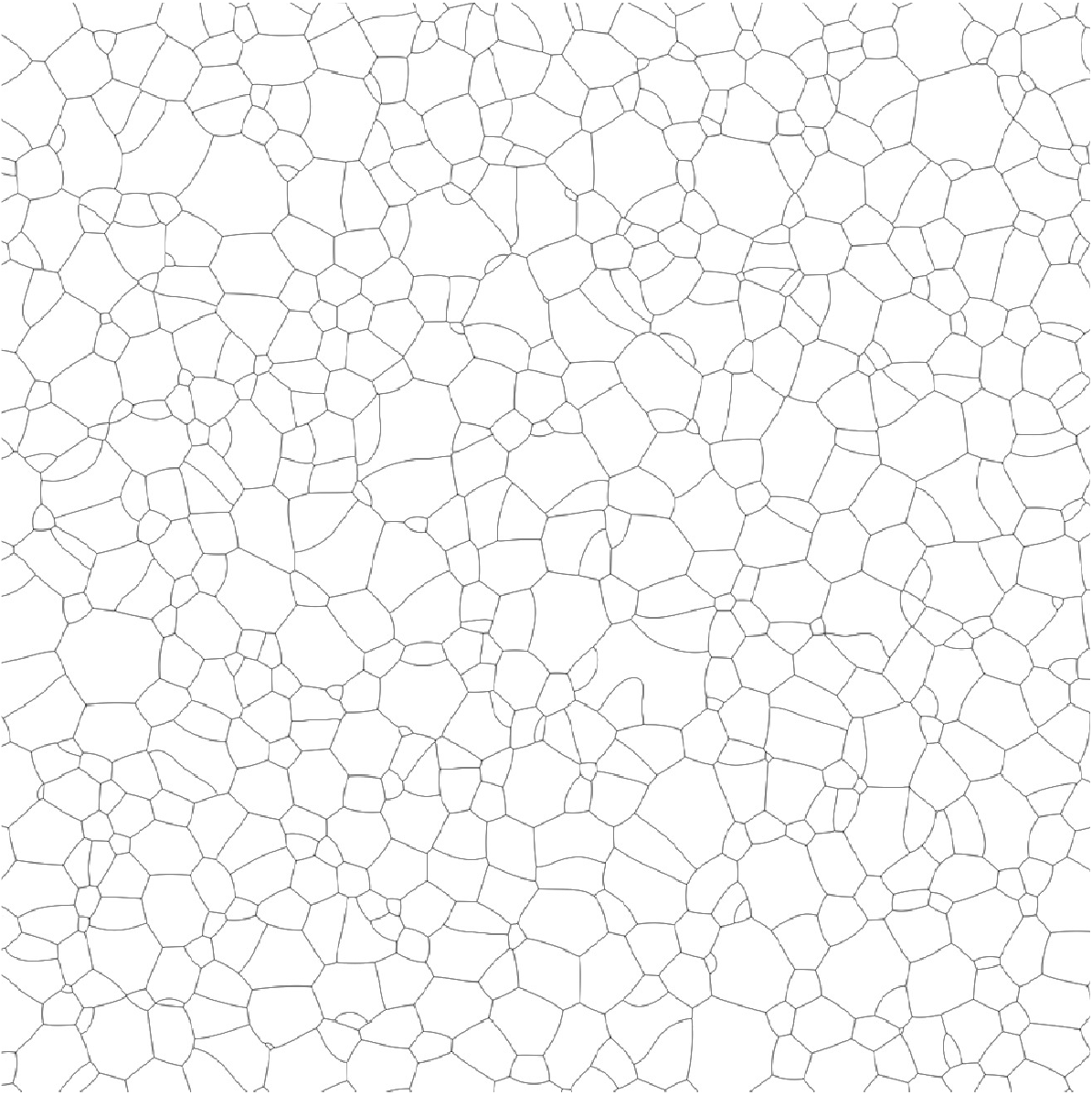} & \includegraphics[width=\widthtwofigures]{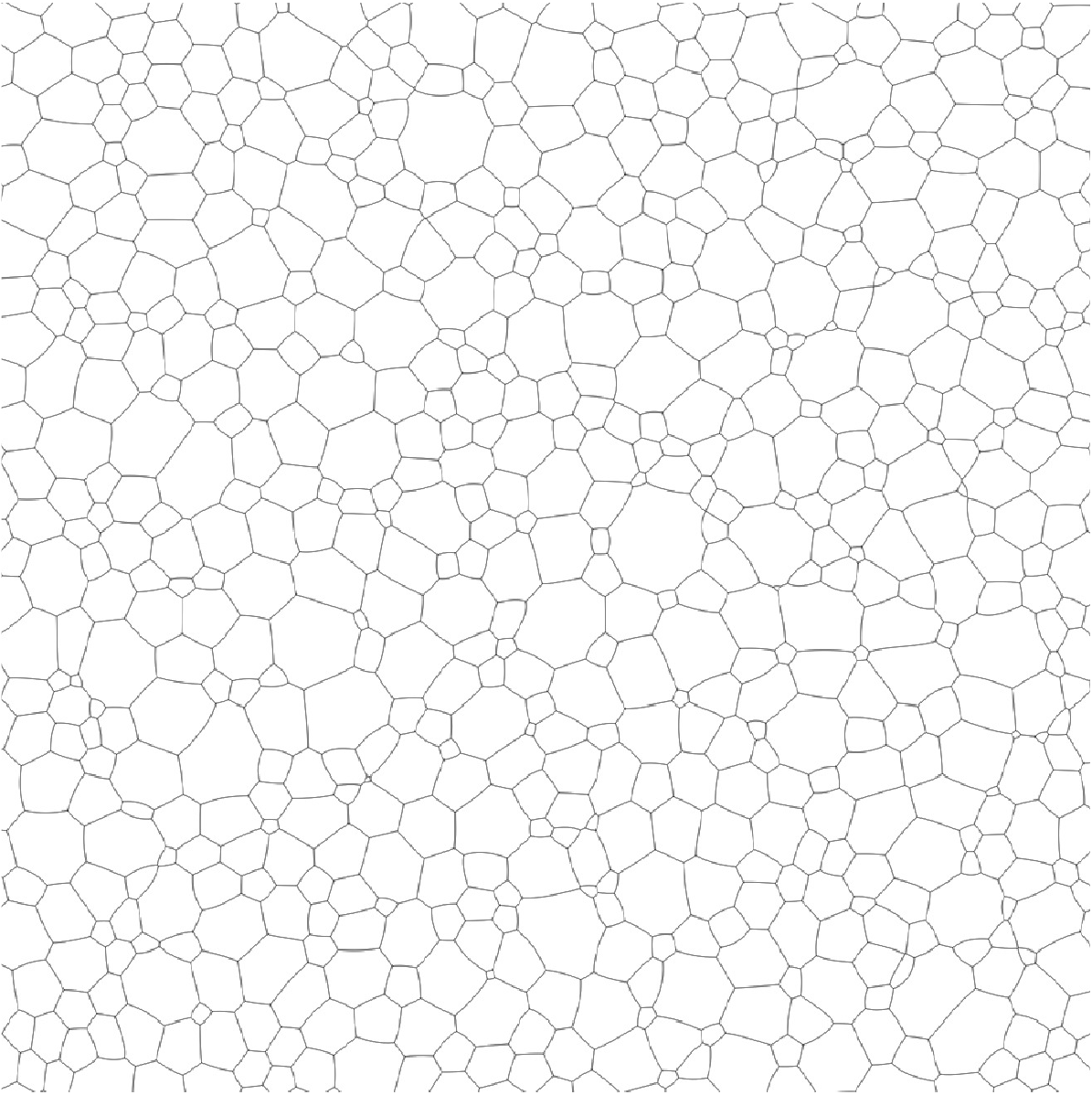}
\end{tabular}
\caption{Close up of the final microstructures for model (i) (left) and model (iii) (right). The former produces more elongated grains with narrow shapes, while the latter more rounded grains.}
\label{fig:microstructures2d}
\end{figure}

\begin{figure}[h]
\centering
\begin{tabular}{ccc}
\includegraphics[width=\widththreefigures]{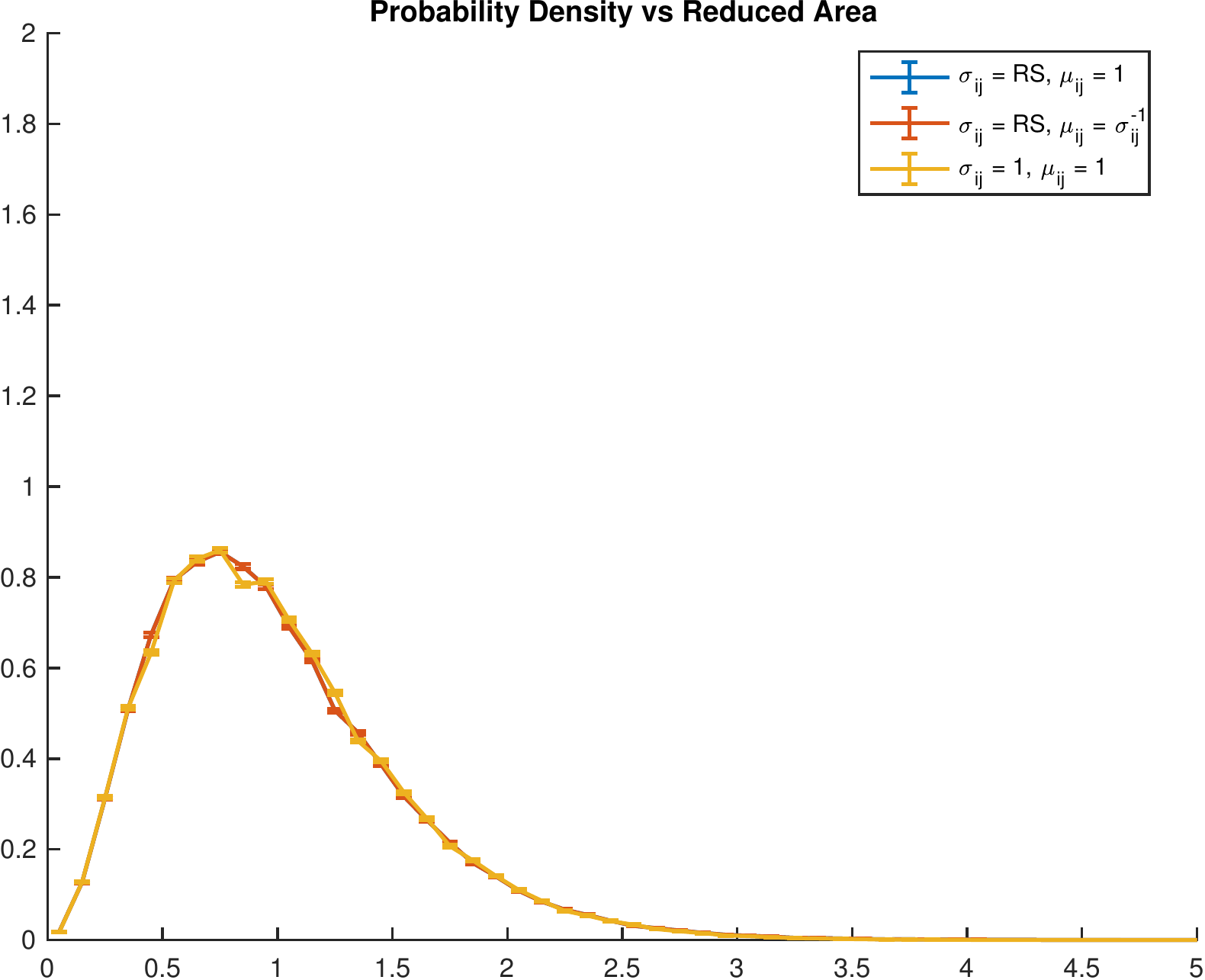} & \includegraphics[width=\widththreefigures]{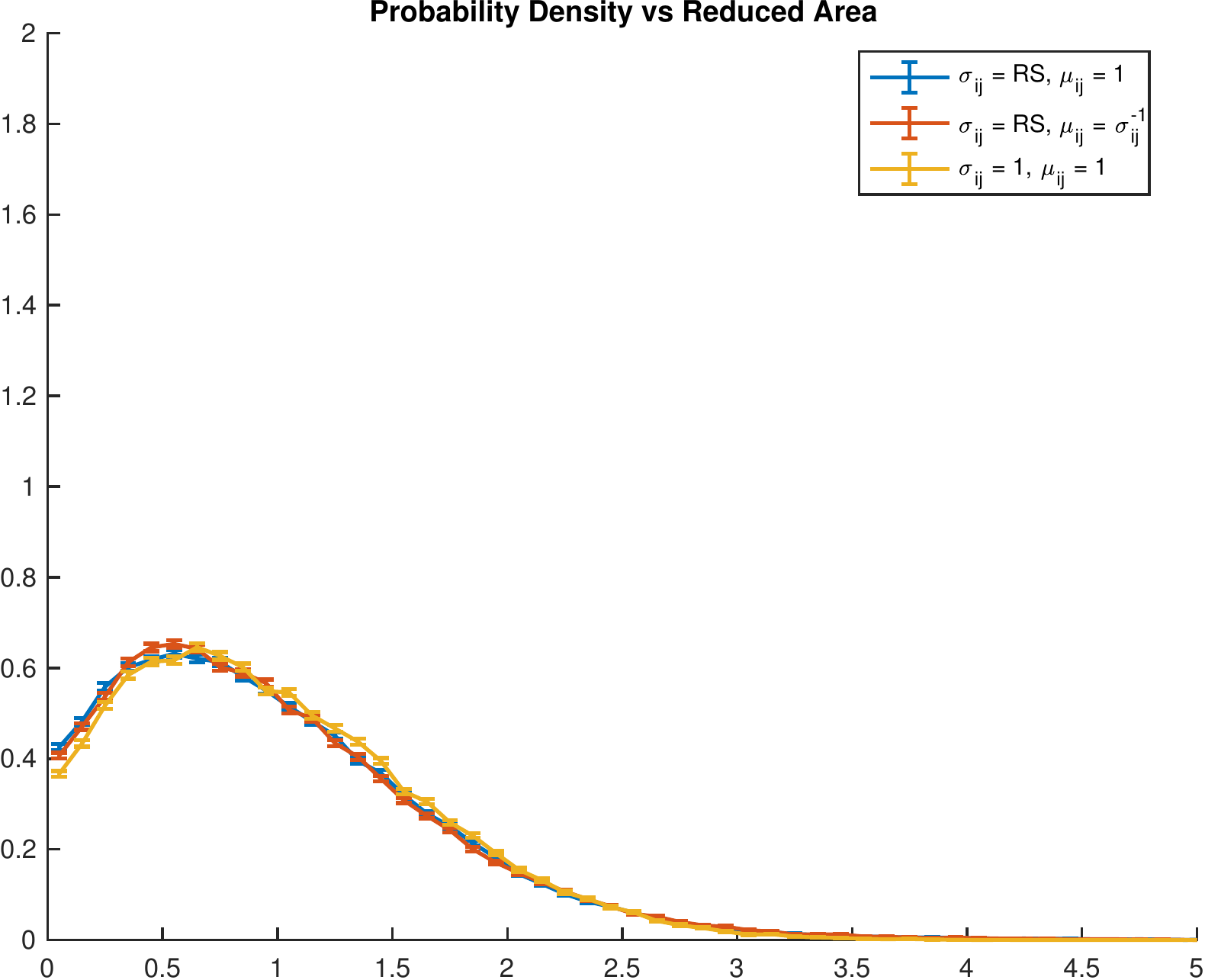} & \includegraphics[width=\widththreefigures]{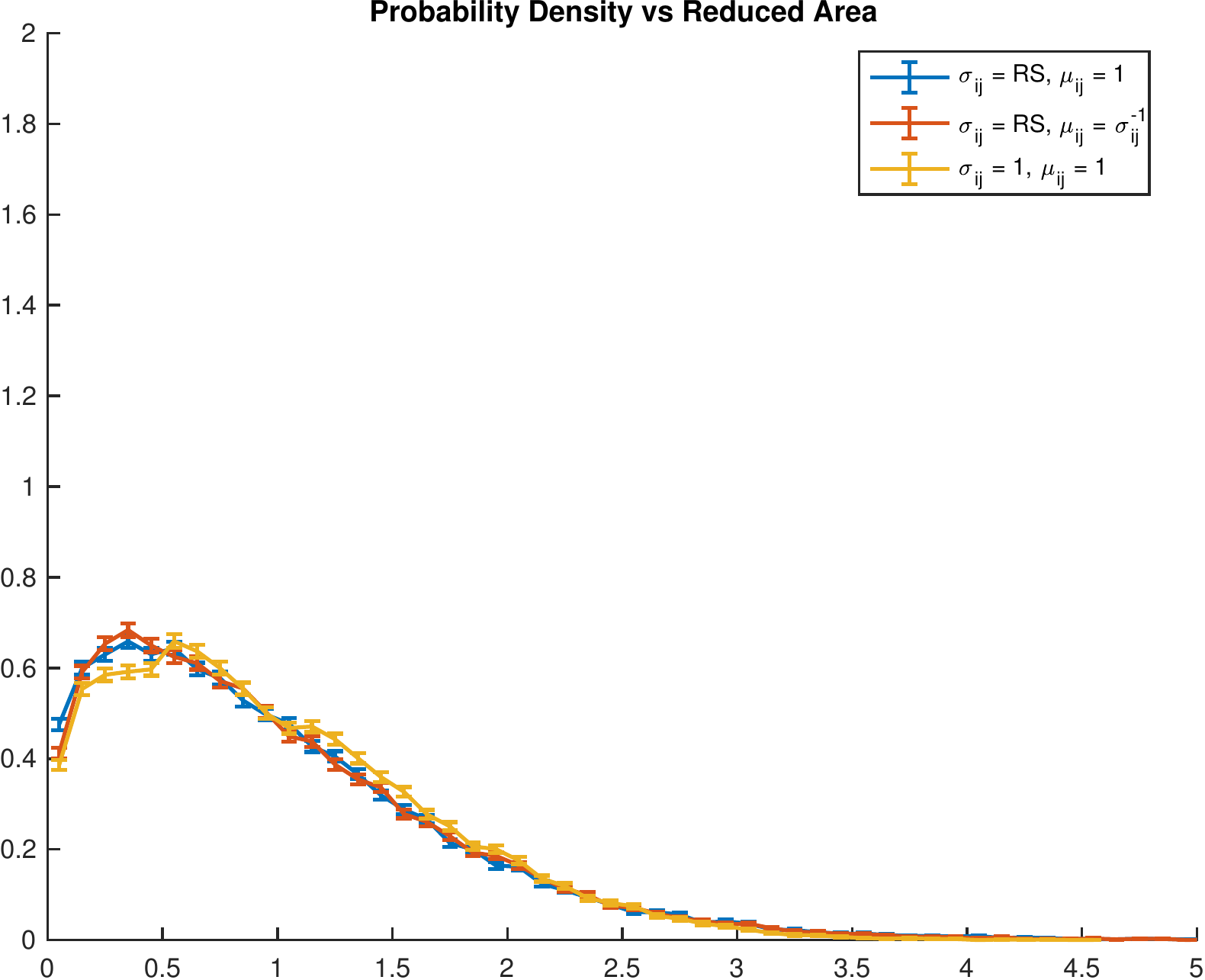}\end{tabular}
\caption{GSD (probability density vs. reduced area $\frac{A}{<A>}$) for the three different models: initial Voronoi data (left), at time $t_i$ when approximately 30\% of grains remain (center) and at time $t_f$ when approximately 10\% of grains remains (right).}
\label{fig:reduced_area2d}
\end{figure}

\begin{figure}[h]
\centering
\begin{tabular}{ccc}
\includegraphics[width=\widththreefigures]{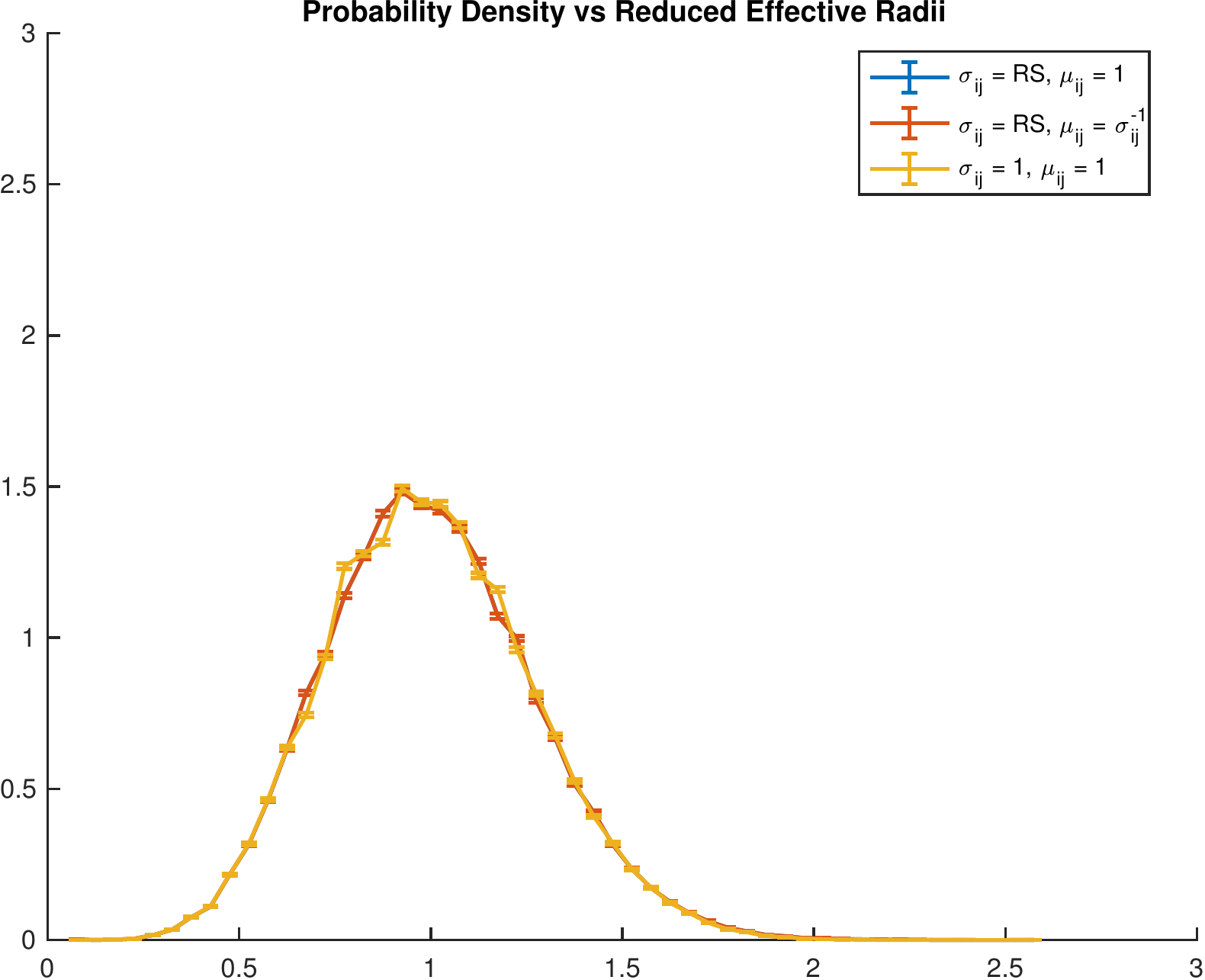} & \includegraphics[width=\widththreefigures]{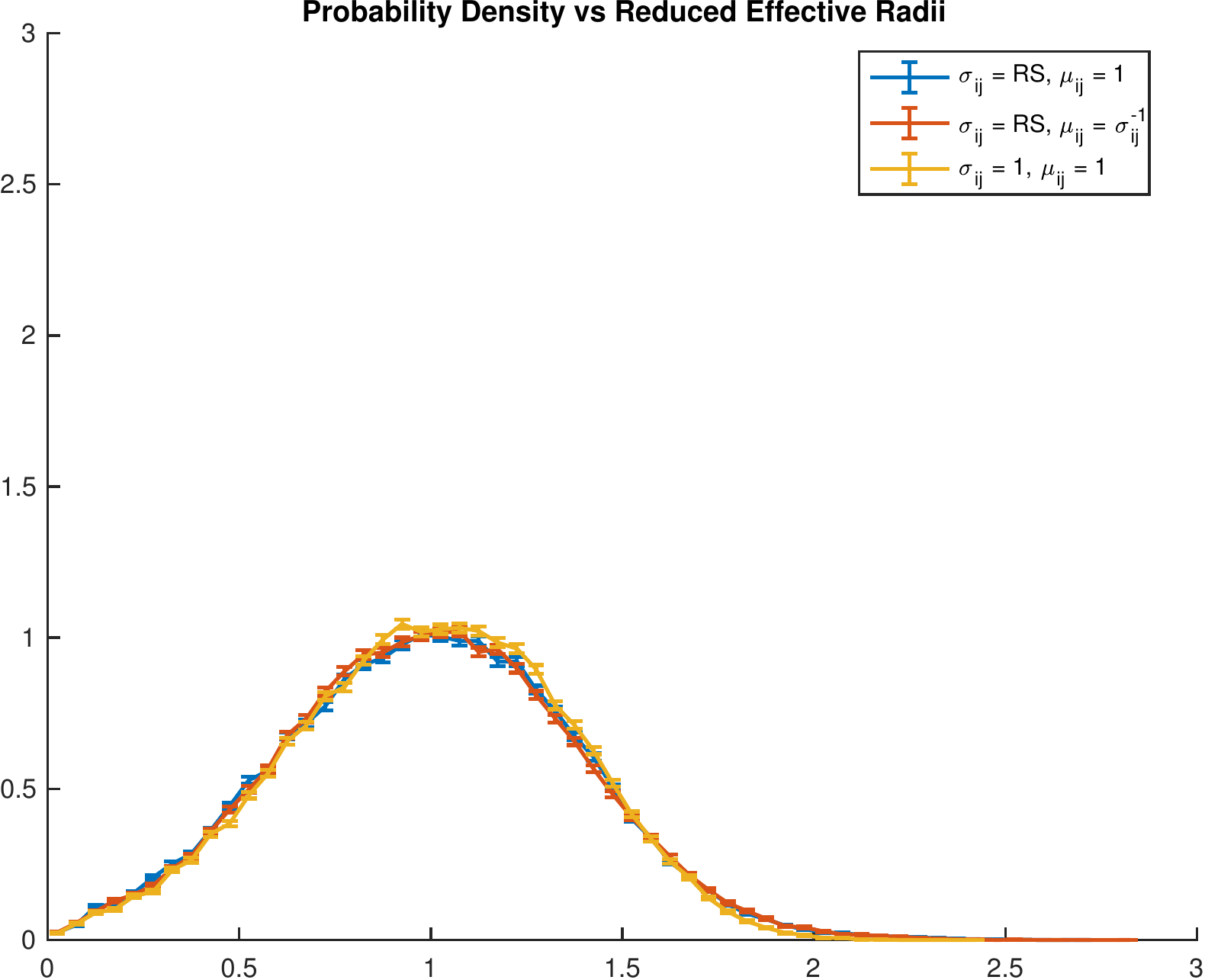} & \includegraphics[width=\widththreefigures]{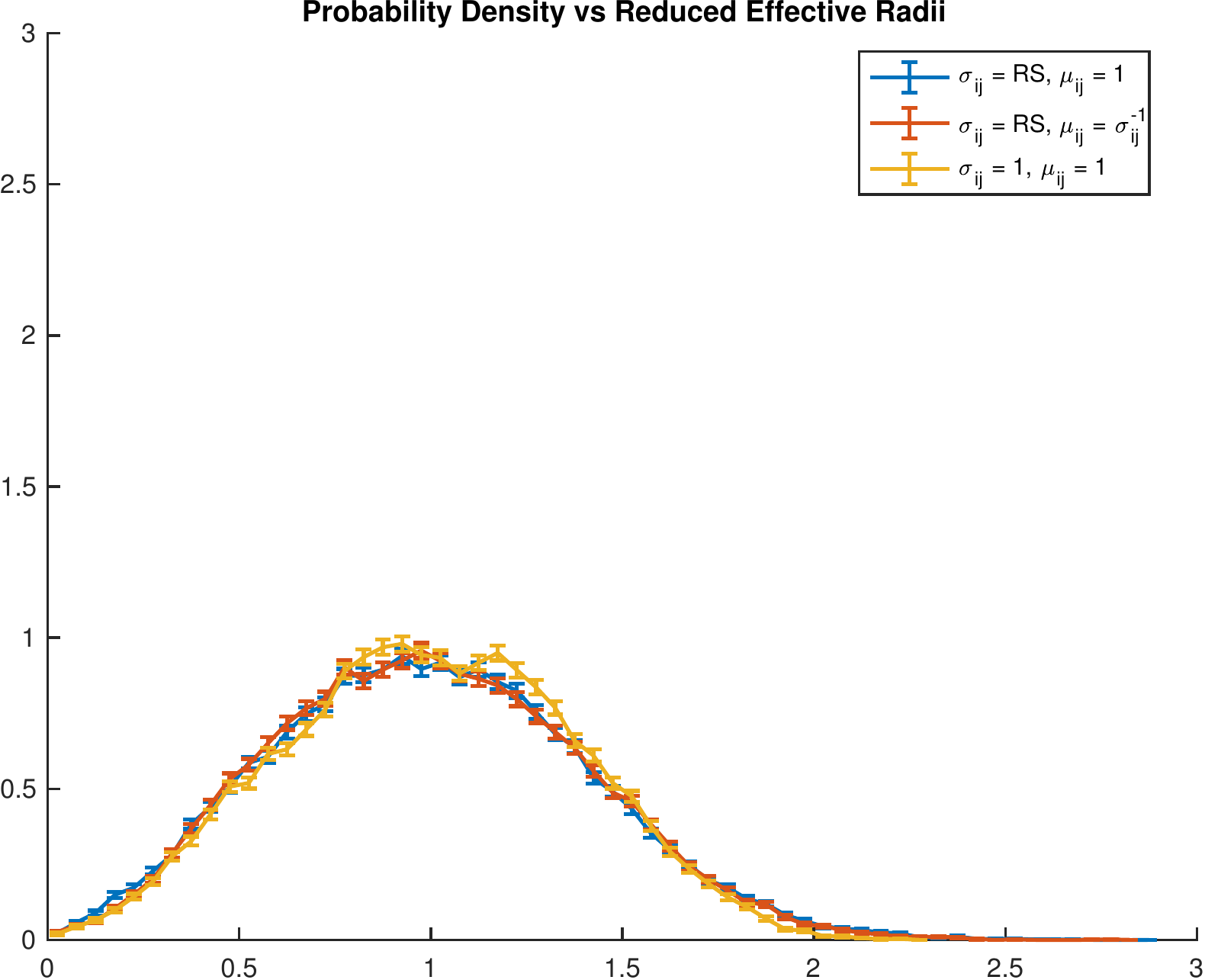}
\end{tabular}
\caption{Probability density vs. reduced effective radii $\frac{\sqrt{A}}{\sqrt{<A>}}$ for the three different models: initial Voronoi data (left), at time $t_i$ when approximately 30\% of grains remain (center) and at time $t_f$ when approximately 10\% of grains remains (right).}
\label{fig:reduced_effective_radii2d}
\end{figure}

Figure \ref{fig:reduced_area2d} shows the time evolution of the GSD for the three different models, while Figure \ref{fig:reduced_effective_radii2d} shows the corresponding distributions of reduced effective radii.
For models (i) \& (ii) with Read-Shockley surface tensions, the distributions appear stationary with no significant difference between the two.
The distributions for the isotropic model (iii) are also very similar, except at final time $t_f$, we see some evidence of the two distinct peaks observed previously in simulations of \cite{MasonStatisticsGBM} with the same isotropic model, but using the very different algorithm of front tracking.
This two peak distribution has not been observed in other recent 2D simulations of the {\em isotropic} (constant) surface tension and mobility model (i) in e.g. \cite{Kinderlehrer2006,Kim2006,Miyoshi2018}, whose results look closer to the ones we obtained with Read-Shockley surface tensions.
On the other hand, the recent study \cite{MiyoshiAnisotropic}, which includes simulations with models (i) \& (iii), also has some evidence of the double peak with model (iii); our results are in close agreement with theirs for both of these models.

\begin{figure}[h]
\centering
\begin{tabular}{ccc}
\includegraphics[width=\widththreefigures]{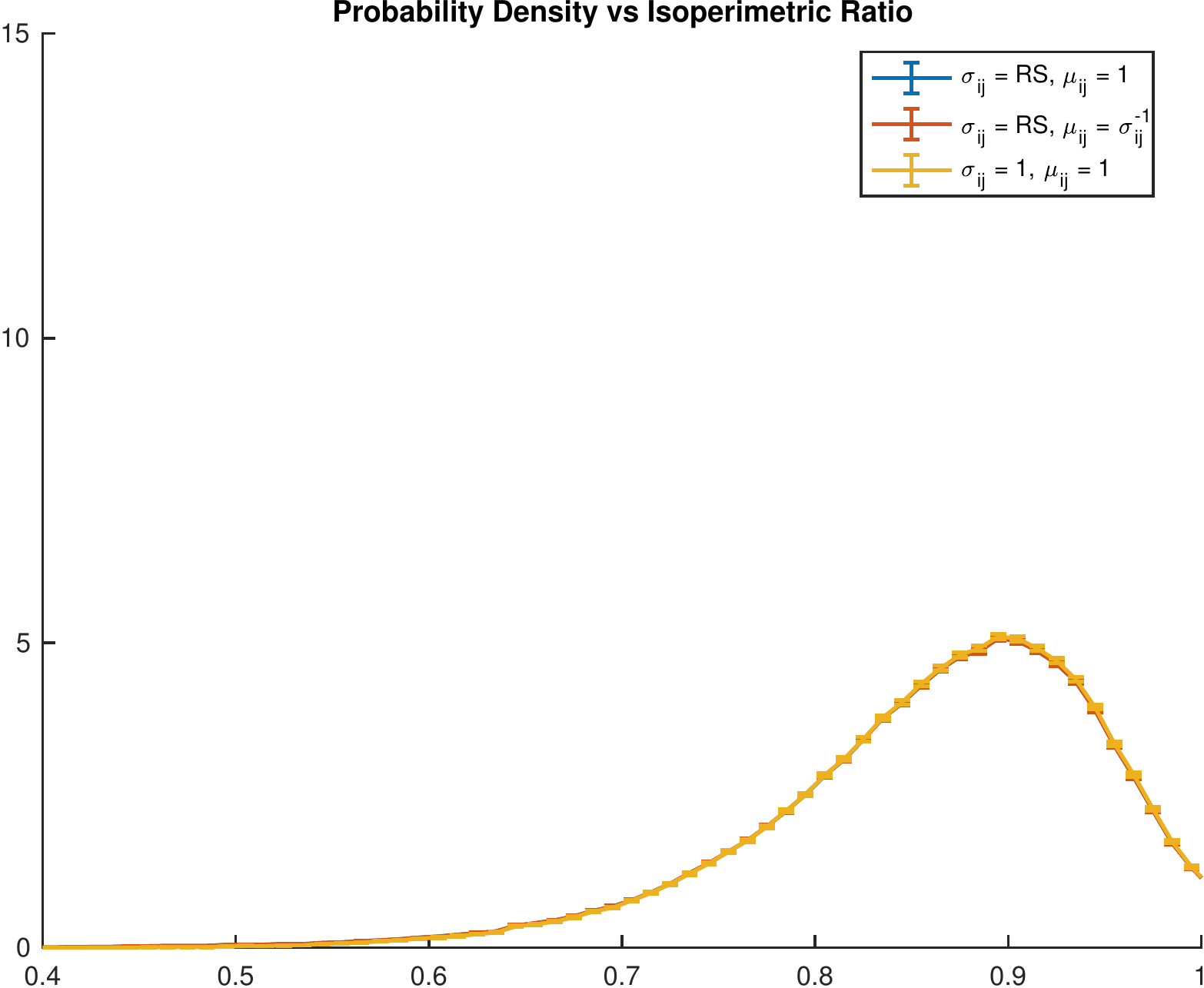} & \includegraphics[width=\widththreefigures]{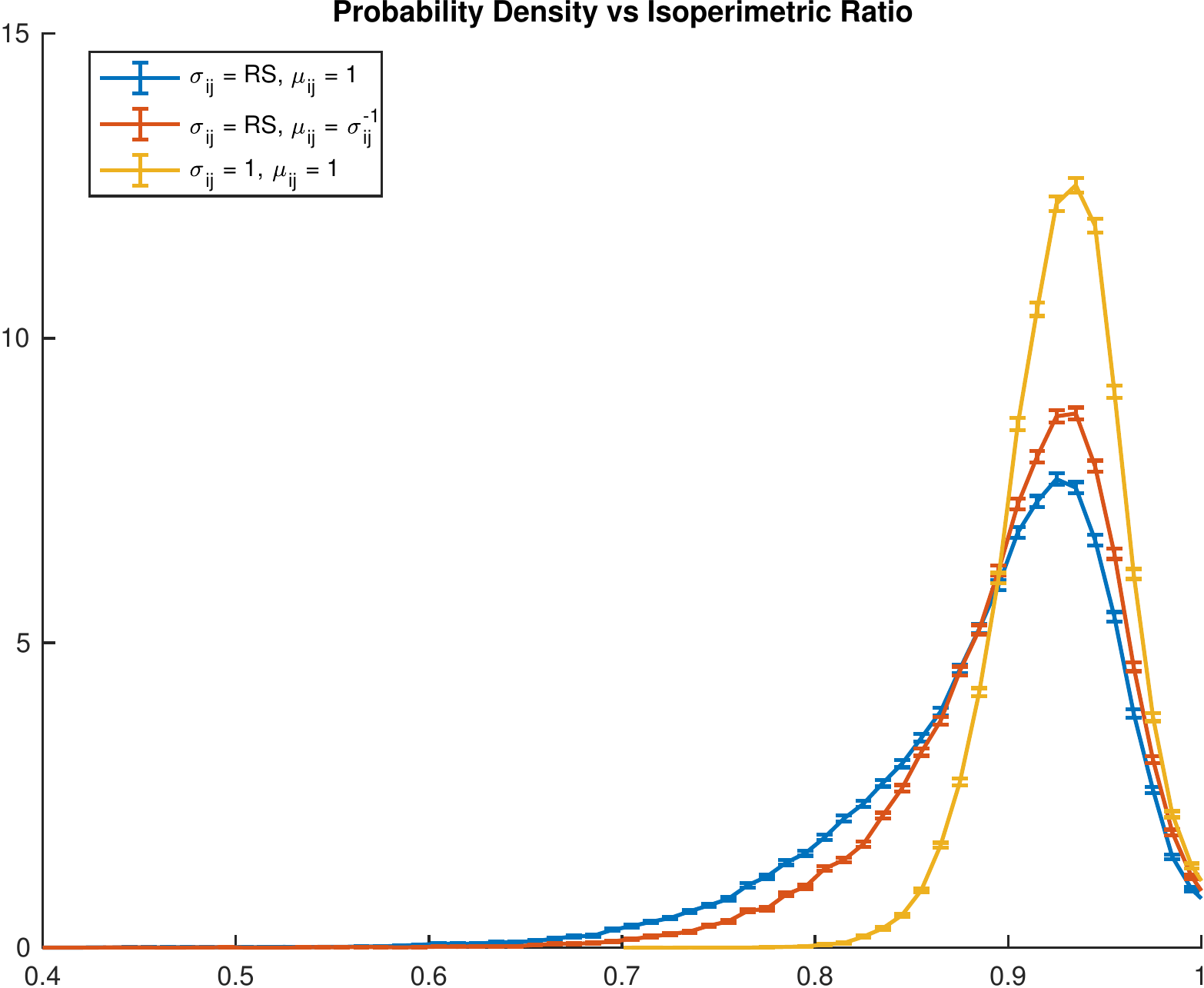} & \includegraphics[width=\widththreefigures]{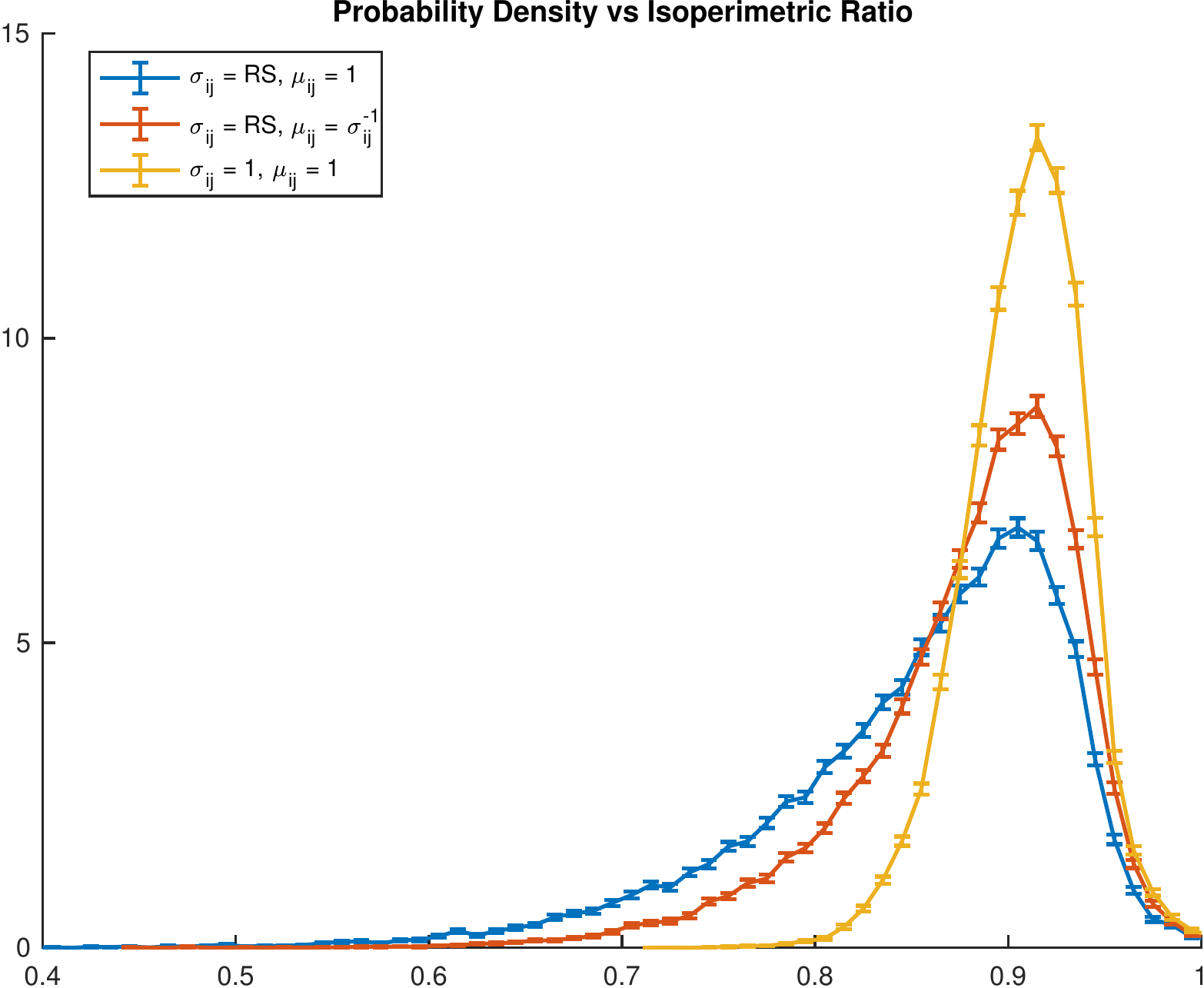}
\end{tabular}
\caption{Probability density vs isoperimetric ratio $\frac{4\pi A}{L^2}$ for the three different models: initial Voronoi data (left), at time $t_i$ when approximately 30\% of grains remain (center) and at time $t_f$ when approximately 10\% of grains remains (right).}
\label{fig:isopratio2d}
\end{figure}

The similarity in the GSDs of the three surface tension / mobility models considered here notwithstanding, it is natural to expect some differences when we move on to statistics of grain {\em shapes}.
After all, it is well known that the microstructure of anisotropic models can be visually quite different than that of the isotropic model (see Figures \ref{fig:microstructures2dzoomin} and \ref{fig:microstructures2d}).
Indeed, the similarities finally cease when we look at the isoperimetric ratio, also called circularity, which is given by $\frac{4\pi A}{P^2}$ where $A$ and $P$ denote the area and perimeter of the grain, respectively.
Figure \ref{fig:isopratio2d} shows that all three surface tension / mobility models lead to peak distributions but with a higher amplitude for the isotropic model. From time $t_i$ to time $t_f$, while $20,000$ disappear, the isoperimetric ratio retains its peaked distribution but with the peak moving towards 1 which indicates that the grains become more rounded.


\begin{figure}[h]
\centering
\begin{tabular}{ccc}
\includegraphics[width=\widththreefigures]{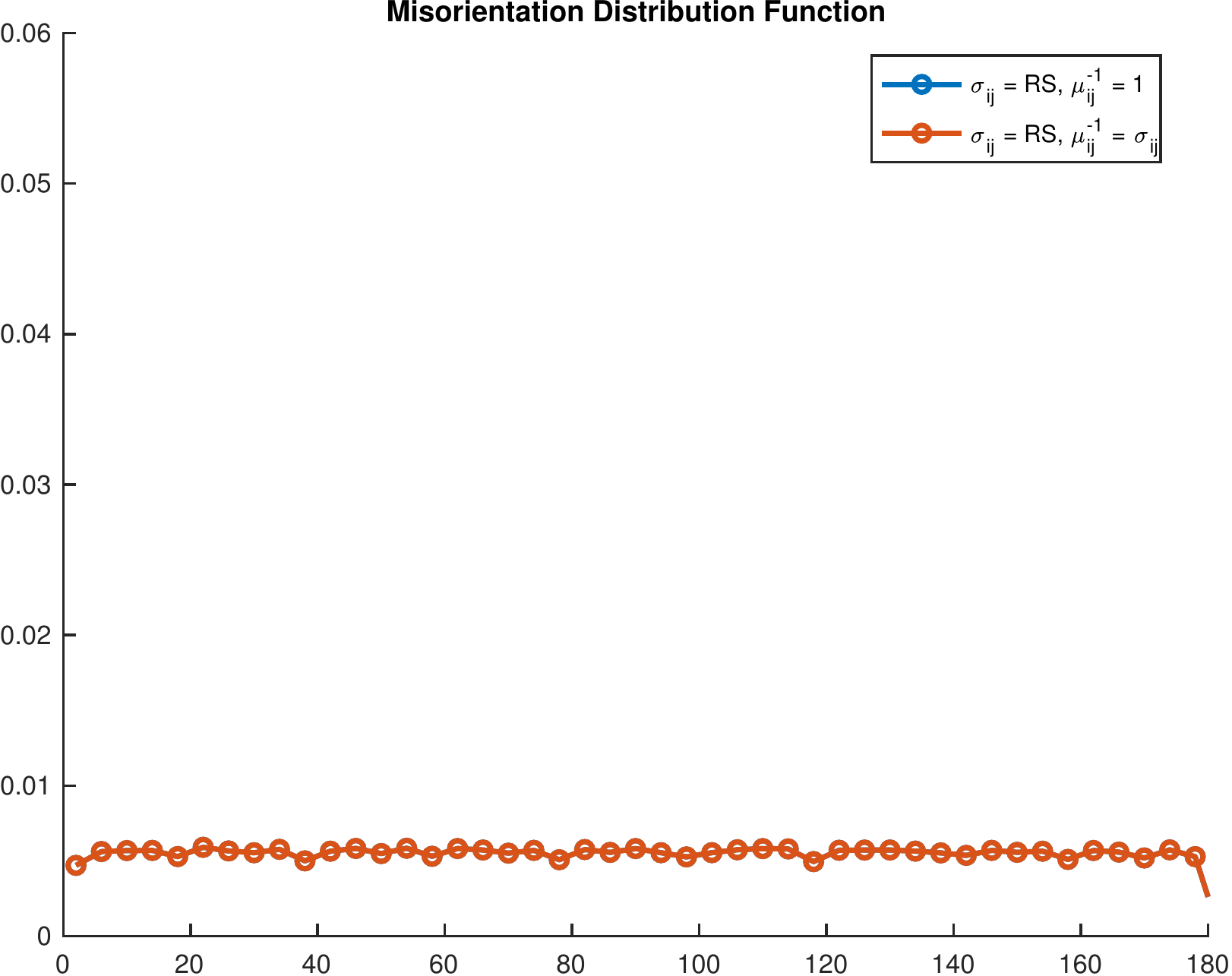} &\includegraphics[width=\widththreefigures]{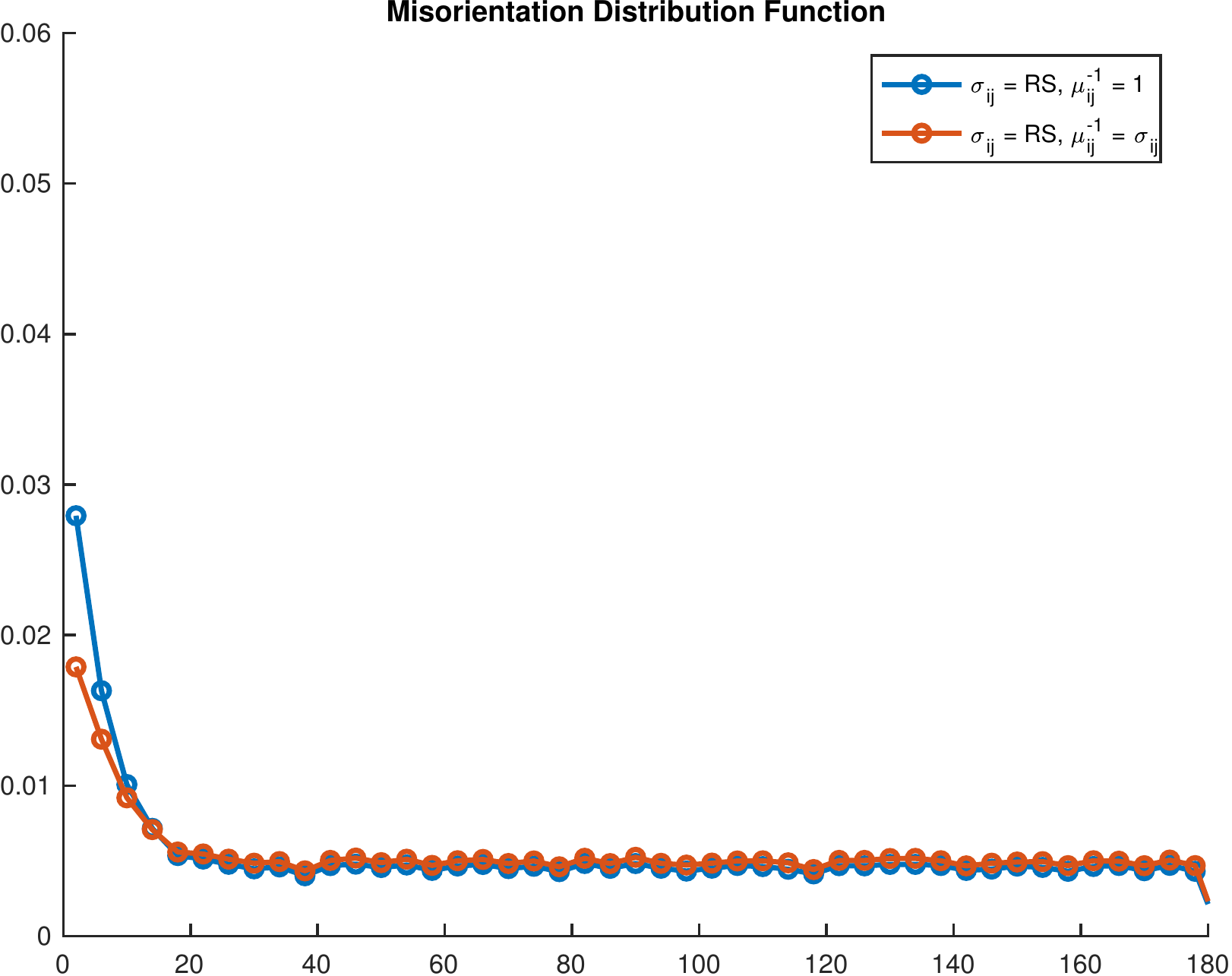} & \includegraphics[width=\widththreefigures]{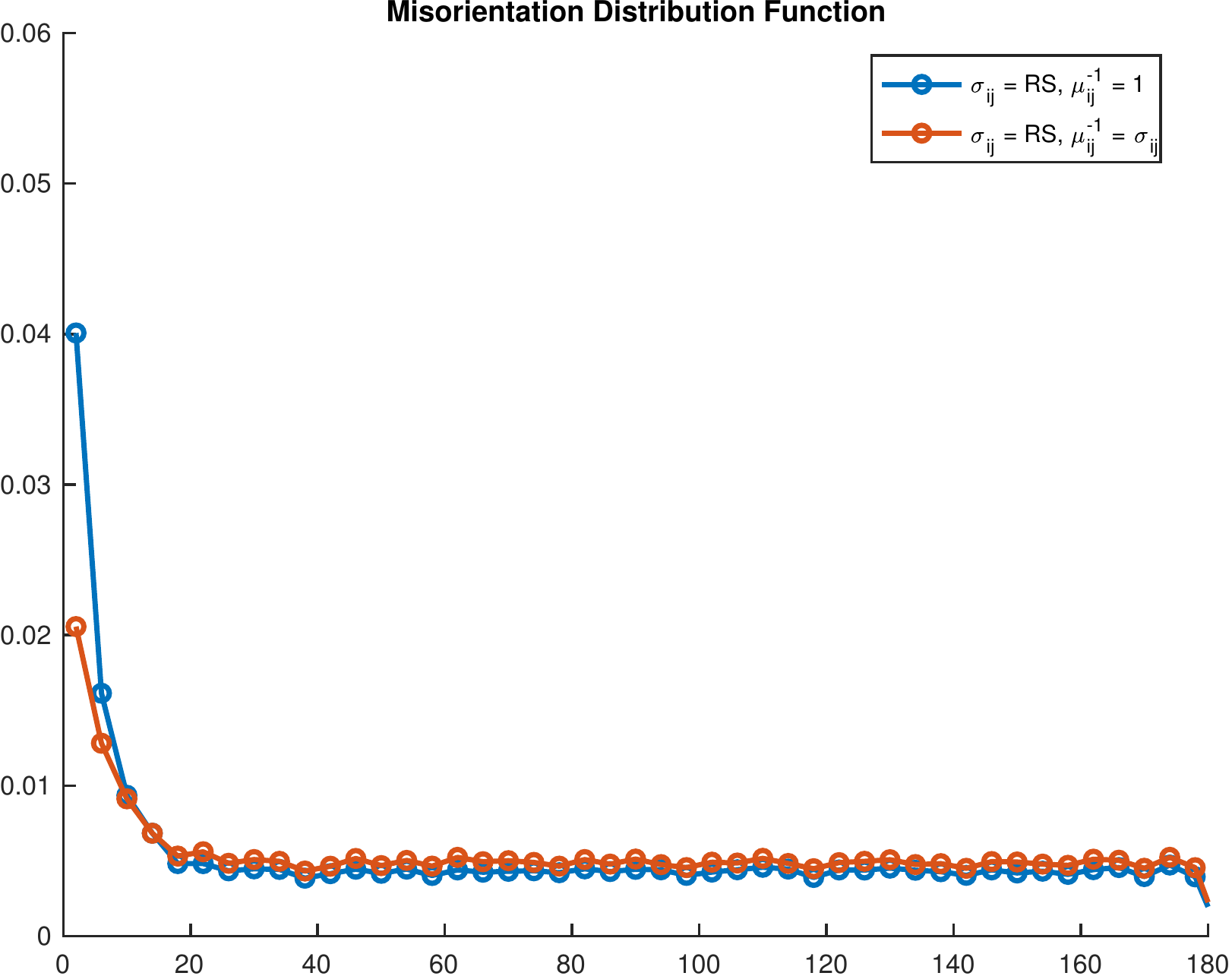}
\end{tabular}
\caption{Misorientation distribution for models (i) and (ii): initial Voronoi data (left), at time $t_i$ when approximately 30\% of grains remain (center) and at time $t_f$ when approximately 10\% of grains remains (right).}
\label{fig:MDF2d}
\end{figure}

Finally, Figure \ref{fig:MDF2d} shows the MDF for the two unequal surface tension models: Read-Shockley with equal and reciprocal mobilities.
Due to its random fiber texture, the MDF of the initial data is uniform.
As time evolves, it concentrates at small misorientations corresponding to small surface tensions.
This is in agreement with previous studies, e.g. 
\cite{HolmMisorientationAngles,HolmDimensionalEffects,Kinderlehrer2004,ElseyEsedogluSmerekaActaMaterialia}.
There is a slight but noticeable difference between the MDFs of the two models, which may in any case keep concentrating at the origin with further evolution: Model (i) has a more concentrated MDF compared to (ii), perhaps because low angle boundaries move slower in model (i) and hence persist longer.

\subsection{Three spatial dimensions}

For the 3D simulations, we initialize the microstructure with approximately 10,000 well-resolved grains, chosen as the approximate Voronoi diagram to points placed uniformly at random in a periodic box that is discretized into a $256 \times 256 \times 256$ uniform grid.
This results in approximately $12$ grid points across each grain in every direction at time $t=0$, which is roughly the same resolution as in our 2D simulations.
Each grain is assigned a fully random orientations from $SO(3)$.
We present results at two distinct times for each model: model (i) at $t_i = 4.578 \times 10^{-3}$ and $t_f = 7.629 \times 10^{-3}$, model (ii) and (iii) at $t_i =  1.831 \times 10^{-3}$ and $t_f = 4.578 \times 10^{-3}$. 
Just like in the 2D simulations, the different stopping times ensure that at the intermediate time $t_i$, around 30\% of all grains remain in each one of our runs, while at the final time $t_f$ around 10\% of all grains remain.
As in our 2D simulations, we impose a minimum misorientation angle, this time of $5^\circ$, between the grains once again in order to keep the quotient $\alpha/\beta$ that appears in our algorithm at a moderate value.
Given that the minimum misorientation angle imposed here is larger than usual, a natural question arises: if it were taken to be smaller, would the different statistics change significantly? In order to investigate this, we run our simulations for model (ii) without doing any merging and observed that the statistics remained essentially the same.


\begin{figure}[h]
\centering
\begin{tabular}{ccc}
\includegraphics[width=\widththreefigures,trim={4cm 1cm 4cm 1cm},clip] {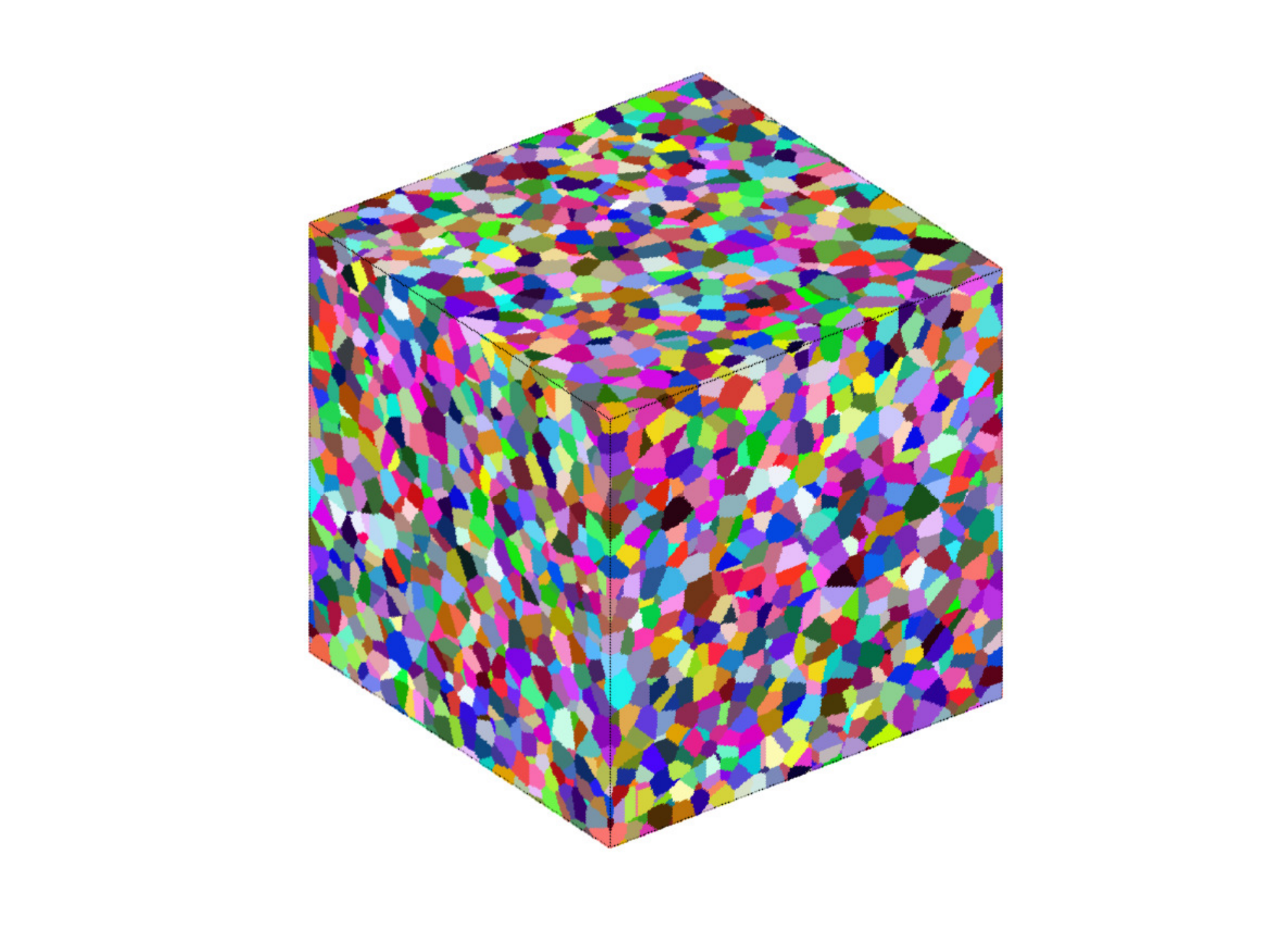} & \includegraphics[width=\widththreefigures,trim={4cm 1cm 4cm 1cm},clip]{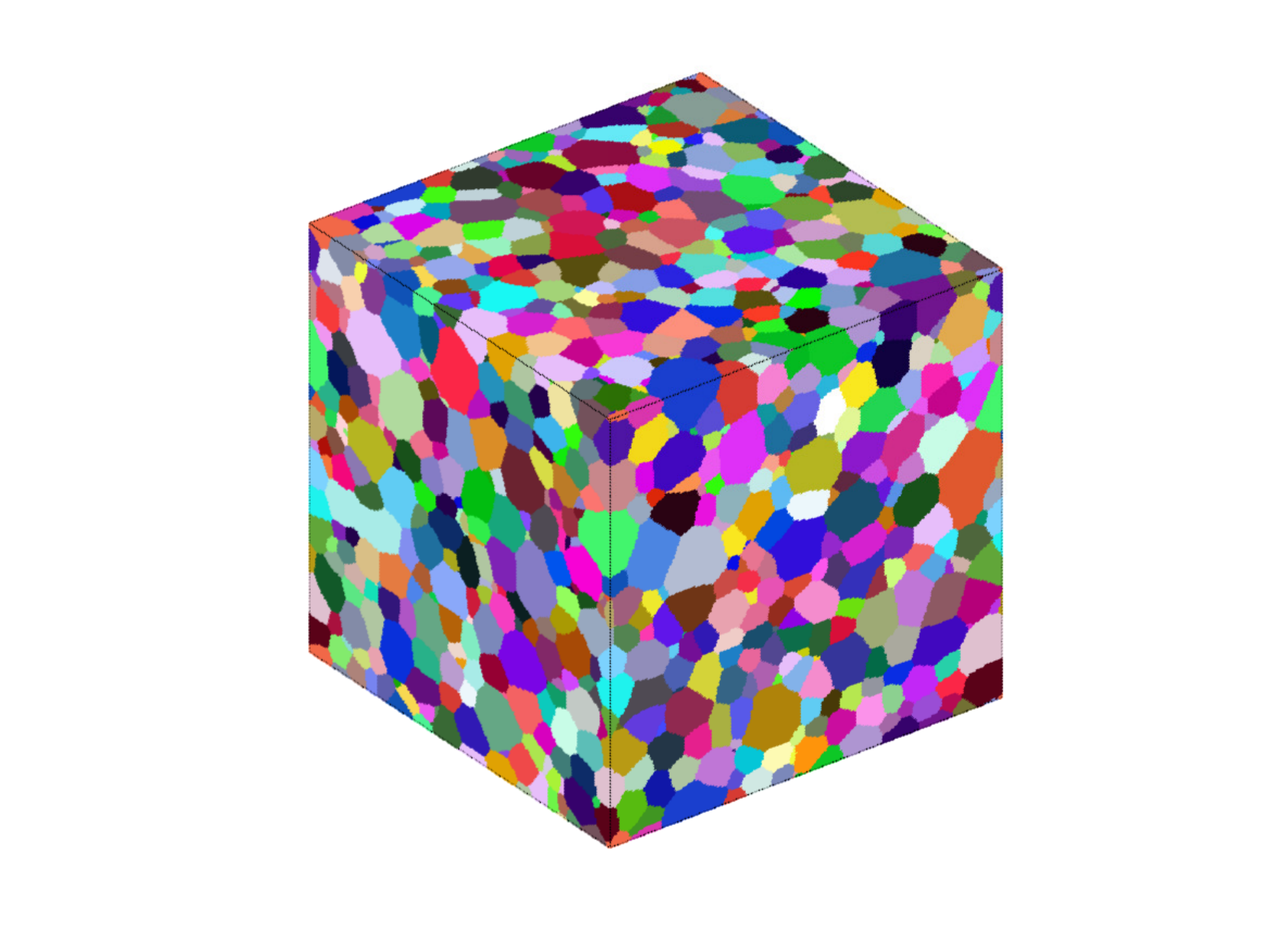} & \includegraphics[width=\widththreefigures,trim={4cm 1cm 4cm 1cm},clip]{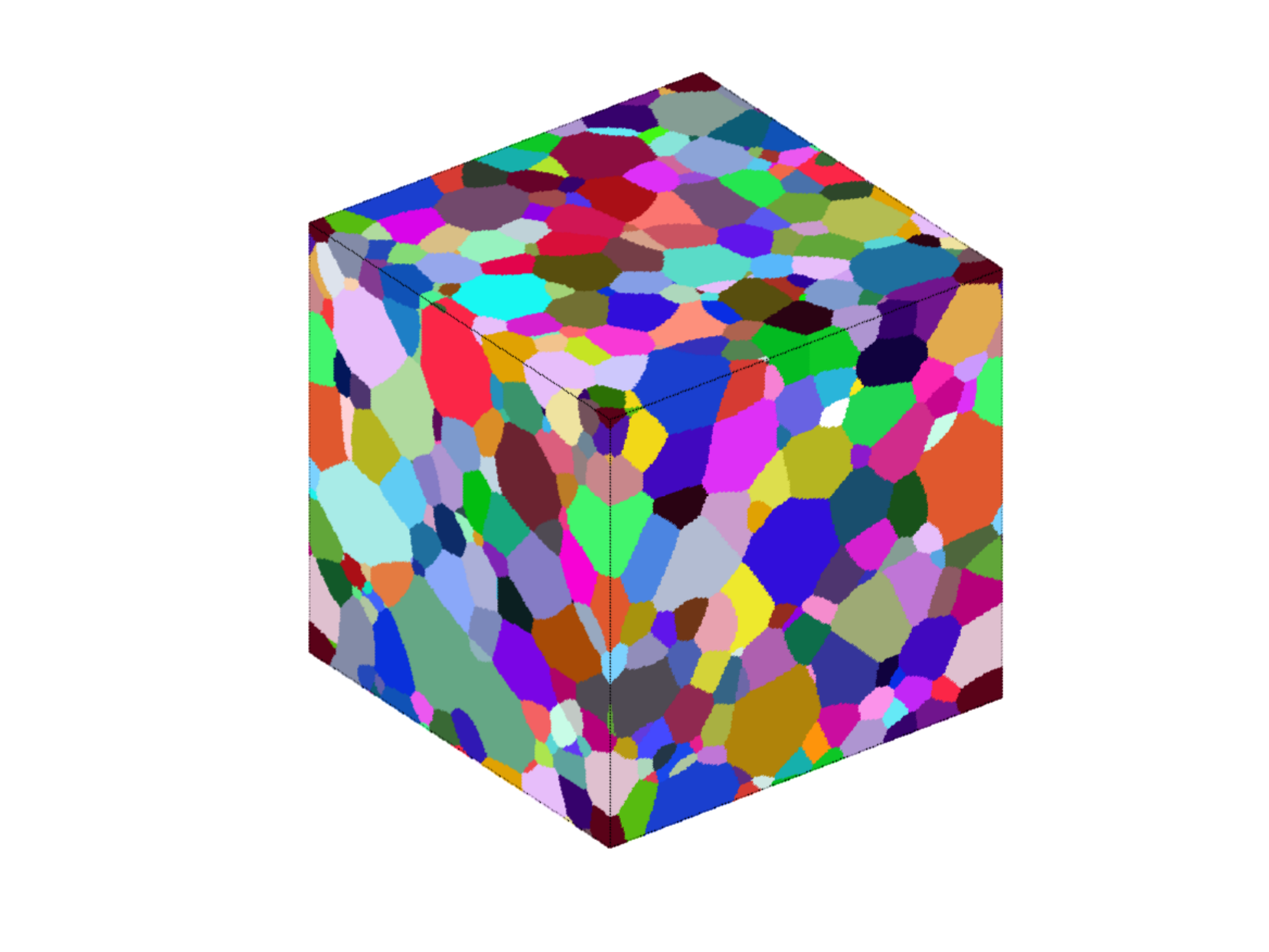}
\end{tabular}
\caption{The initial microstructure contains $9881$ grains (left). At the intermediate time $t_i = 4.578 \times 10^{-3}$, $3031$ grains remain in full microstructure while at the final time $t_f = 7.629 \times 10^{-3}$, 998 grains remain (right). These microstructures were obtained with model (i).}
\label{fig:microstructures3d}
\end{figure}

\begin{figure}[h]
\centering
\begin{tabular}{ccc}
\includegraphics[width=\widththreefigures]{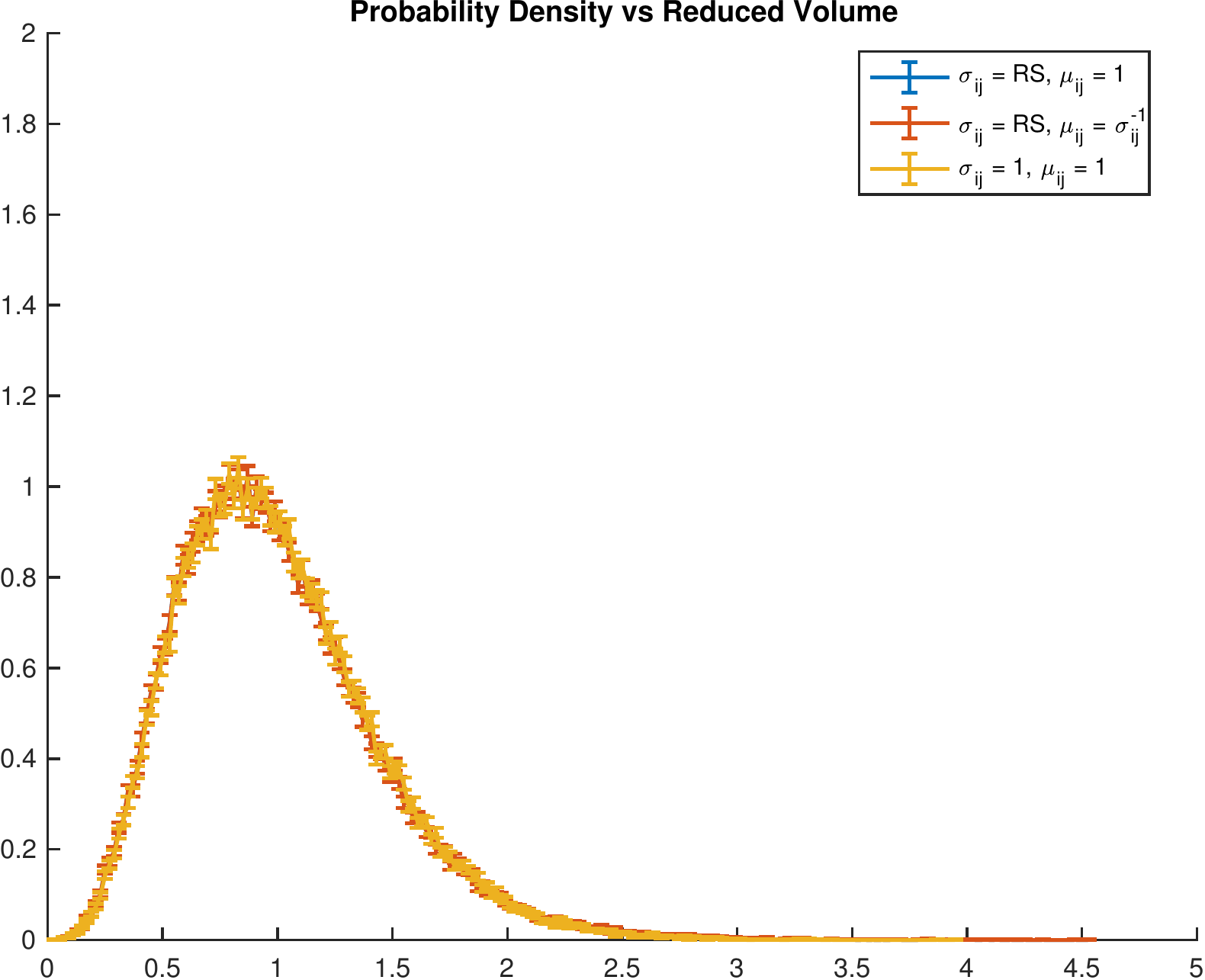} & \includegraphics[width=\widththreefigures]{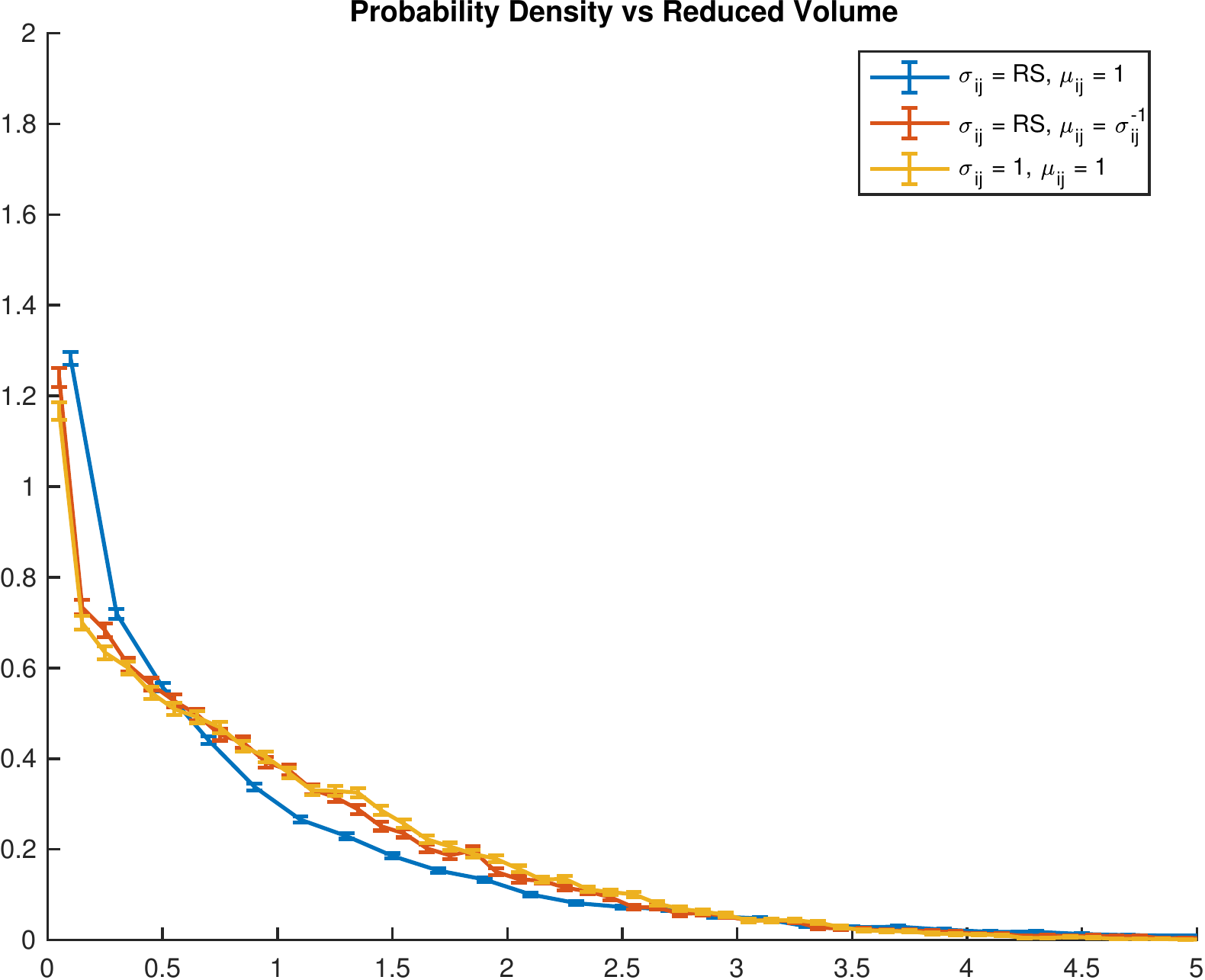} & \includegraphics[width=\widththreefigures]{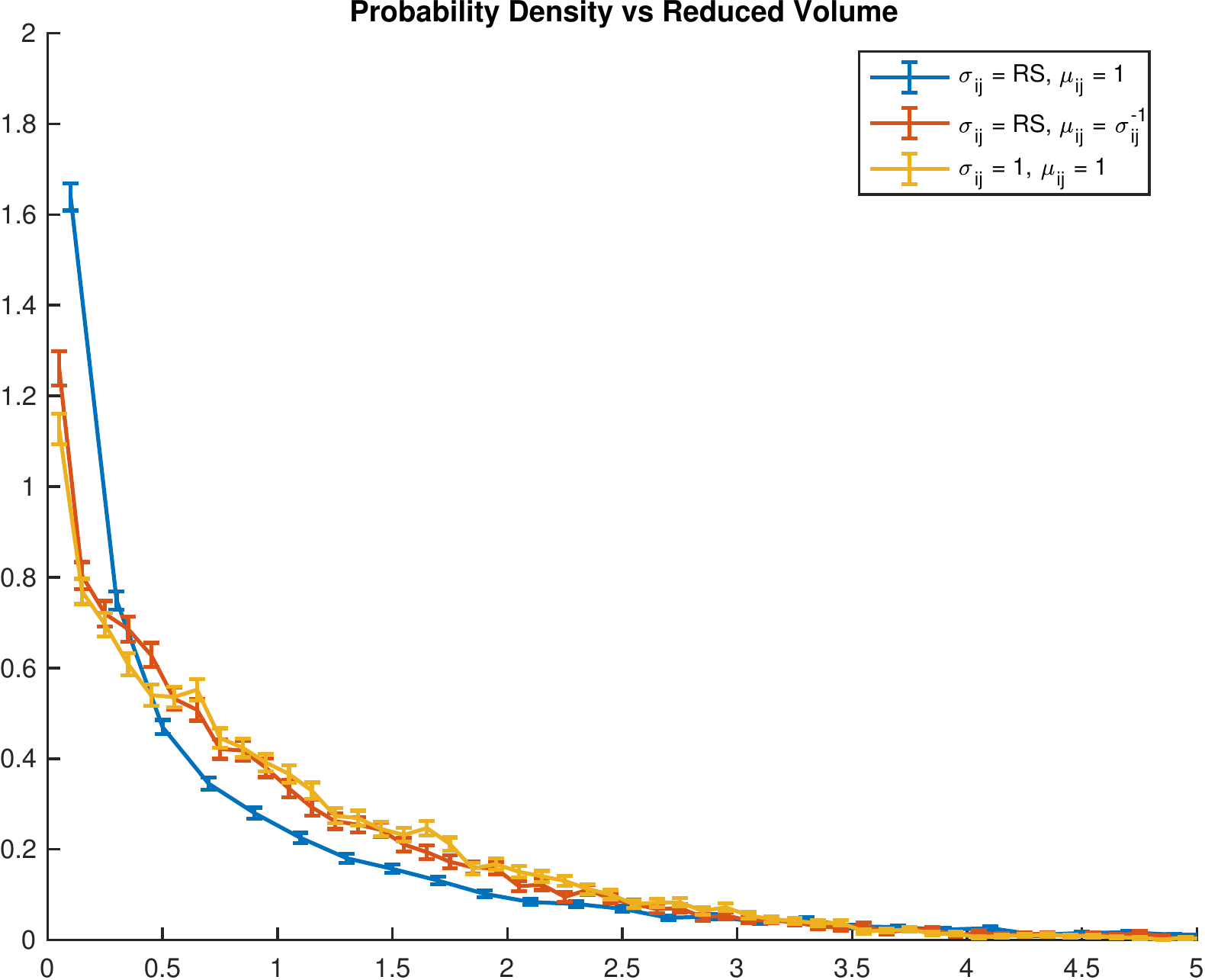}
\end{tabular}
\caption{GSD (probability density vs. reduced volume $\frac{V}{<V>}$) for the three different models: initial Voronoi data (left), at time $t_i$ when approximately 30\% of grains remain (center) and at time $t_f$ when approximately 10\% of grains remains (right).}
\label{fig:reduced_volume3d}
\end{figure}

\begin{figure}[h]
\centering
\begin{tabular}{ccc}
\includegraphics[width=\widththreefigures]{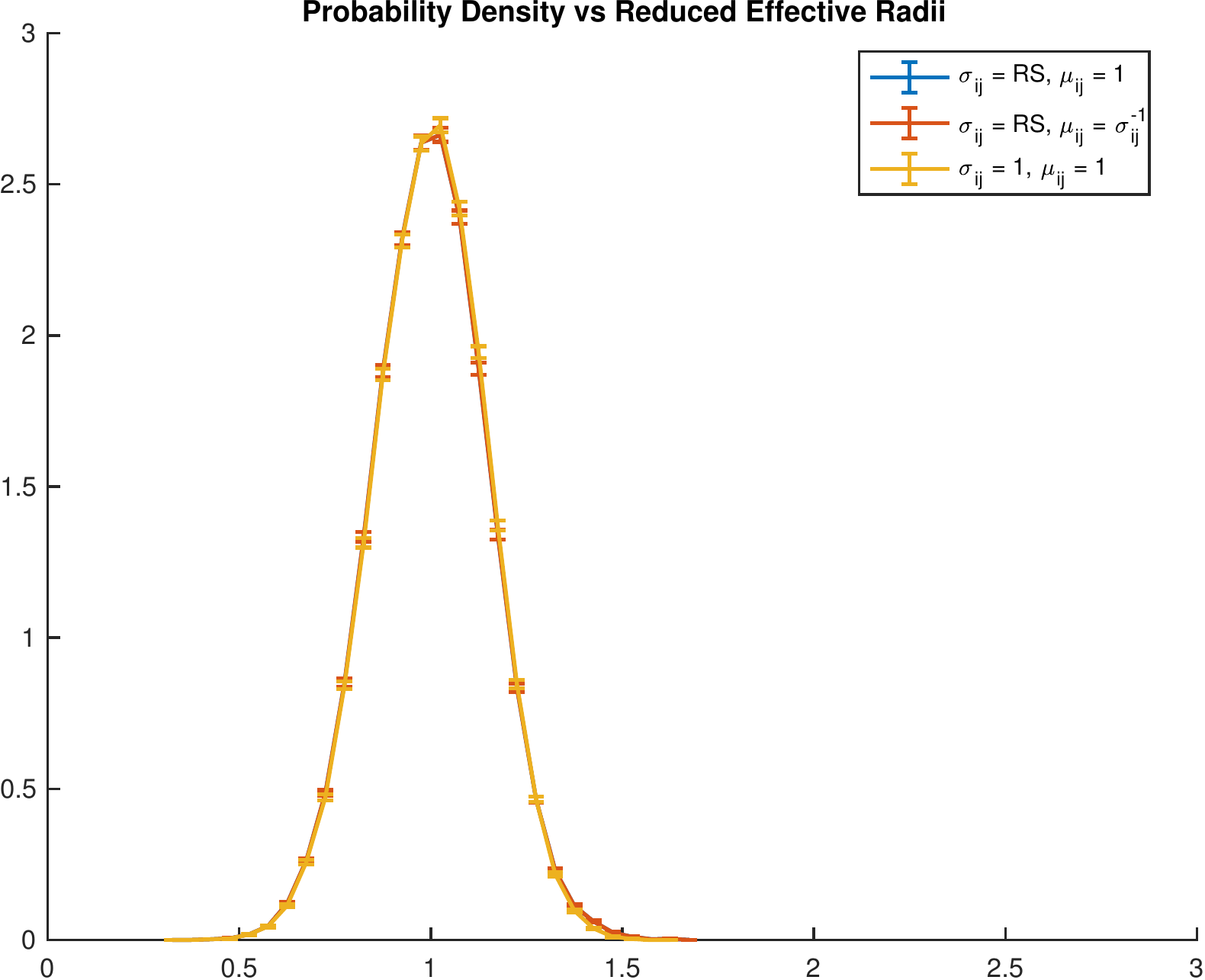} & \includegraphics[width=\widththreefigures]{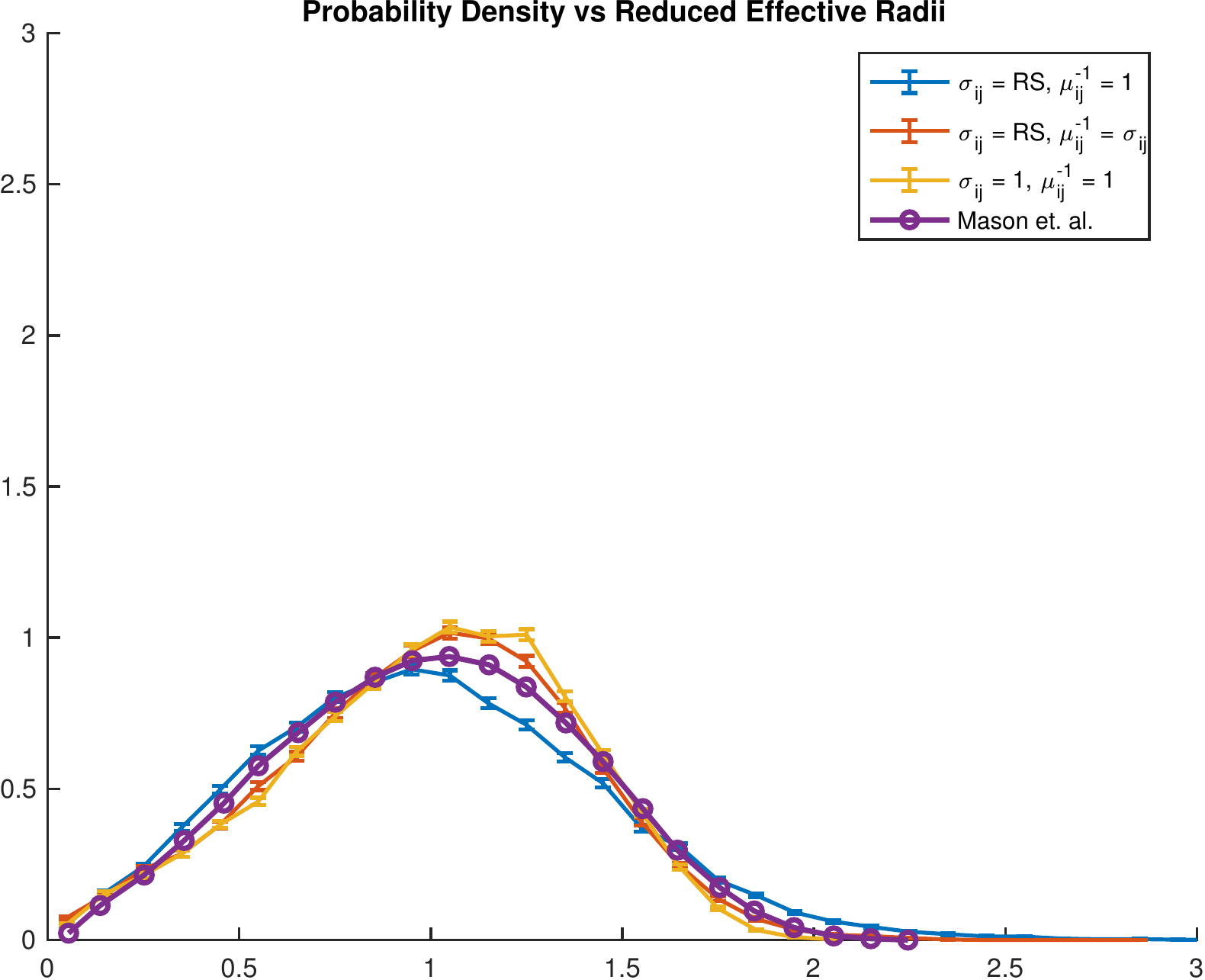} & \includegraphics[width=\widththreefigures]{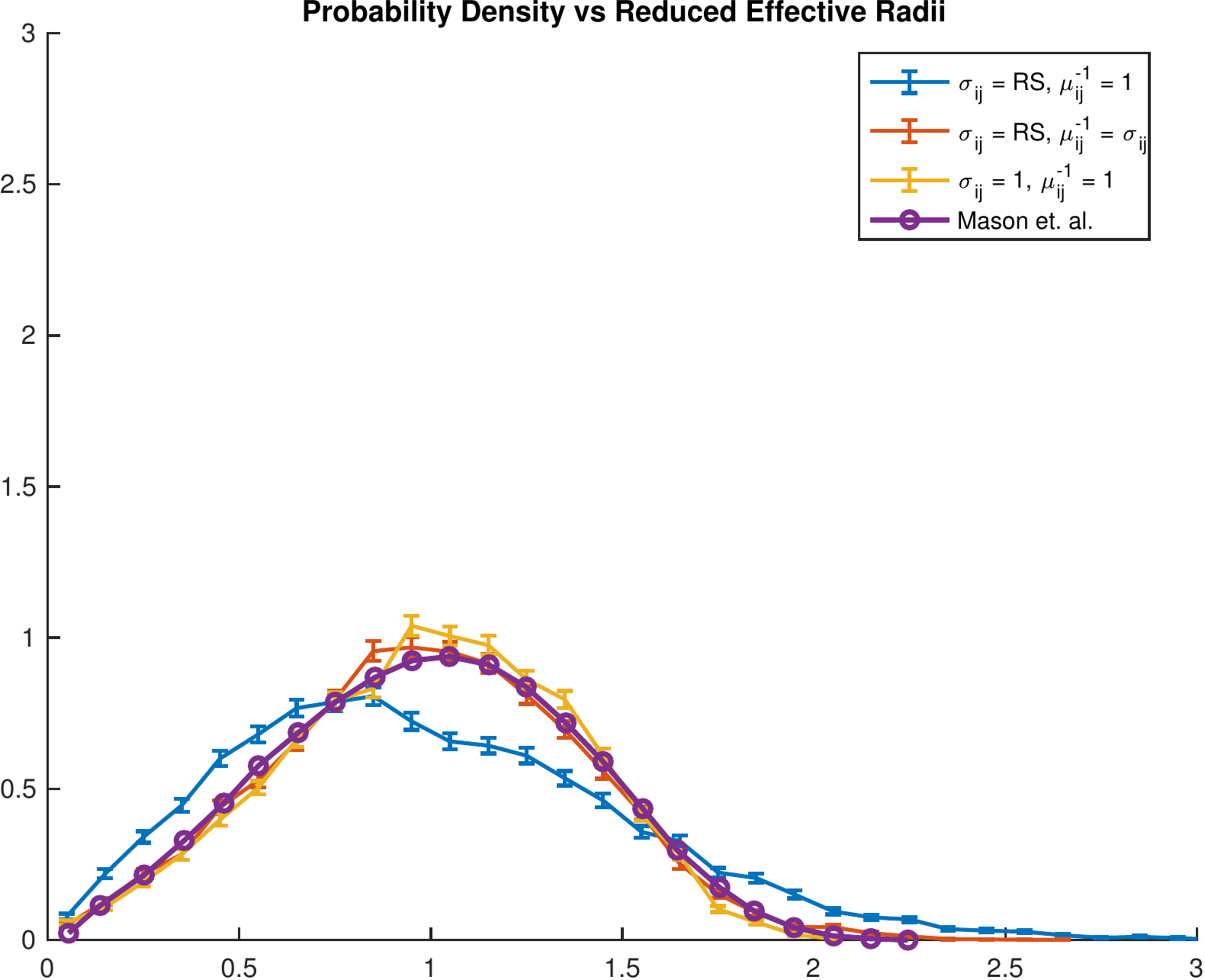}
\end{tabular}
\caption{Probability density vs. reduced effective radii $\frac{V^{1/3}}{<V^{1/3}>}$ for the three different models: initial Voronoi data (left), at time $t_i$ when approximately 30\% of grains remain (center) and at time $t_f$ when approximately 10\% of grains remains (right).}
\label{fig:reduced_effective_radii3d}
\end{figure}

\begin{figure}[h]
\centering
\begin{tabular}{cc}
\includegraphics[width=\widththreefigures]{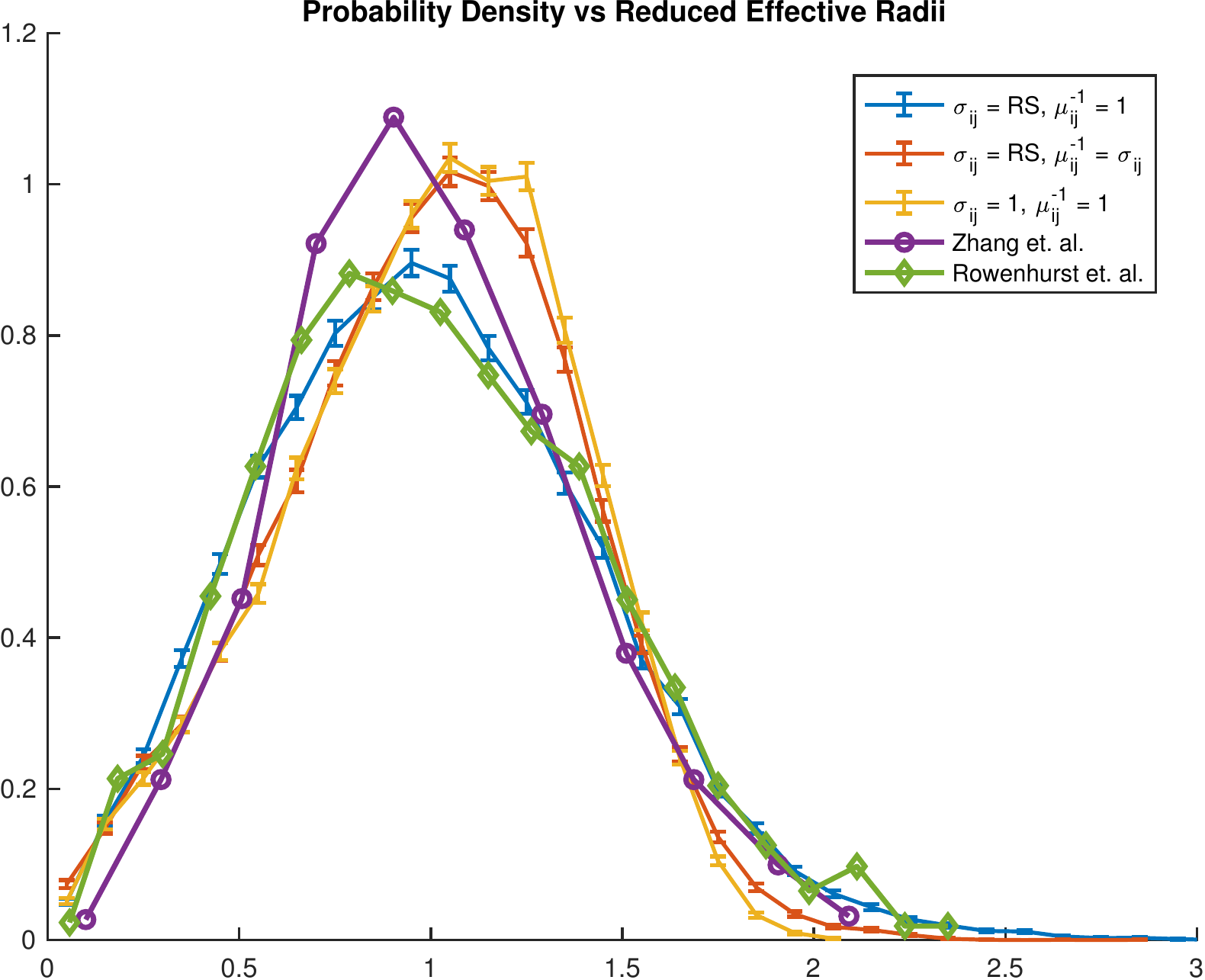} & \includegraphics[width=\widththreefigures]{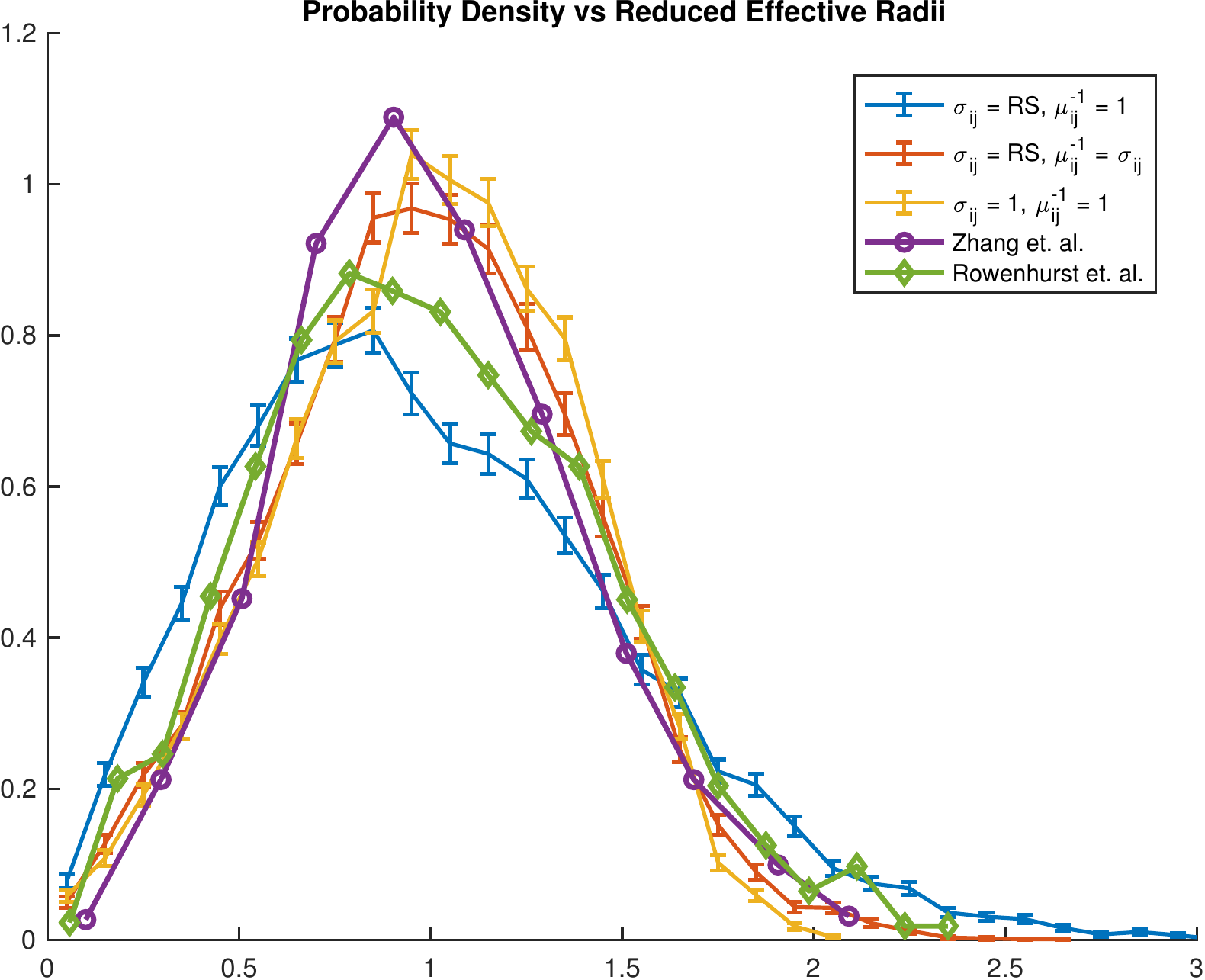}
\end{tabular}
\caption{Comparison of the GSD in terms of the reduced effective radii distributions in models (i), (ii) and (iii) at time $t_i$ when approximately 30\% of grains remain (left) and time $t_f$ when approximately 10\% of grains remain (right) with the experimental data from \cite{zhang} and \cite{rowenhurst}.}
\label{fig:reduced_effective_radii3dexperimental}
\end{figure}

Unlike in our 2D simulations, we see substantial differences in the eventual GSD reached by surface tension / mobility model (i) vs. models (ii) \& (iii).
Figures \ref{fig:reduced_volume3d} and Figure \ref{fig:reduced_effective_radii3d} show the time evolution of the GSD in terms of the reduced volume and corresponding effective radii distributions, respectively.
We note that while the effective radii distributions of (ii) \& (iii) remain very close to each other throughout the evolution (and agree with previous 3D simulations of the same model via different numerical methods, e.g. \cite{Kim2006,ElseyEsedogluSmerekaRoyal,Miyoshi2018,MasonStatisticsGBM}), that of (i) is more spread out, and its peak is shifted in the direction of smaller grains -- a difference not seen in the 2D simulations with the same algorithms, and at very similar resolutions.
Figure \ref{fig:reduced_effective_radii3dexperimental} suggests the different eventual GSD observed with surface tension / mobility model (i) may not be inconsistent with experimental data:
Two experimental distributions available in existing literature, namely \cite{zhang} and \cite{rowenhurst}, display substantial deviation from the GSD seen in simulations of the isotropic model (iii) in our and previous simulations such as \cite{Kim2006,ElseyEsedogluSmerekaRoyal,Miyoshi2018,MasonStatisticsGBM}, in the same direction as our simulations with model (i), namely in the direction of smaller grains and more spread out distribution.

\begin{figure}[h]
\centering
\begin{tabular}{ccc}
\includegraphics[width=\widththreefigures]{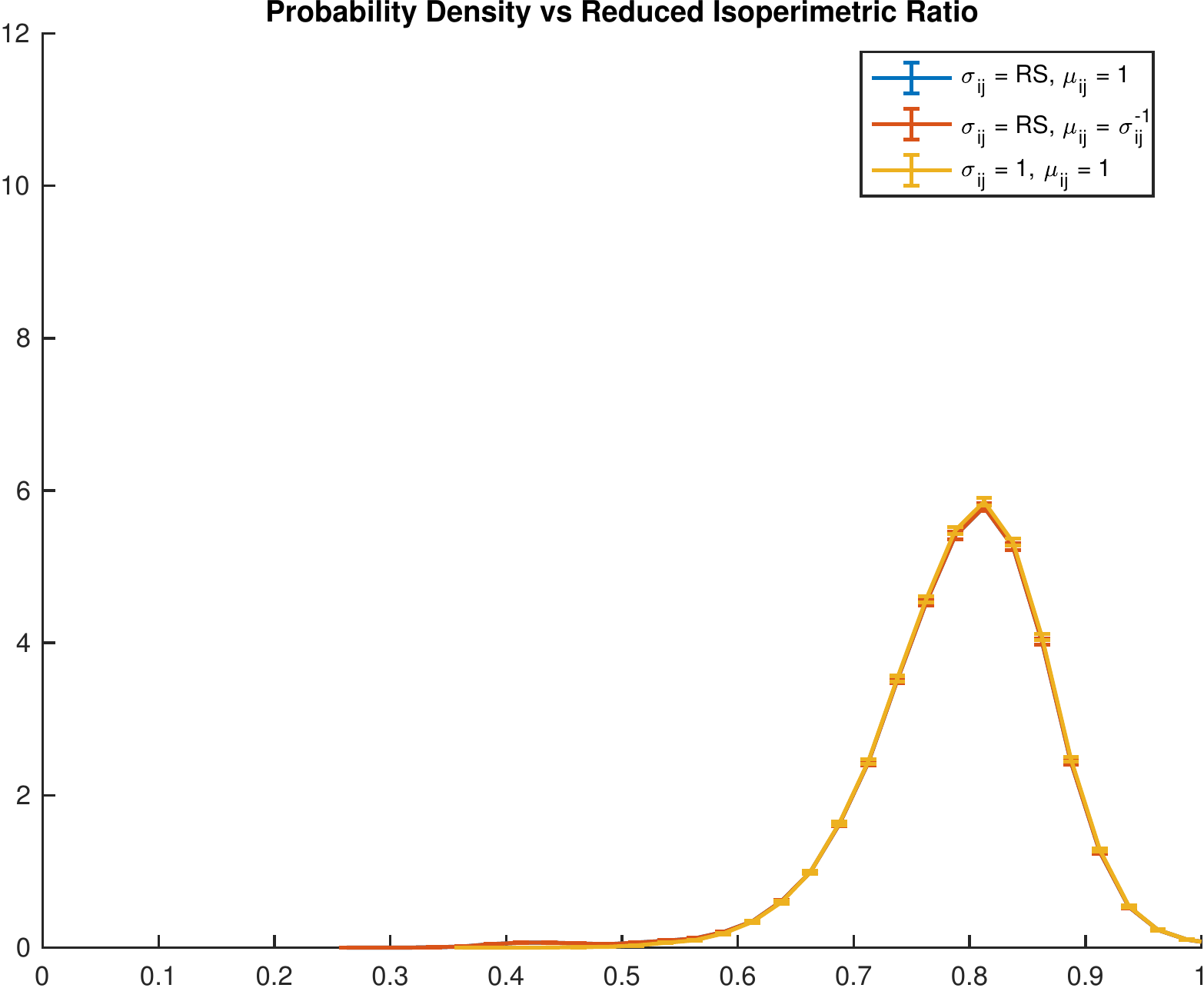} & \includegraphics[width=\widththreefigures]{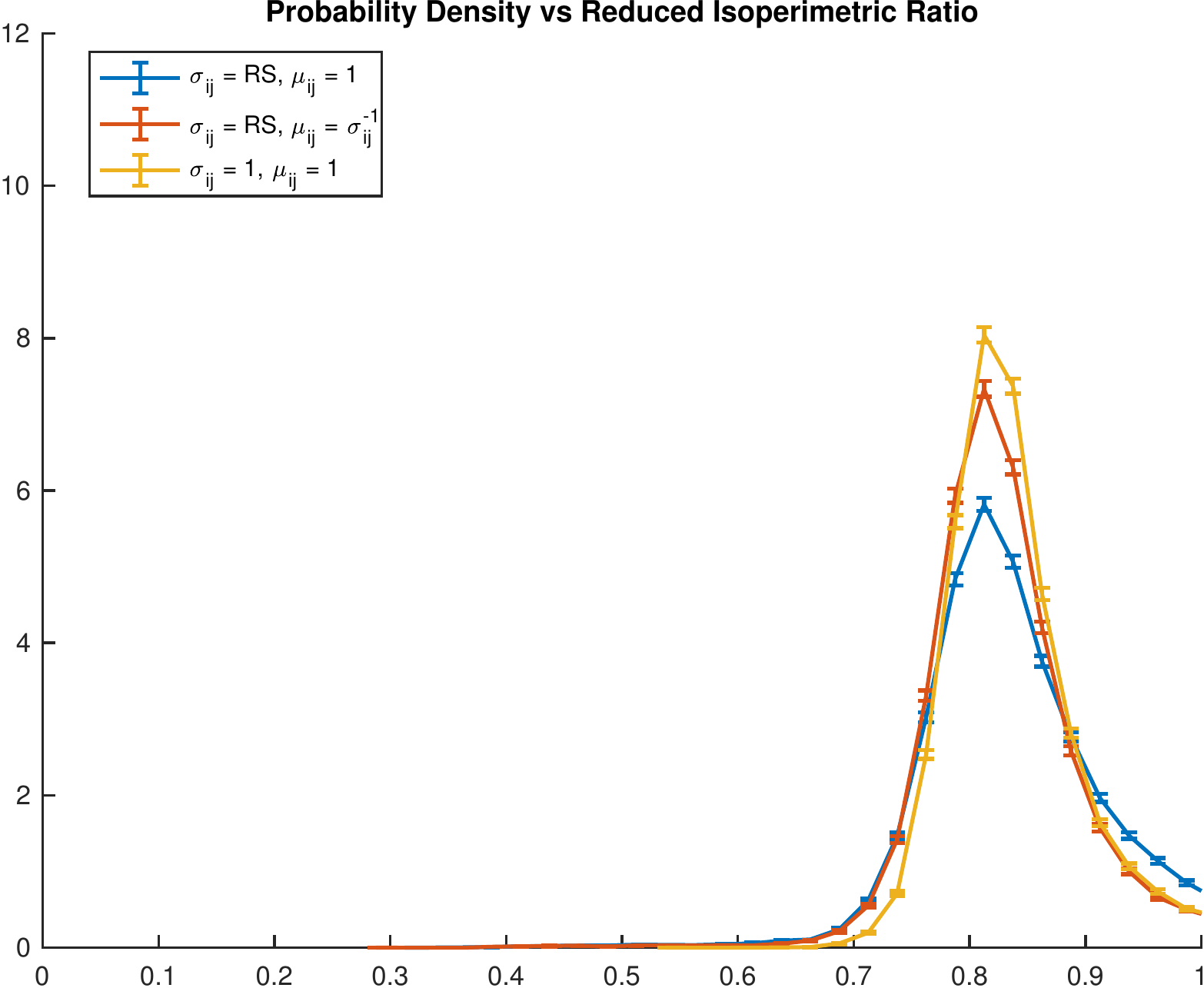} & \includegraphics[width=\widththreefigures]{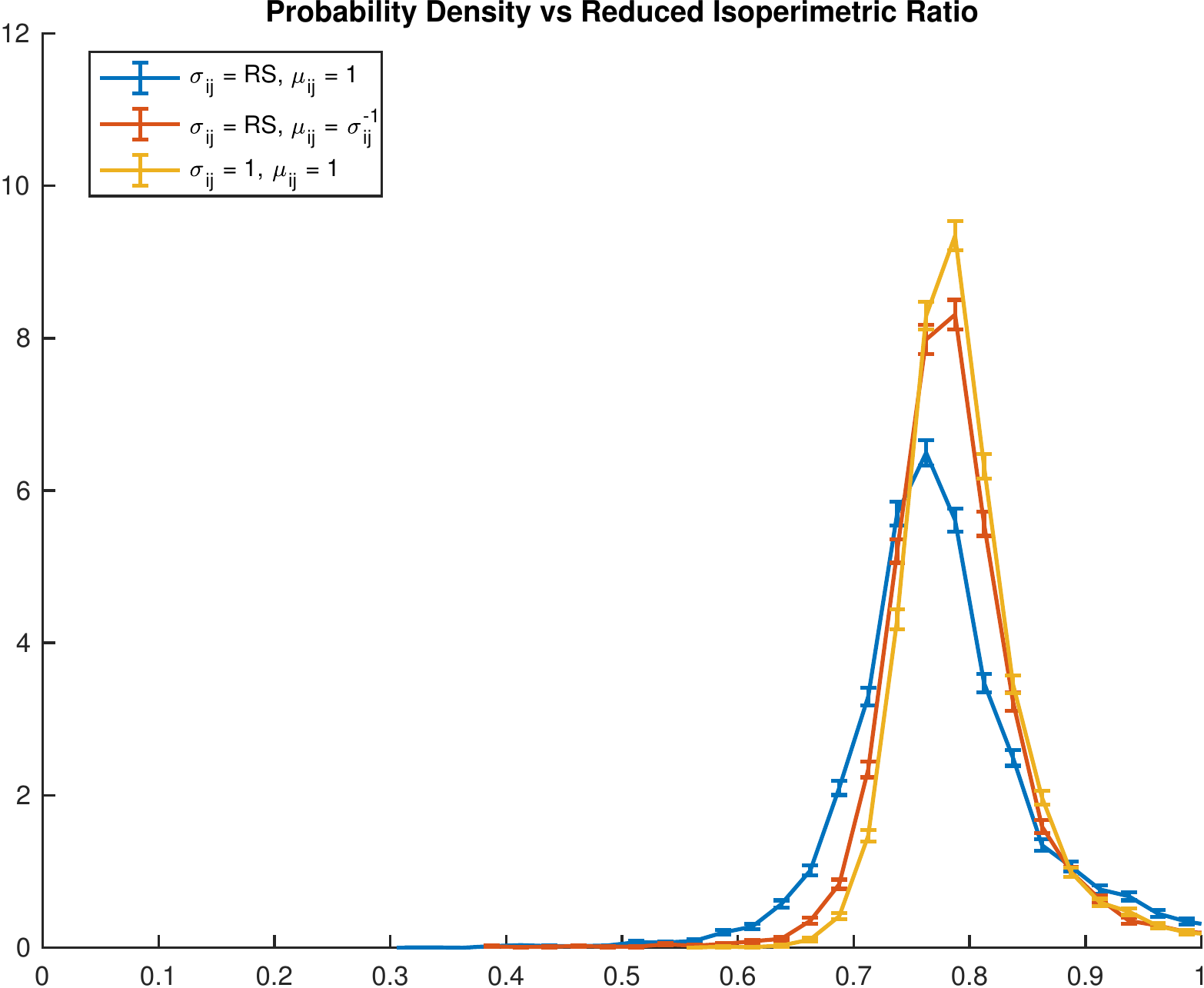}
\end{tabular}
\caption{Probability Density vs Isoperimetric Ratio $\frac{36\pi V^2}{S^3}$ for the three different models: initial Voronoi data (left), at time $t_i$ when approximately 30\% of grains remain (center) and at time $t_f$ when approximately 10\% of grains remains (right).}
\label{fig:isopratio3d}
\end{figure}

Figure \ref{fig:isopratio3d} shows the distribution of the isoperimetric ratio, namely $\frac{36\pi V^2}{S^3}$, where $V$ denotes the volume and $S$ the surface area of the grain.
Remarkably, we see less difference between surface tension / mobility models (ii) and (iii) in 3D than we did in 2D simulations: their stationary isoperimetric ratio distributions are almost identical.
The distribution for model (i), however, is broader, indicating a greater presence of eccentric grains compared to the isotropic model (iii), as in our 2D results show in Figure \ref{fig:isopratio2d}.


\begin{figure}[h]
\centering
\begin{tabular}{ccc}
\includegraphics[width=\widththreefigures]{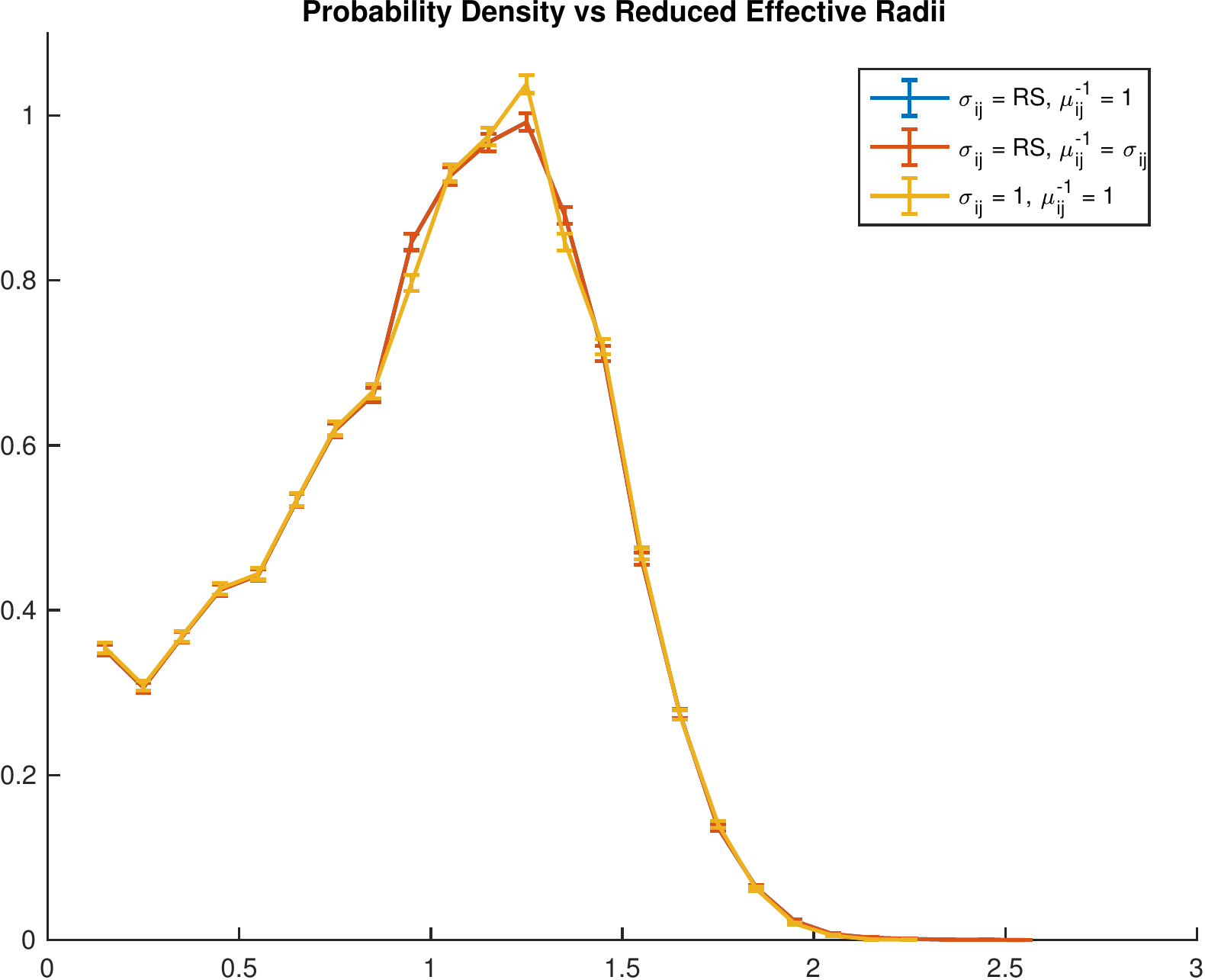} & \includegraphics[width=\widththreefigures]{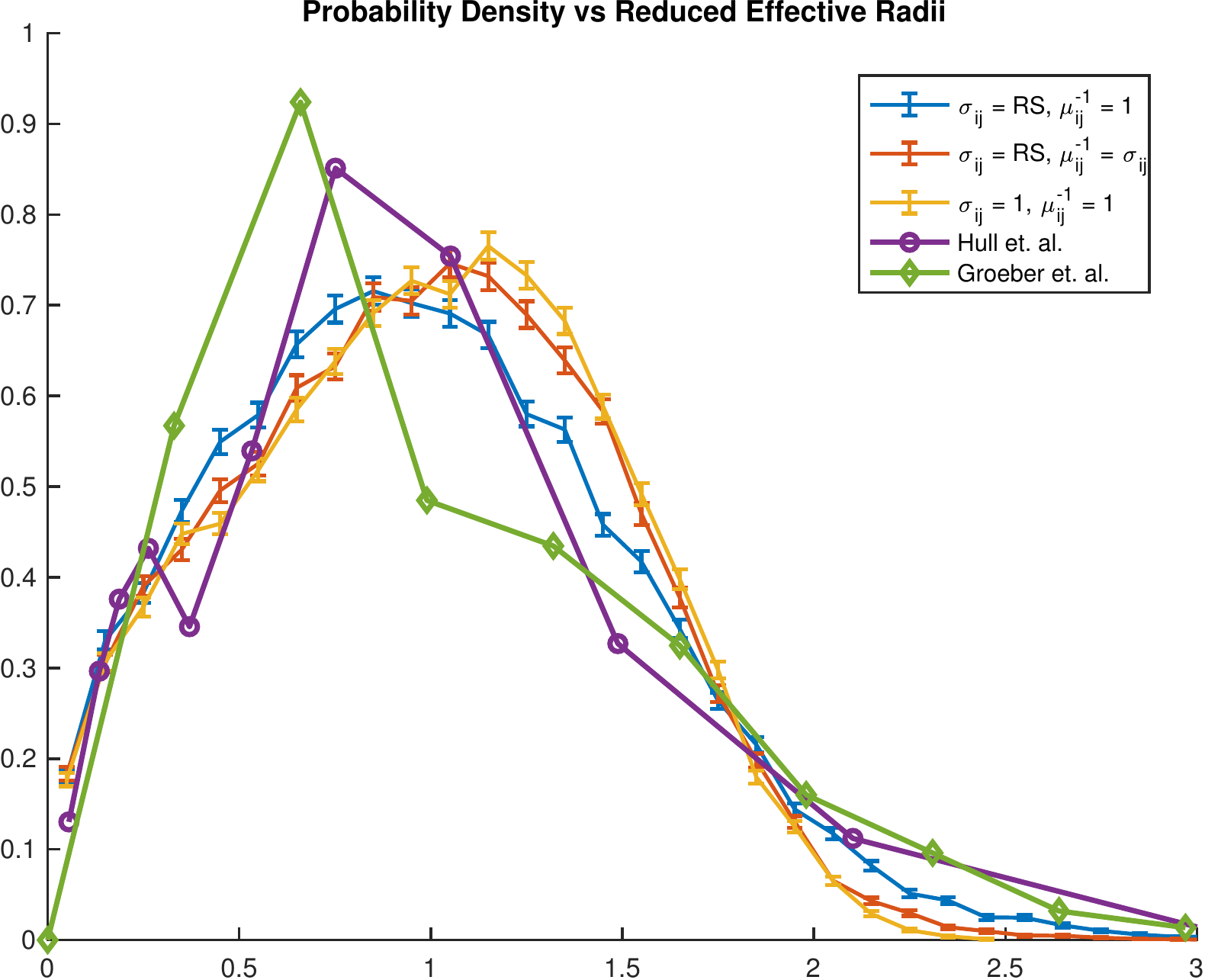} & \includegraphics[width=\widththreefigures]{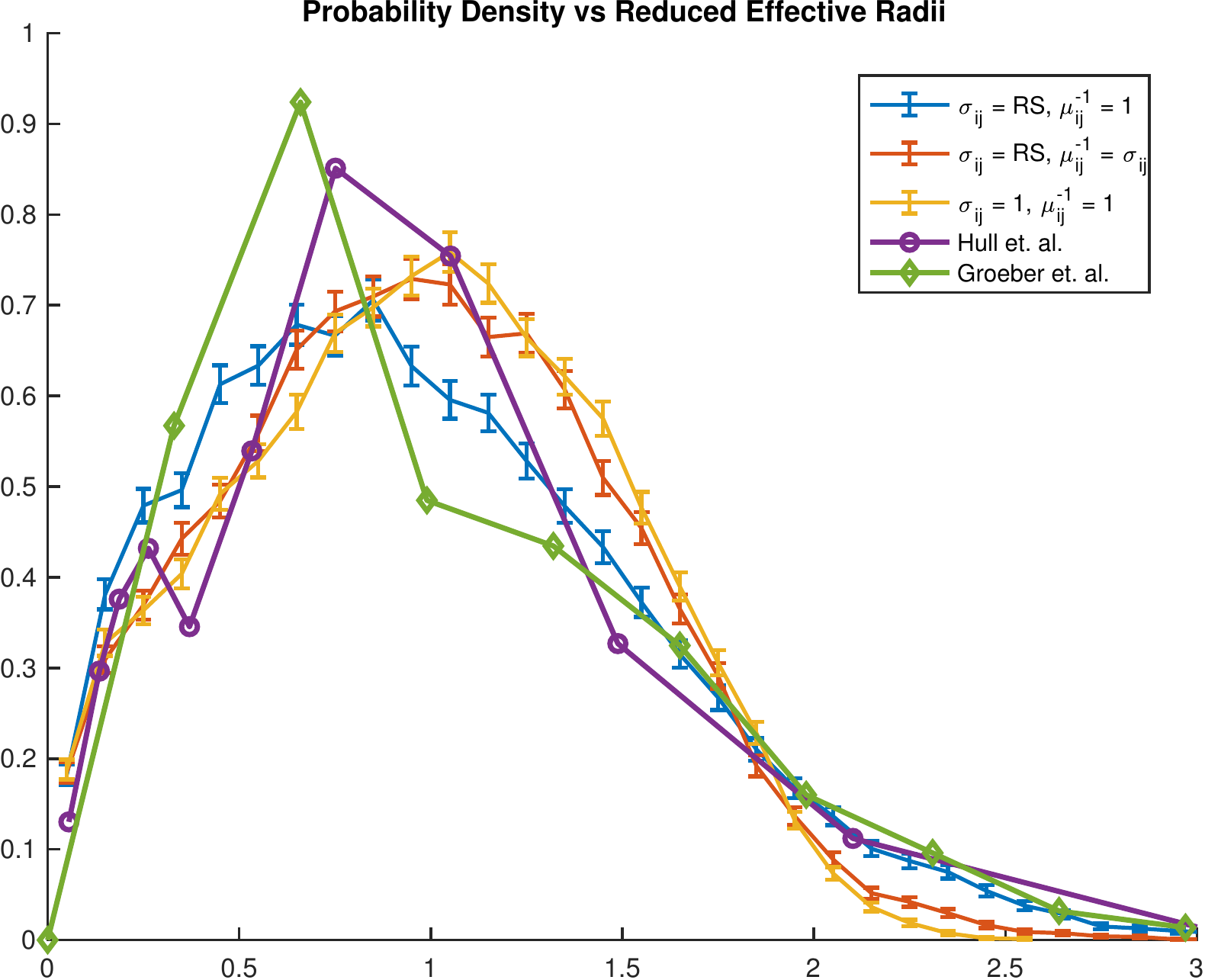}
\end{tabular}
\caption{Probability density vs reduced effective radii distributions $\frac{\sqrt{A}}{\sqrt{<A>}}$ taken from the cross-sections of constant $z$-value from the initial Voronoi data (left), at time $t_i$ when approximately 30\% of grains remain (center) and $t_f$ when approximately 10\% of grains remain (right). Experimental data from \cite{hull} and \cite{groeber} is included for comparison.}
\label{fig:reduced_effective_radii3DX}
\end{figure}

\begin{figure}[h]
\centering
\begin{tabular}{ccc}
\includegraphics[width=\widththreefigures]{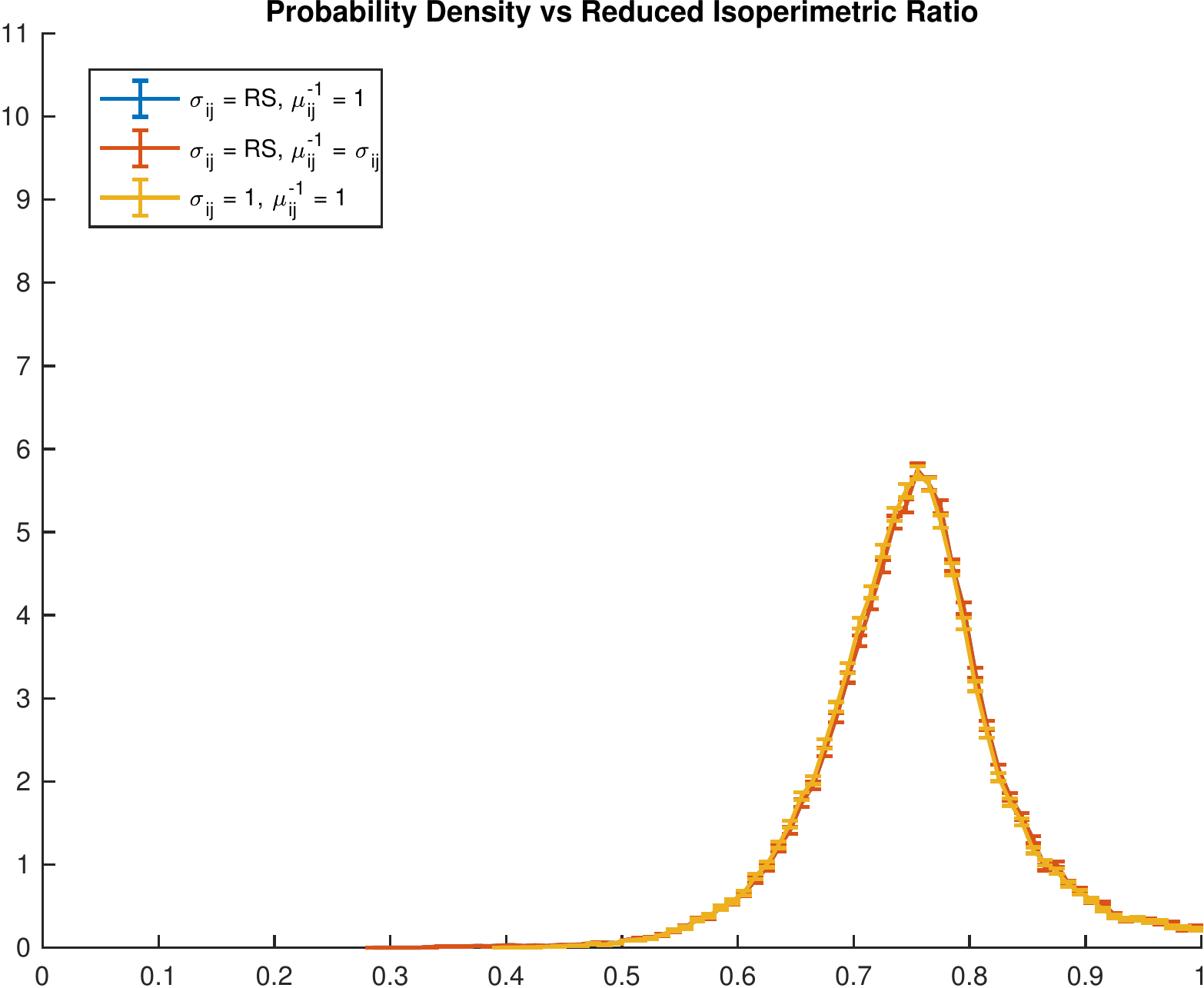} & \includegraphics[width=\widththreefigures]{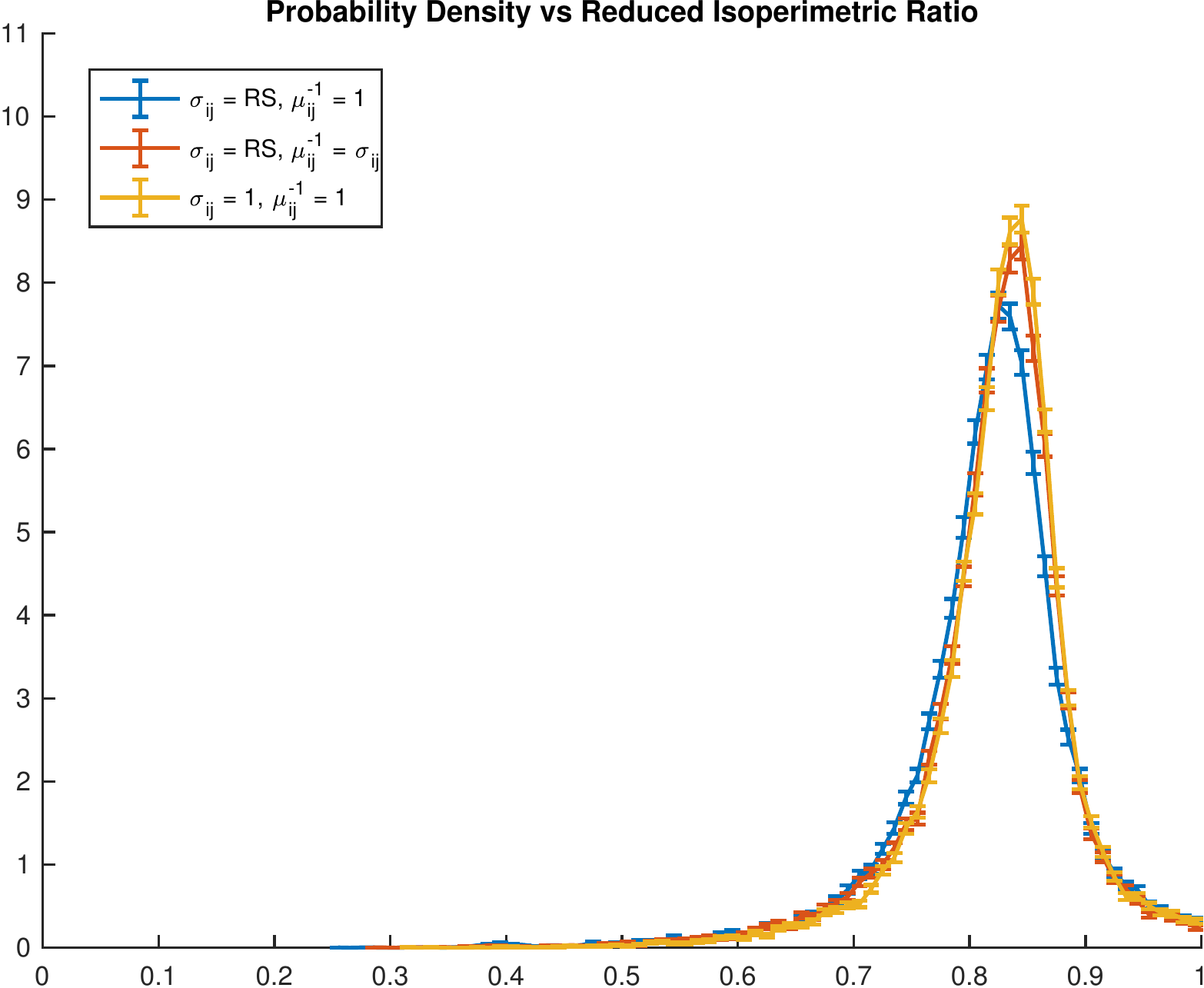} & \includegraphics[width=\widththreefigures]{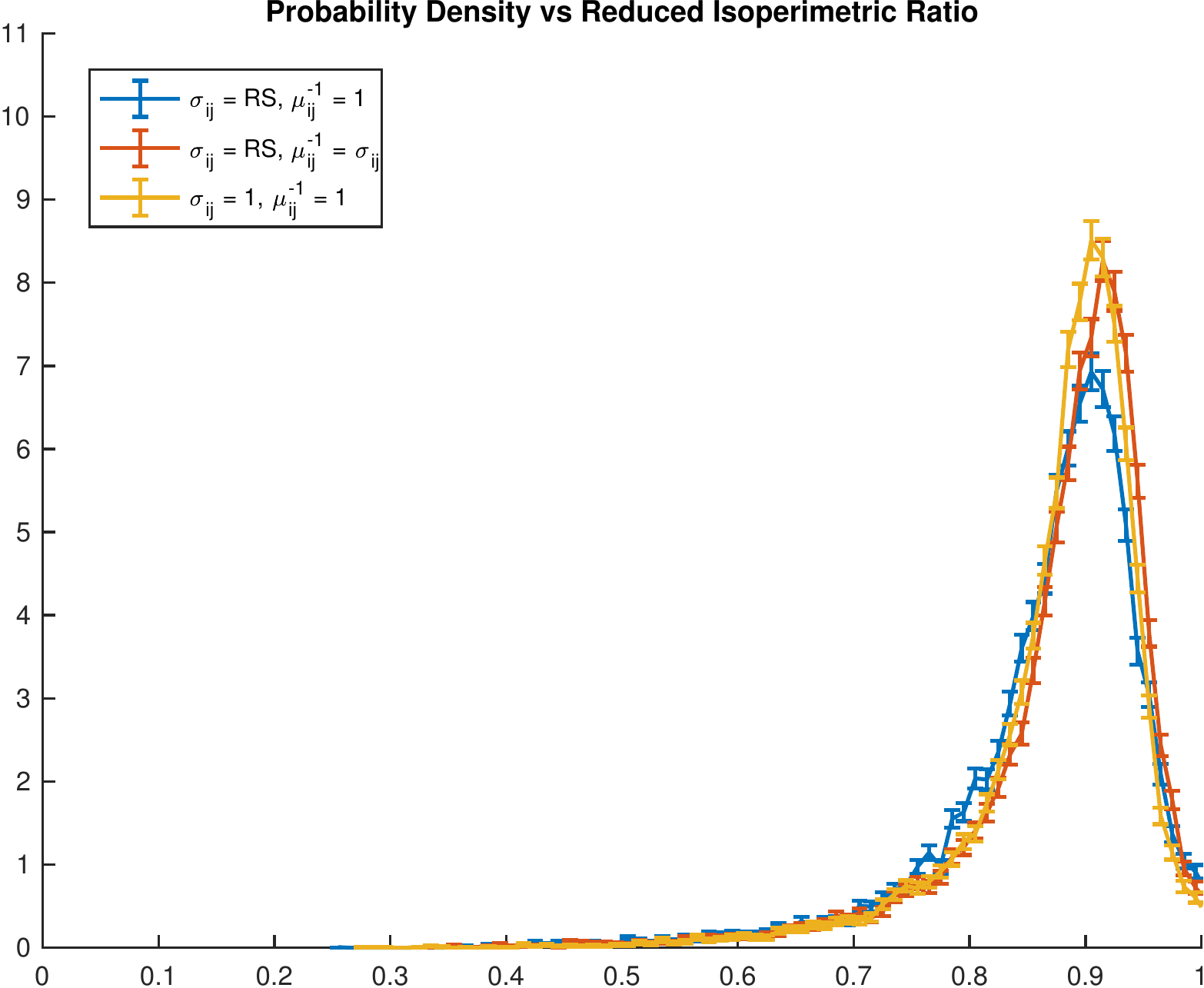}
\end{tabular}
\caption{Probability density vs isoperimetric ratio $\frac{4\pi A}{L^2}$ taken from the cross-sections of constant $z$-value from the initial Voronoi data (left), at time $t_i$ when approximately 30\% of grains remain (center) and $t_f$ when approximately 10\% of grains remain (right).}
\label{fig:isopratio3DX}
\end{figure}

We now turn to statistics of grain shapes and sizes in 2D slices of 3D data, often called 3DX statistics in the literature.
To this end, we took a series of cross sections parallel to the faces of the cubic computational domain $[0,1]^3$.
They were spaced roughly five average grain diameters apart to reduce correlations between neighboring cross sections. Figure \ref{fig:reduced_effective_radii3DX} shows the distribution of GSD in terms of the reduced effective radii distributions.
The results are similar to the fully 3D grains: the surface tension / mobility models (ii) and (iii) are close to each other, while the distribution for model (i) is more spread out and its peak is achieved at smaller grains. 
In Figure \ref{fig:isopratio3DX} we compare the distribution of the isoperimetric ratio, namely $\frac{4\pi A}{P^2}$, where $A$ denotes the area and $P$ the perimeter of the grain in the 2D slice. All surface tension / mobility models present almost identical peak distributions, with models (ii) and (iii) attaining a higher peak.

\begin{figure}[h]
 \centering
 \begin{tabular}{ccc}
 \includegraphics[width=\widththreefigures]{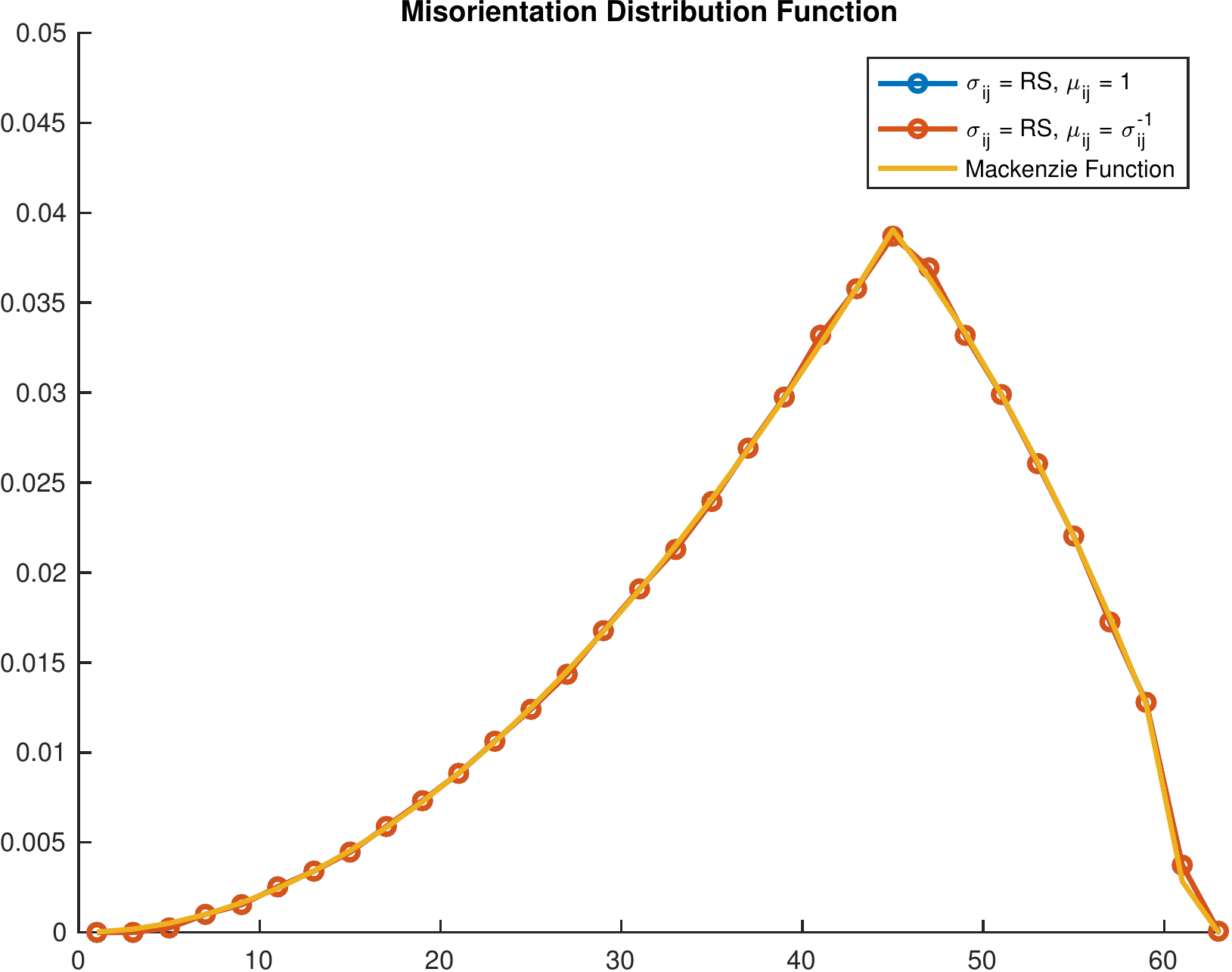} &\includegraphics[width=\widththreefigures]{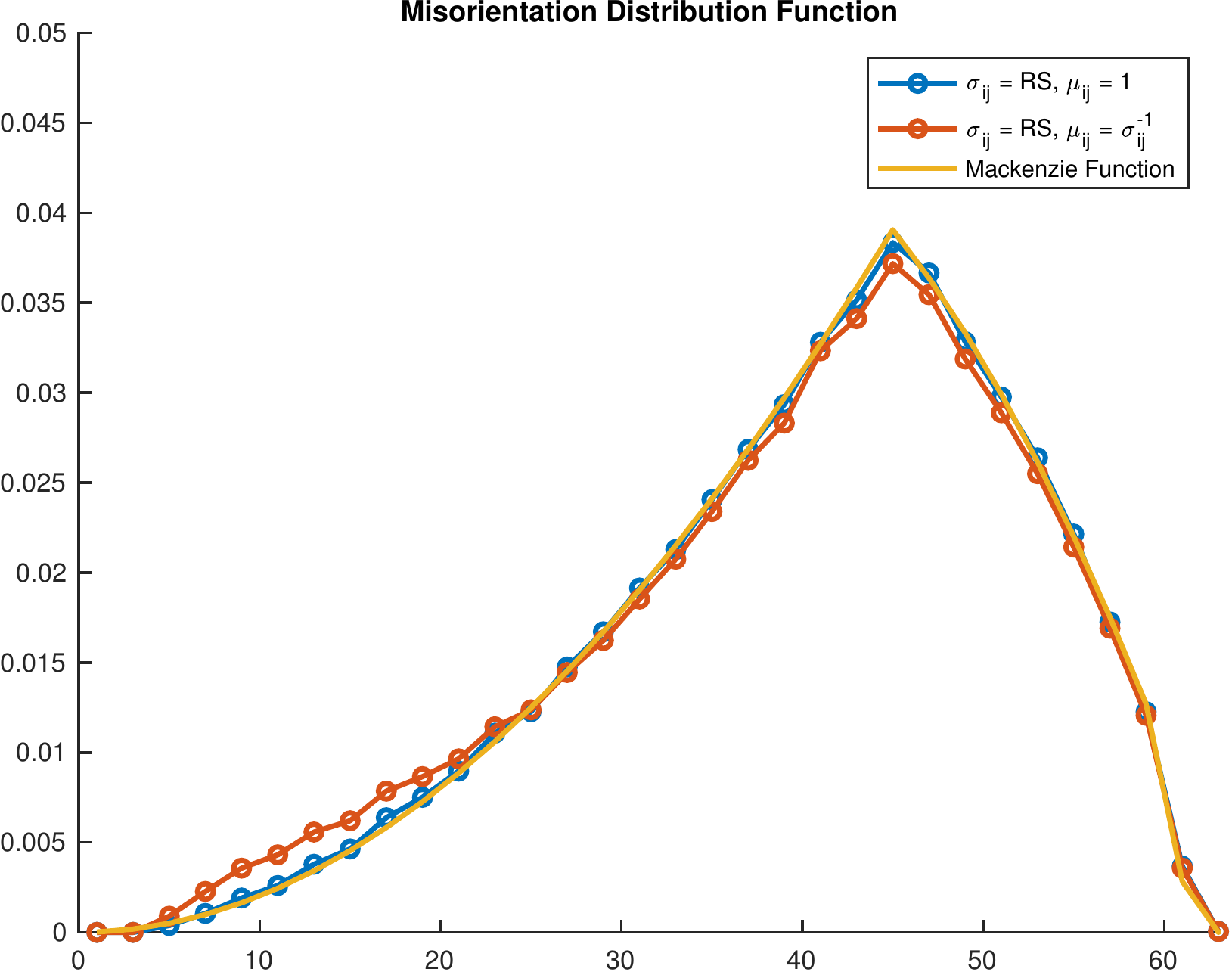} & \includegraphics[width=\widththreefigures]{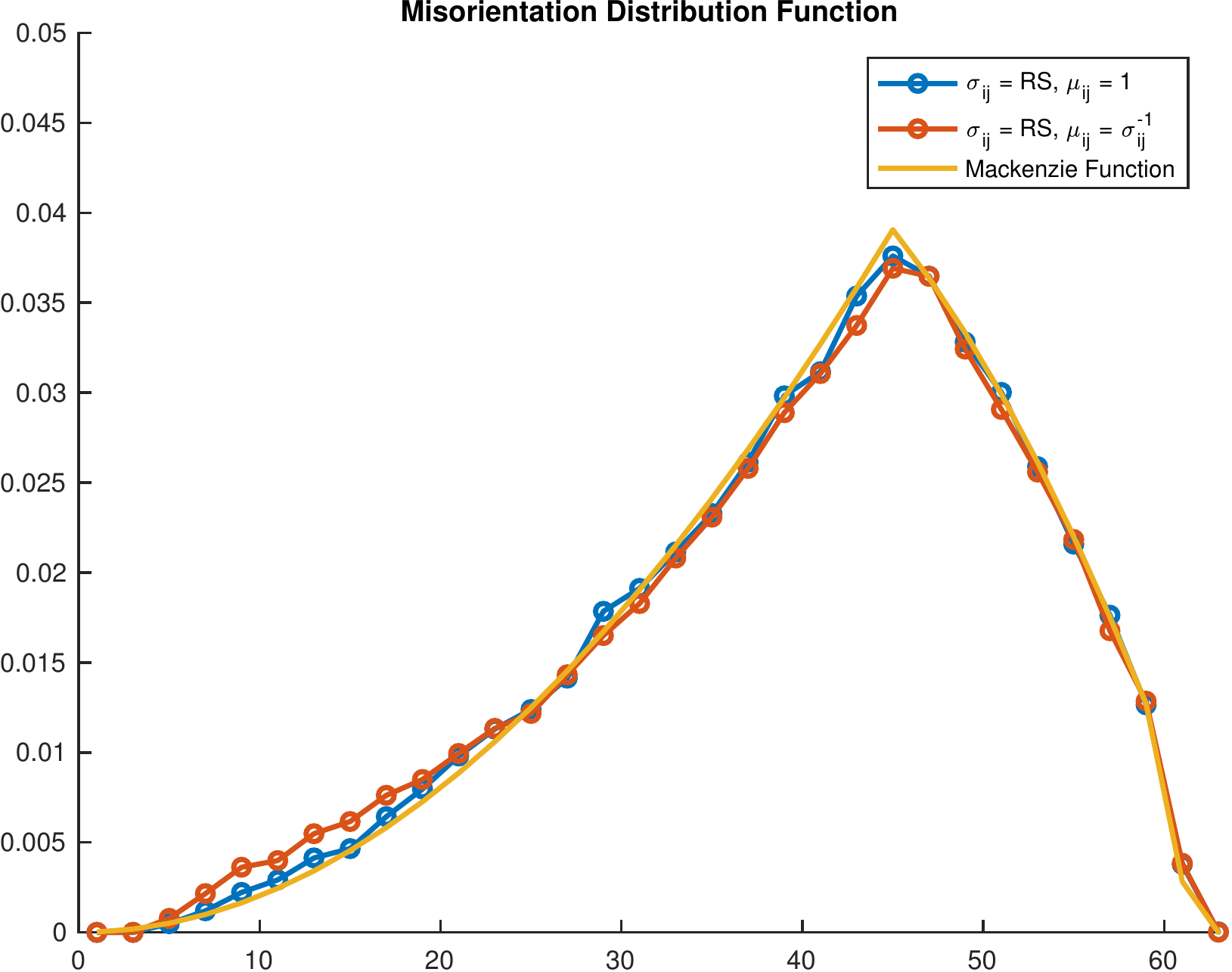}\\
 \includegraphics[width=\widththreefigures]{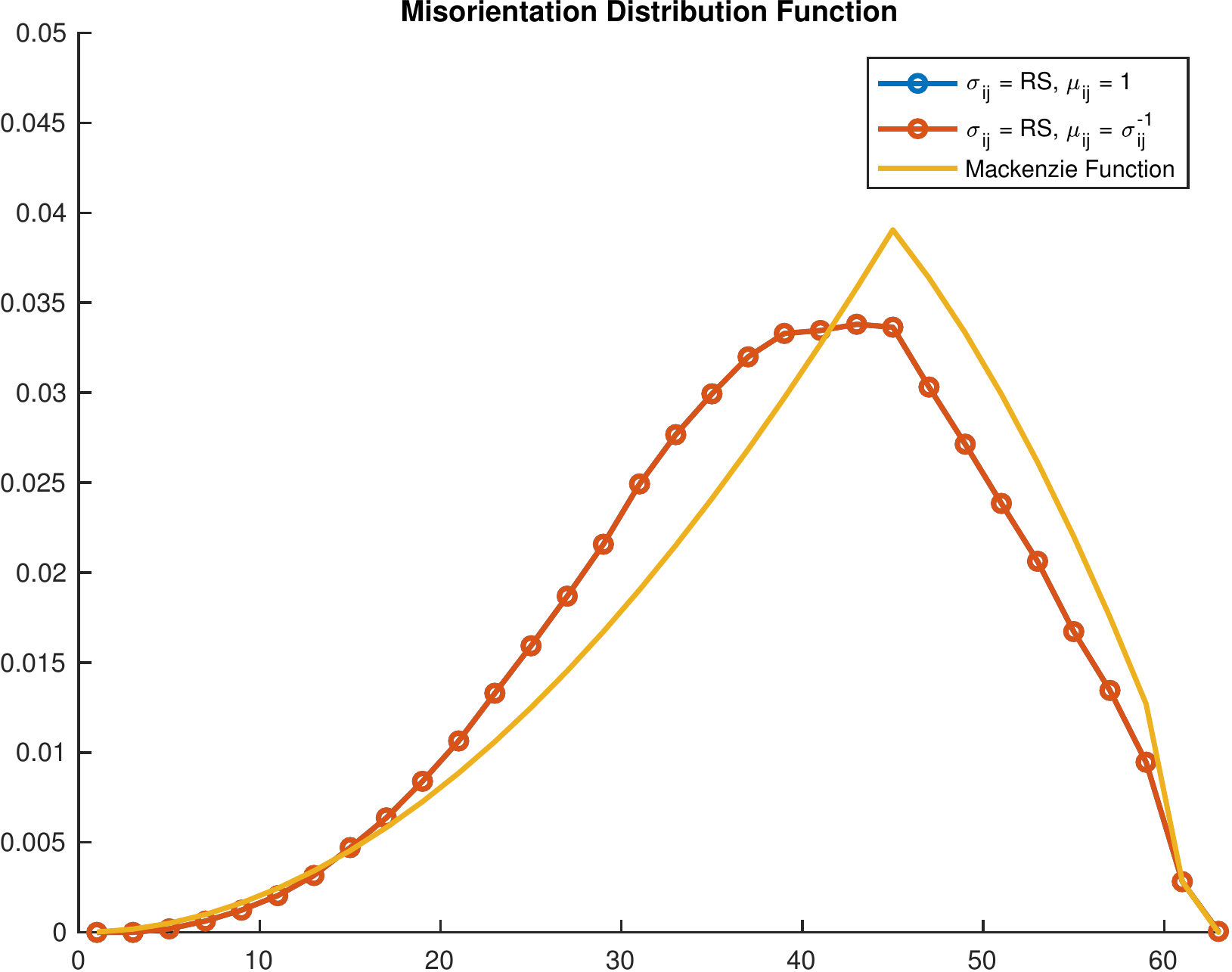} &\includegraphics[width=\widththreefigures]{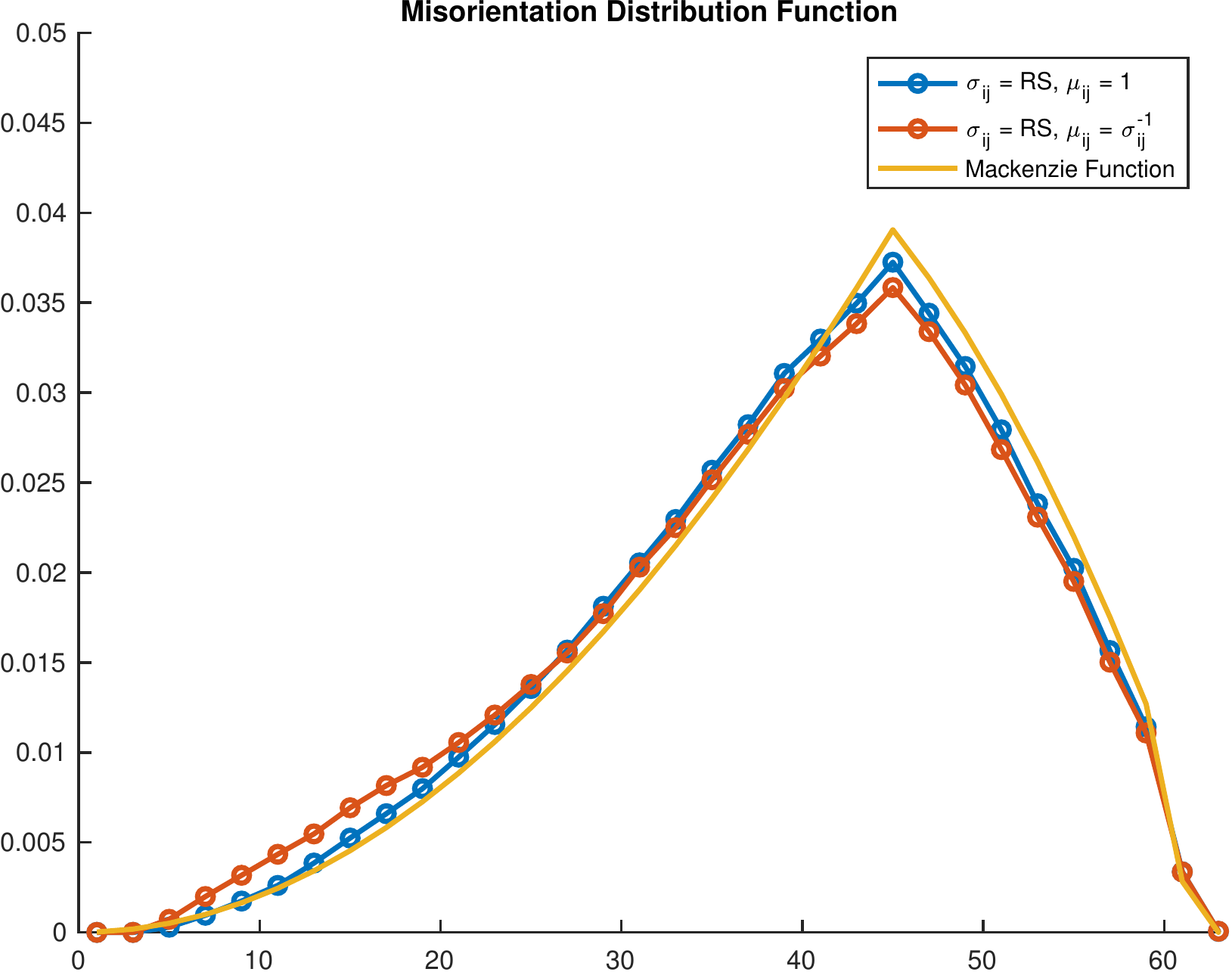} & \includegraphics[width=\widththreefigures]{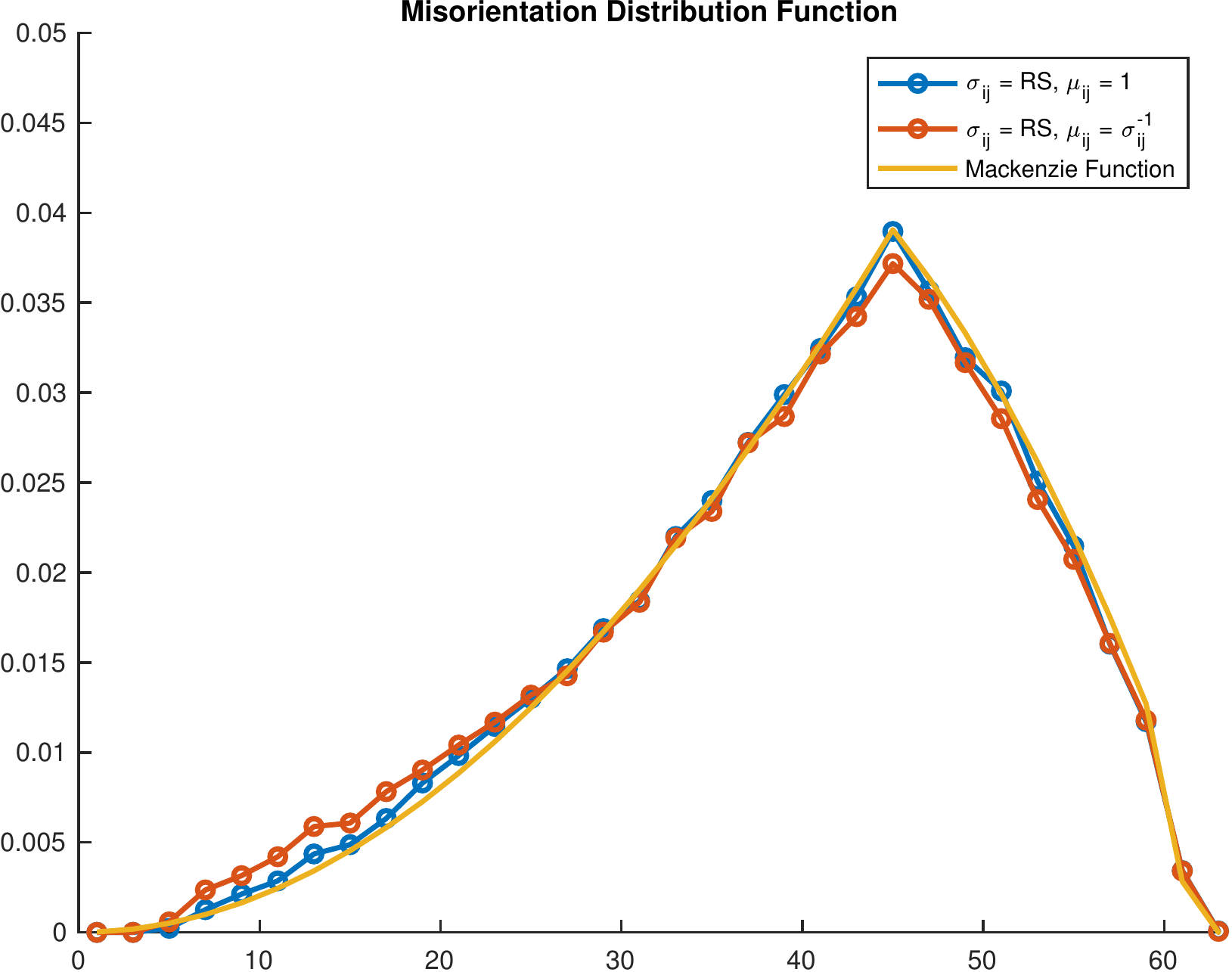}\\
 \includegraphics[width=\widththreefigures]{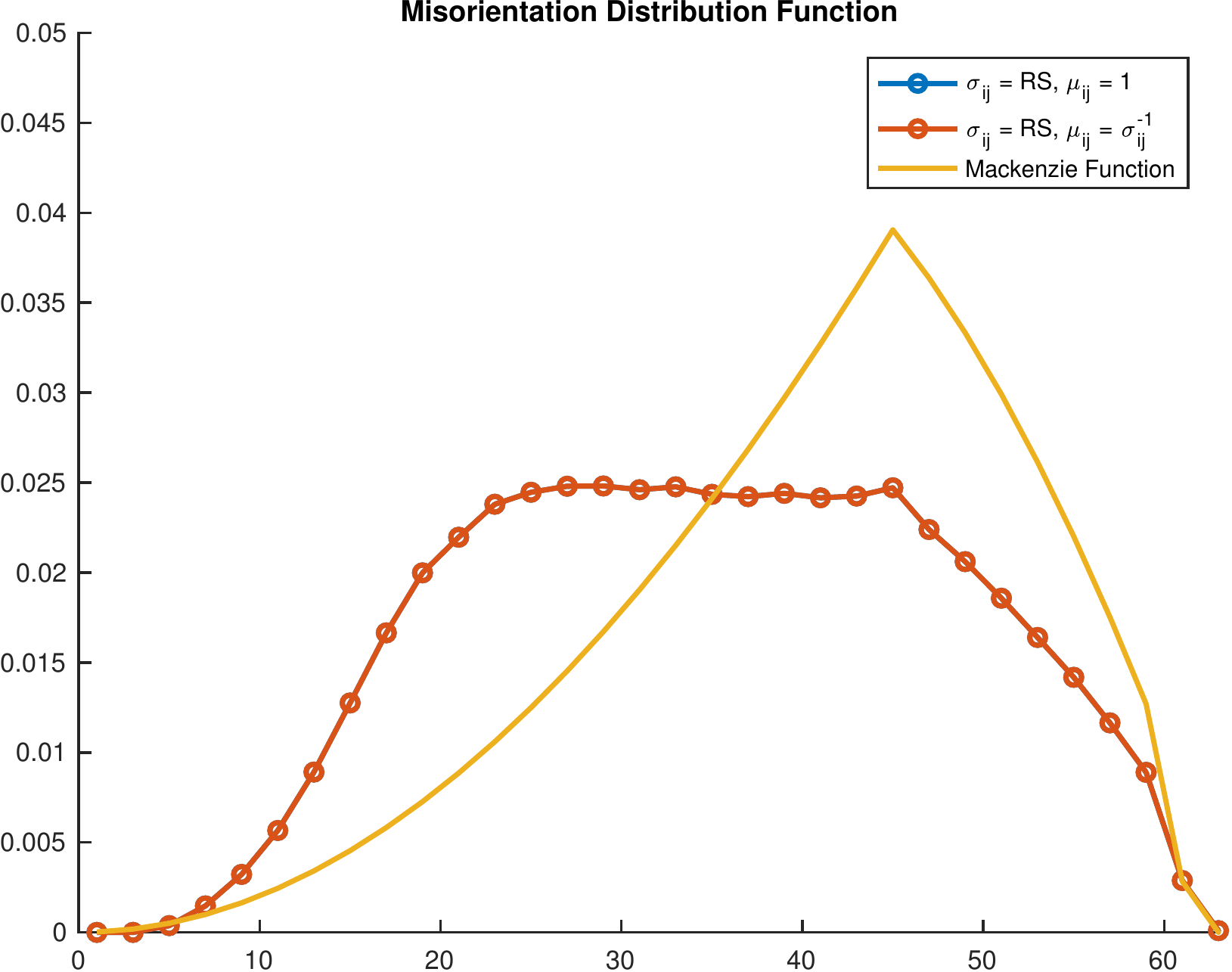} &\includegraphics[width=\widththreefigures]{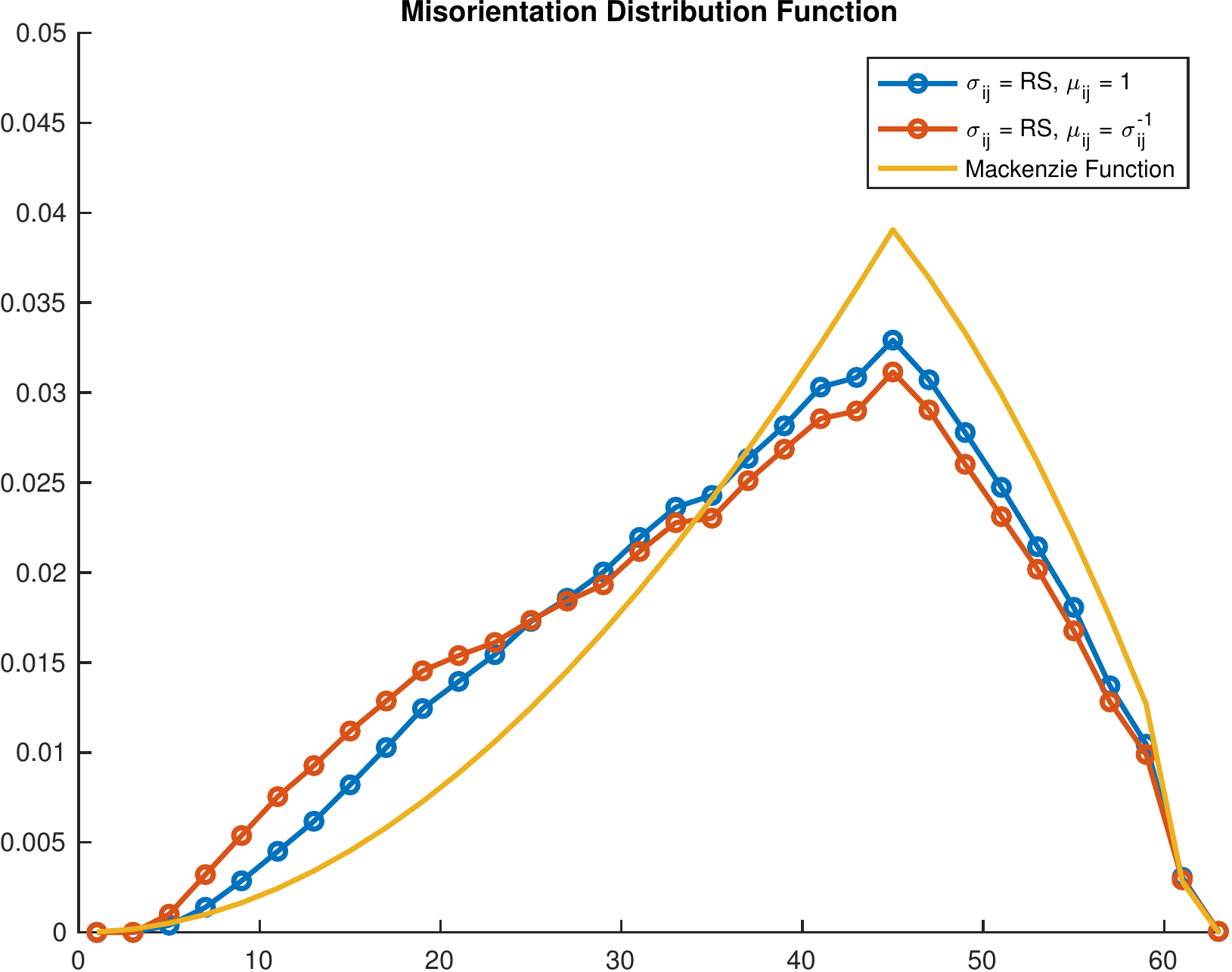} & \includegraphics[width=\widththreefigures]{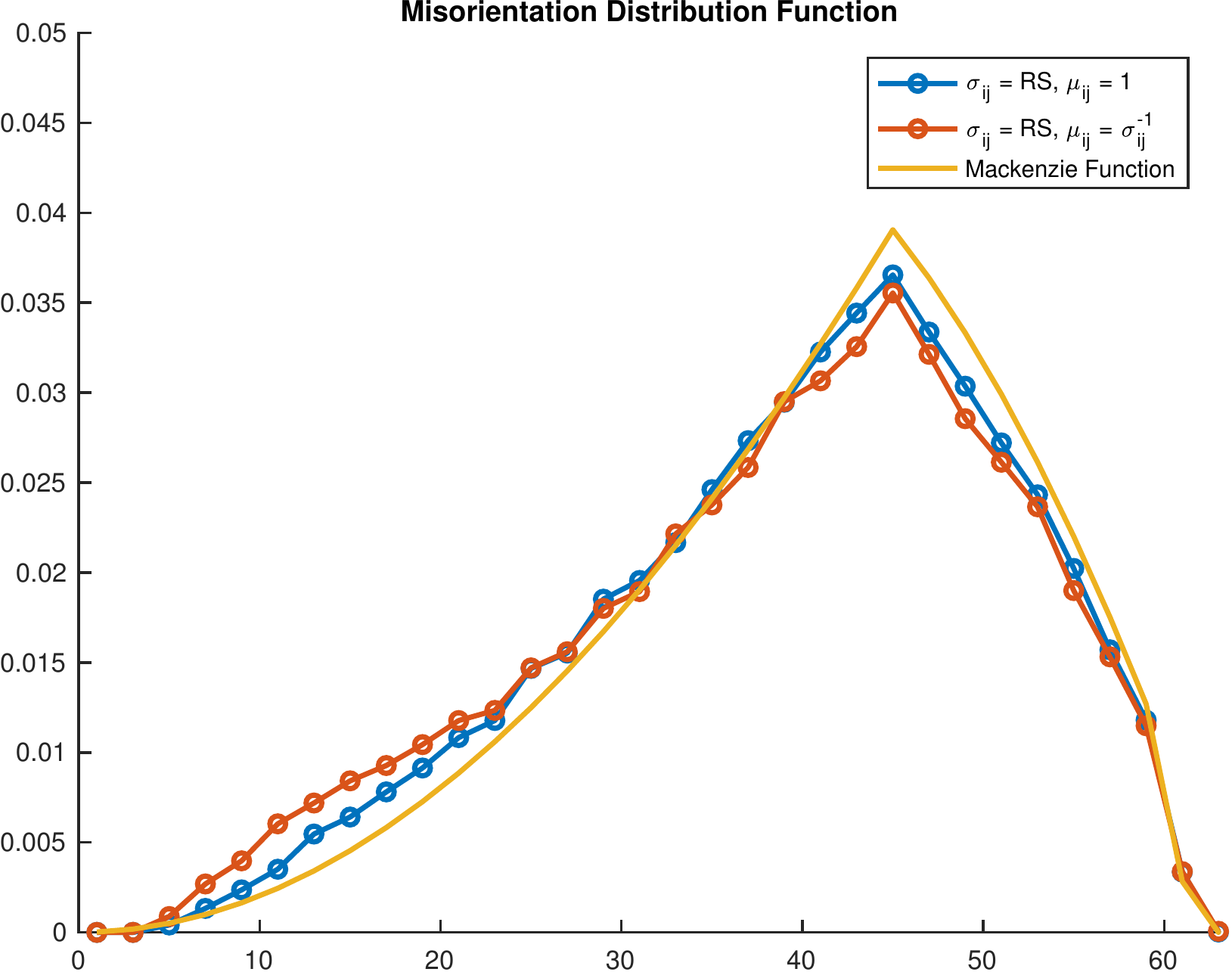}
 \end{tabular}
 \caption{Misorientation distribution function for models (i) and (ii) with different initial MDFs at the initial time (left column), at time $t_i$ when approximately 30\% of grains remain (middle column) and time $t_f$ when approximately 10\% of grains remain (right column).}
 \label{fig:MDF3d}
 \end{figure}

Finally, we report asymptotic behavior of the MDF in our 3D simulations with surface tension / mobility models (i) \& (ii).
When the initial data is generated by assigning orientations at random sampled from the uniform distribution on $SO(3)$ to grains from a Voronoi construction, it is well known that the initial MDF is given by the Mackenzie distribution \cite{Mackenzie}.
With this initial data, both models (i) \& (ii) behave very similarly: Their MDFs stay in close vicinity of the Mackenzie distribution; see the first row of plots in Figure \ref{fig:MDF3d}.
This observed behavior is in close agreement with previous simulation results, e.g. \cite{HolmDimensionalEffects, HolmDimensionalEffects, GruberMCPart1}.
The very slight deviation from the Mackenzie distribution (observed in these previous studies as well) appears to be present mainly in model (ii), and not so much in (i).

We also explore the asymptotic behavior of the MDF when the initial configuration deviates substantially from the Mackenzie distribution.
To that end, we devised and implemented an algorithm to modify the randomly assigned initial orientations so that the initial MDF is perturbed substantially away from the Mackenzie distribution.
We leave the shapes of the initial grains unchanged, as the random Voronoi diagram.
To be specific, consider the function
%
\[
 \text{MDF}_\delta (x) = \sum_{i<j} \text{Area}(\Gamma_{ij}) b_\delta(x-\theta_{ij})
 \]
 for $\delta>0$ with
 \[
 b_\delta(x) = \frac{1}{\delta\sqrt{2\pi}}e^{-\frac{x^2}{2\delta^2}}.
 \]
 Without loss of generality assume that
 \[
 \sum_{i<j} \text{Area}(\Gamma_{ij}) = 1.
 \]
Then $\text{MDF}_\delta$ is a regularized approximation of the MDF of the grain network.
Let $T(x)$ denote the desired initial MDF.
Based on \eqref{eq:misorentation_angle3D}, $\text{MDF}_\delta$ can be treated as a function of $g_1,\ldots,g_N \in SO(3)$, the orientations of each grain, by expressing the misorientation angles $\theta_{ij}$ in terms of these.
We apply a steepest descent procedure on the $L^2$ norm of the difference between $\text{MDF}_\delta$ and $T$ with respect to the orientations $g_i$ of the grains; this updates the initial orientations $g_j$ until the initial MDF closely matches the desired distribution $T(x)$.
As this procedure has no preference for a particular orientation, we expect that the initial orientation distribution remains uniform.
In other words, out approach allows perturbing the MDF without introducing texture.

\begin{figure}[h]
\centering
\begin{tabular}{ccc}
\includegraphics[width=\widththreefigures]{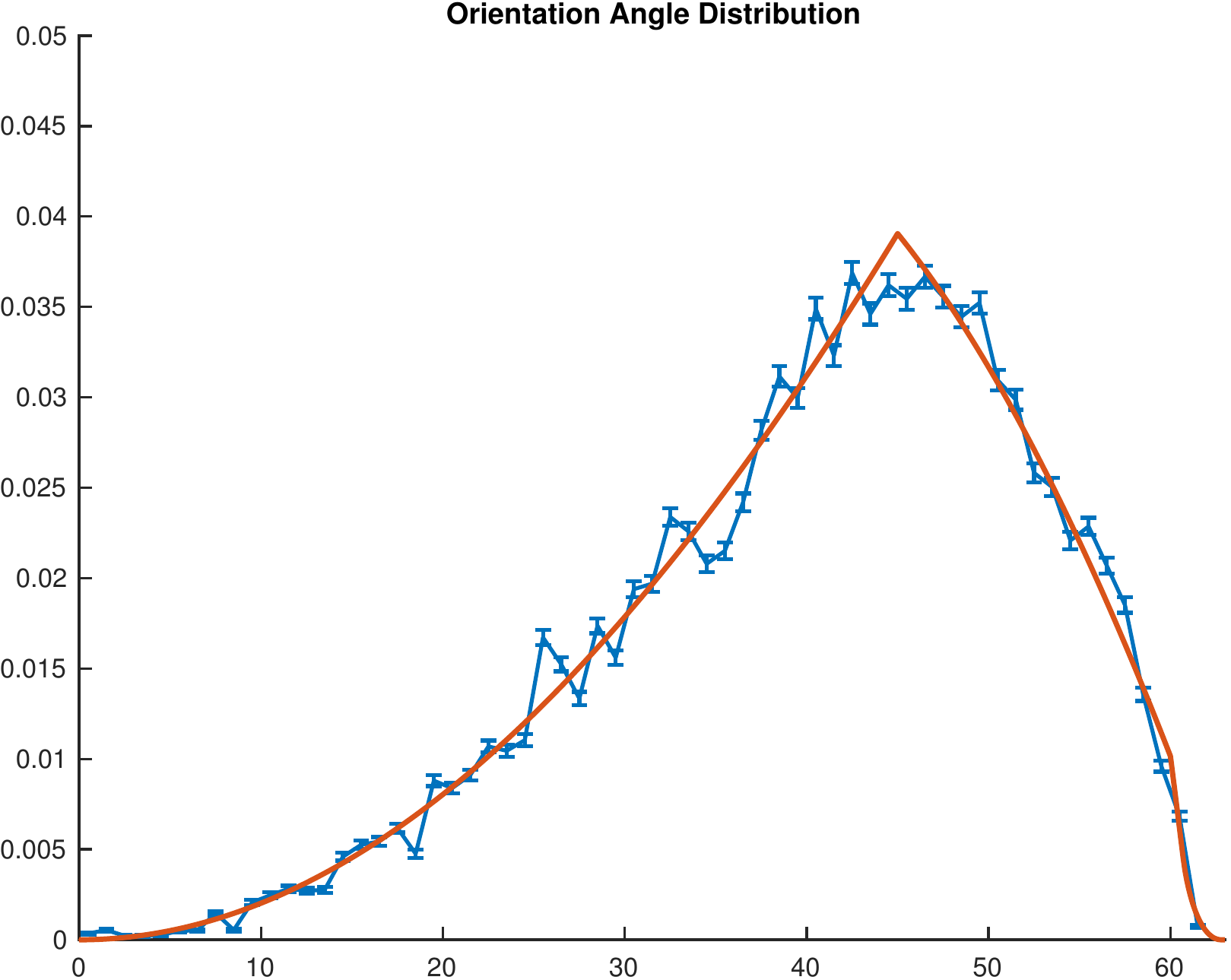} & \includegraphics[width=\widththreefigures]{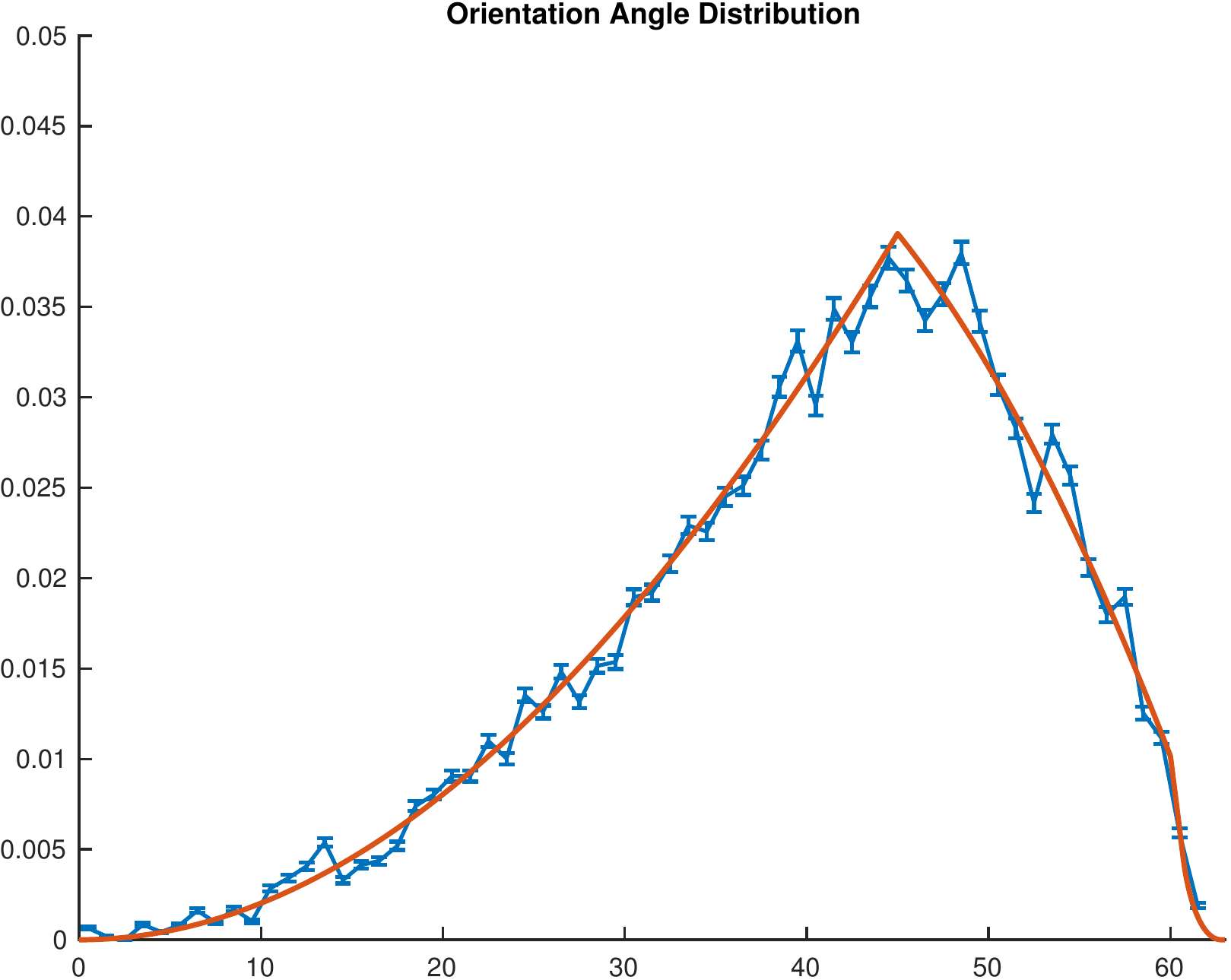} & \includegraphics[width=\widththreefigures]{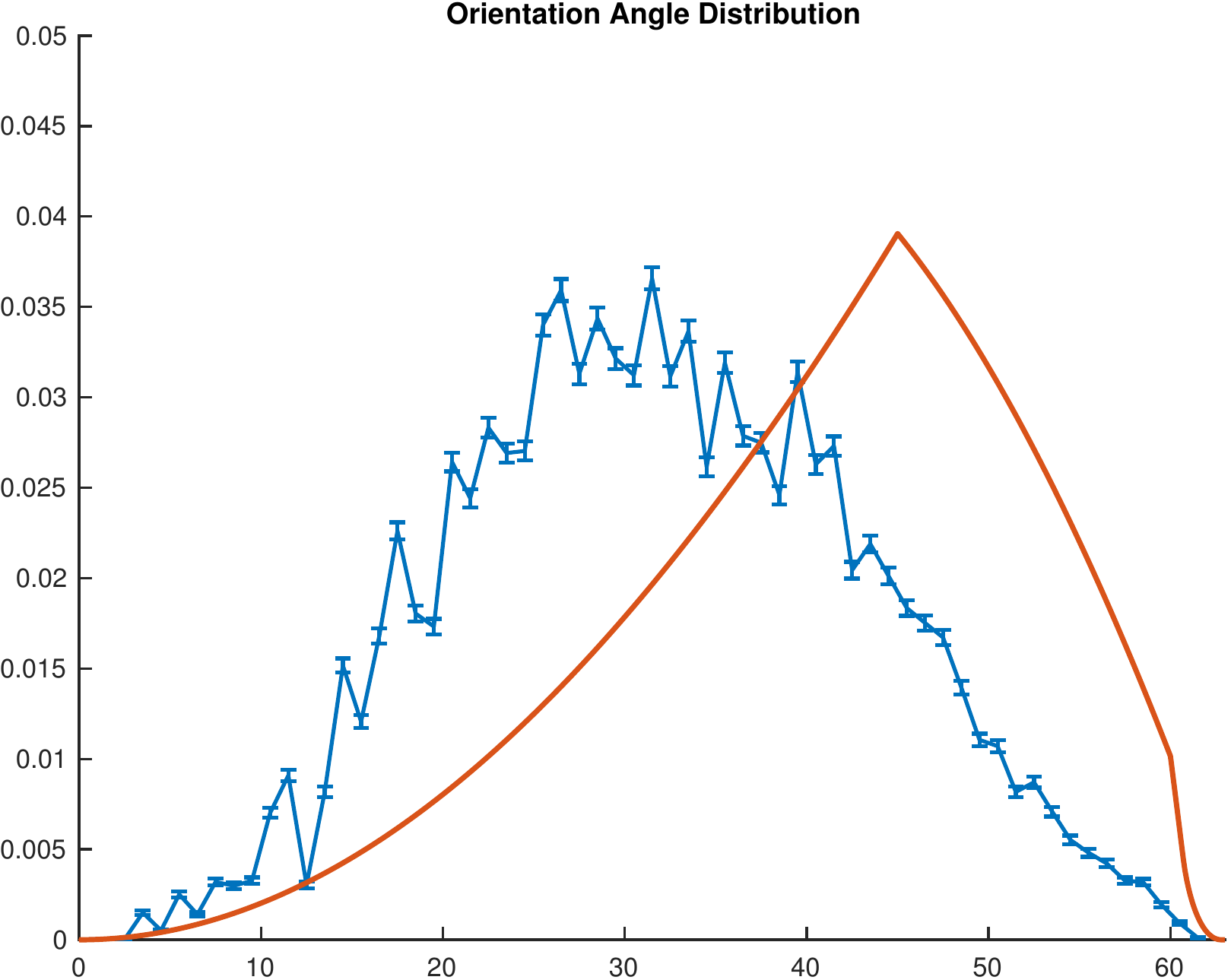}
\end{tabular}
\caption{Orientation angle distribution (probability density vs orientation angle) for the initial Voronoi data when the MDF is perturbed with the steepest descent procedure described here (left and center graphs whose corresponding MDFs displayed in the second and third row of Figure \ref{fig:MDF3d} left column, respectively) and with the method of \cite{GruberMCPart1}.}
\label{fig:ODF3d}
\end{figure}

The second and third rows of Figure \ref{fig:MDF3d} show the evolution of the MDF when it is initially perturbed away from the Mackenzie distribution by this procedure, by two different sized (and quite large) perturbations.
In each case, the MDF evolves back to a very close vicinity of the Mackenzie distribution, which is quite different from earlier results in \cite{GruberMCPart1}, where no steady state MDF is observed in simulations with initial data whose MDFs differ from the Mackenzie distribution.
We expect the difference is due to the different ways of generating non-Mackenzie initial data: The method of \cite{GruberMCPart1} introduces texture (non-uniform orientation distribution), while ours does not. This is illustrated in Figure \ref{fig:ODF3d} where we plot the orientation angle distribution. Here, by orientation angle of a grain we mean its misorientation angle when measured against the sample reference grain, the one that is completely aligned with the axes. In the presence of a random texture the orientation angle distribution will match the Mackenzie distribution.

\section{Conclusions}

We investigated, via large scale simulations, the role that surface tension and mobility models play in continuum descriptions of grain growth.
This was accomplished by applying in the context of very large numbers of grains the new, extremely simple and streamlined threshold dynamics algorithm recently introduced in \cite{SalvadorEsedogluSimplified} together with the algorithm previously introduced in \cite{SelimFelix} to perform 2D and 3D simulations.
Grain statistics, such as the GSD and the MDF, for three distinct choices of surface tension / mobility models were compared to each other and to available experimental data in existing literature.
The 2D simulations did not reveal a strong dependence of the GSD on the surface tension / mobility model used.
However, in 3D, a clear distinction emerged between the stationary GSDs of the model with anisotropic reduced mobilities and the other two models considered that had isotropic reduced mobilities.
The asymptotic behavior of the MDF was also studied, both in 2D and 3D, using 2D and 3D crystallography, respectively.
In 3D, it was observed that as long as the orientation texture of the initial data is random, even an initial MDF that is very far from the Mackenzie distribution eventually enters a very close vicinity of it.

\section{Acknowledgements}
\noindent The authors gratefully acknowledge support from the NSF grant DMS-1719727.

\bibliographystyle{amsalpha}
\bibliography{bibliography}

\end{document}